\documentclass[10pt,hidelinks]{article}

\usepackage{amssymb} 
\usepackage{amsmath} 

\usepackage[margin=1in]{geometry}

\usepackage{graphicx,ctable,booktabs}
\usepackage{verbatim}
\usepackage{enumerate}
\usepackage{float}
\usepackage{multicol}

\usepackage{epstopdf}
\usepackage{url}
\usepackage{xcolor}

\newcolumntype{P}[1]{>{\centering\arraybackslash}p{#1}}
\newcolumntype{M}[1]{>{\centering\arraybackslash}m{#1}}

\begin{document}

\title{Comparison of Immersed Boundary Simulations of Heart Valve Hemodynamics against In Vitro 4D Flow MRI Data}
\author{Alexander D. Kaiser$^{1,2,*}$, 
Nicole K. Schiavone$^{3,*}$,   
Christopher J. Elkins$^{3}$, \\ 
Doff B. McElhinney$^{1,2,4}$   
John K. Eaton$^{3}$, 
Alison L. Marsden$^{1,2,3,5,6}$}
\date{
{\small
*Authors contributed equally \\ 
$^{1}$Department of Pediatrics (Cardiology), Stanford University; \\
$^{2}$Stanford Cardiovascular Institute; \\
$^{3}$Department of Mechanical Engineering, Stanford University; \\
$^{4}$Department of Cardiothoracic Surgery, Stanford University; \\
$^{5}$Institute for Computational and Mathematical Engineering, Stanford University; \\
$^{6}$Department of Bioengineering, Stanford University \\
\vspace{10pt}
\today
}
}

\maketitle

\thispagestyle{empty}

\begin{abstract}

The immersed boundary (IB) method is a mathematical framework for fluid-structure interaction problems (FSI) that was originally developed to simulate flows around heart valves. 
Direct comparison of FSI simulations around heart valves against experimental data is challenging, however, due to the difficulty of performing robust and effective simulations, the complications of modeling a specific physical experiment, and the need to acquire experimental data that is directly comparable to simulation data. 
Such comparators are a necessary precursor for further formal validation studies of FSI simulations involving heart valves. 
In this work, we performed physical experiments of flow through a pulmonary valve in an in vitro pulse duplicator, and measured the corresponding velocity field using 4D flow MRI (4-dimensional flow magnetic resonance imaging). 
We constructed a computer model of this pulmonary artery setup, including modeling valve geometry and material properties via a technique called design-based elasticity, and simulated flow through it with the IB method. 
The simulated flow fields showed excellent qualitative agreement with experiments, excellent agreement on integral metrics, and reasonable relative error in the entire flow domain and on slices of interest. 
These results illustrate how to construct a computational model of a physical experiment for use as a comparator.

\end{abstract}

\section{Introduction}

Fluid-structure interaction (FSI) simulations of flow around heart valves have been used to study cardiac and vascular hemodynamics and disease.
Robust valve simulations are needed to fully simulate flow in the ventricles and atria or vascular flows immediately downstream of the heart in the aorta or pulmonary artery.
Well-validated and accurate valve simulation methods could be used for patient-specific modeling to optimize surgical procedures for valve replacement or repair, or to design and test new prosthetic valves in realistic, anatomical geometries.
These methods could be applied to congenital and acquired heart diseases such as tetralogy of Fallot, hypertrophic cardiomyopathy and aortic regurgitation or stenosis.
As of this writing, however, relatively few FSI studies of heart valves have directly validated their results against experimental data, which is an essential step towards integrating computational methods in clinical care. 
Thus, there is a critical need for validation of FSI simulations of heart valves that can accurately address a wide range of clinical conditions and anatomies.

Direct validation of simulations of flows around heart valves against experimental data remains challenging for three major reasons. 
First, simulations around heart valves are difficult to perform. 
To simulate valve dynamics over multiple cardiac cycles requires achieving precise balance of tension in the valve leaflets and loading pressures, as well as a numerical method that handles contact and changes in topology in the fluid domain while still allowing the leaflets to reopen on subsequent cardiac cycles \cite{kheradvar2015emergingIV}. 
Second, matching the physics of a particular experiment is complex, with myriad parameters to specify and tune including valve geometry and material properties, geometric representation of the fluid domain of interest and appropriate boundary conditions. 
Third, it is difficult to collect experimental data that are directly comparable to simulations, and many experimental techniques do not allow access to the local fluid velocity. 

Physical experiments to study flows around heart valves or to test heart valve prostheses are frequently performed in vitro in a pulse duplicator, also referred to as a flow loop. 
A pulse duplicator is an apparatus that, via pumps, tubing, reservoirs and other elements, creates flow and pressure through a chamber in which a valve is mounted. 
The pressure waveforms and flow rates generated typically mimic the loading and flow conditions of the living heart. 
Such a setup, however, is considerably more controlled than the beating heart of a living person. 
This provides an opportunity to validate the FSI simulations for flow around heart valves by simulating a physical experiment in a pulse duplicator. 

A variety of FSI methods have been used to study heart valve hemodynamics. 
The immersed boundary (IB) method is a framework for fluid-structure interaction simulations \cite{ib_acta_numerica} that was originally developed to study flow patterns around heart valves \cite{PESKIN1972252}. 
The IB method has been used repeatedly for studies of valves \cite{kaiser2019modeling,Griffith_aortic,MA2013417,gao_chordae}, simulations of cardiac flows \cite{mcqueen2001heart,mcqueen2015constructing}, studies of prosthetic designs \cite{watton2007dynamic}, and flows through  bicuspid aortic valves \cite{kaiser2022controlled}.
Other highly-effective approaches to simulating heart valve hemodynamics include the immersogeometric extension to the IB method \cite{hsu2015dynamic}, smooth particle hydrodynamics \cite{Toma2016,sun_left_heart}, fictitious domain methods \cite{astorino2009fluid,shadden2010computational} and lattice Boltzmann methods \cite{yun2014highly}.

Existing studies have compared heart valve simulations against experimental data, but typically without direct comparisons to the fluid velocity field. 
A sequence of studies used the IB method (within the same solver framework that we use in this work) with a finite element formulation for the aortic valve and modeled flow in an in vitro pulse duplicator \cite{lee2020fluid,lee2021bioprosthetic}. 
These studies achieved excellent qualitative and quantitative agreement in leaflet kinematics and flow and pressure waveforms. 
Their flow loop setup did not allow for pointwise measurement of the velocity field, however, and thus they did not make direct comparisons of flow fields. 
Watton et al. achieved good qualitative agreement on leaflet motion and quantitative agreement on comparisons with quantities such as forward pressure difference across the valve, but prescribed flow rates and so did not fully simulate the force balance of valve closure \cite{watton2007dynamic}.
Sig{\"u}enza et al. simulated flow through the aortic valve and compared to velocity fields measured with particle image velocimetry and achieved some qualitative and quantitative agreement, but used isotropic leaflet material properties and prescribed flow rates, again not fully simulating closure \cite{siguenza2018fluid}. 
For flows involving structures simpler than a heart valve, the IB method has been successfully validated against experimental data on integrated quantities for flows around a cylinder \cite{lai2000immersed,kim2007penalty}. 
Using a partitioned approach to FSI, Emendi et al. achieved good results on qualitative comparisons with 4D flow MRI (4-dimensional flow magnetic resonance imaging) data taken in vivo \cite{emendi2021patient}.

In this work, we directly compared immersed boundary simulations of flow through the pulmonary valve with experimental velocities measured from a physical pulse duplicator using 4D flow MRI, also known as magnetic resonance velocimetry. 
Data collected via 4D flow MRI have substantial advantages over other techniques for measuring flow: it provides high-resolution, three-dimensional, three-component, time-resolved measurements of the fluid velocity field on a Cartesian grid over the entire relevant domain. 
Other techniques such as particle image velocimetry are typically limited to a single plane and may not give information about the out-of-plane components of velocity.  
In our experimental setup, a commercially available, bioprosthetic aortic/pulmonary valve, which was constructed from a porcine aortic valve mounted to a scaffold and comprised of three thin, flexible leaflets, was mounted into a 3D printed model of the pulmonary artery. 
Next, we simulated flow with the IB method in an analogous setup, constructing the model pulmonary valve using methods we recently developed \cite{kaiser2019modeling,kaiser2020designbased,kaiser2022controlled}. 
We then compared results from the physical experiment and IB simulations.
Simulations showed excellent qualitative and acceptable quantitative agreement with experimental results. 

For future use by other researchers, our experimental datasets will be made publicly available for use in comparison and validation studies via \url{http://vascularmodel.com/additionaldata.html}, and code for model generation and FSI simulations is available at \url{http://github.com/alexkaiser/heart_valves}.

\section{Experimental methods}
\label{Experimental_methods}

\subsection{4D Flow Magnetic Resonance Imaging}

\begin{figure*}
    \centering
    \includegraphics[width=0.95\textwidth]{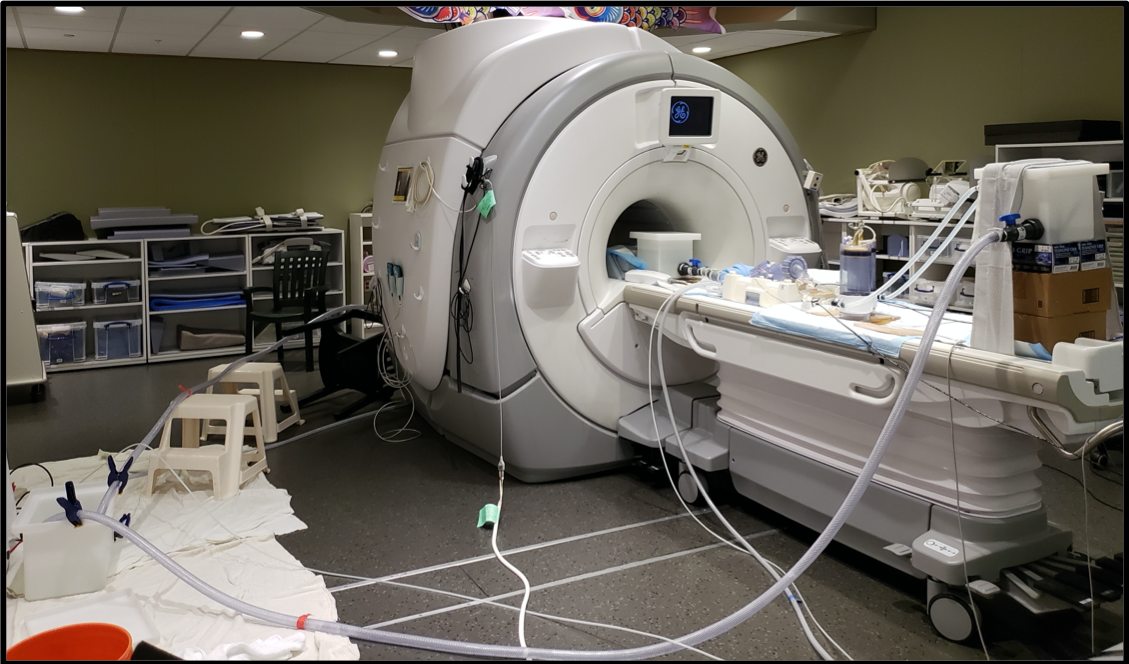}
    \caption{Experimental flow loop setup on the MRI magnet table in the Richard M. Lucas Center for Imaging at Stanford University.} 
    \label{fig-magnetroom}
\end{figure*}

The experimental data in this study were obtained using the 4D flow MRI sequence developed by Markl et al. \cite{markl2003time}, which produced full 3D, three component, phase-averaged velocity fields in a physical model of the right ventricular outflow tract (RVOT) and pulmonary arteries. 
The experiments were performed in a whole body, 3 Tesla General Electric MRI scanner at the Richard M. Lucas Center for Imaging at Stanford University (Figure \ref{fig-magnetroom}), and these data were discussed in a previous publication \cite{schiavone2021vitro}. 
The gating signal for the 4D flow sequence was provided by the data acquisition system that was also used to control the ventricle pump. 
The cardiac cycle duration was 0.8325 s or approximately 72 beats per minute. 
Based on a retrospective gating scheme, the velocity data were acquired with a temporal resolution of approximately 83 ms, providing 10 time phases for the final data. 
The velocity at a point in the flow domain for a given phase represents the mean velocity during the time interval of that phase. 
Therefore, the velocity for each phase represents the  phase-average of the velocity during an 83 ms interval during the cardiac cycle. 
This phase average is constructed from the flow fields of over hundreds of cardiac cycles and is therefore expected to smooth out small variations between cycles. 
The 4D flow MRI scans used 0.9 mm thick sagittal slices and spatial resolutions of 0.894 mm and 0.897 mm in the coronal and axial directions, respectively.

Multiple 4D flow MRI scans were conducted and averaged to increase signal strength and reduce noise. Additionally, pump-off scans were averaged and subtracted from the pump-on scans to reduce various sources of error, including eddy current effects \cite{elkins_magnetic_2007}. To account for drift in the eddy currents and other signal properties, pump-on and pump-off scans were alternated. All of the pump-on scans, with the pump-off data already subtracted, were then averaged to comprise the final dataset. Four pump-on scans and three pump-off scans were measured, with each individual scan taking approximately 20 minutes. This 4D flow MRI procedure results in time-resolved, three-component, phase-averaged velocity data over the entire 3D test section volume.
The following references have further background on MRI \cite{elkins_4d_2003,pelc1994quantitative,pelc1991phase}.

The expected uncertainty of 4D flow MRI velocity measurements is given by 
\begin{equation}
\delta_u = \frac{\sqrt{2}}{\pi} \frac{\mathrm{VENC}}{\mathrm{SNR}}\,,
\end{equation}
where SNR is the signal to noise ratio and VENC is the maximum velocity that can be measured by the 4D flow MRI sequence, known as the velocity encoding value \cite{elkins_4d_2003}. The VENC was set at 250 cm/s for all runs. The signal to noise ratio was calculated as the ratio of the signal in the flow region to the signal in the solid wall, as done by Banko et al. \cite{banko_three-dimensional_2015}. The SNR was 23.7, resulting in an uncertainty of \textpm 4.7 cm/s throughout the domain.

Note that this uncertainty was calculated using the SNR for the entire experimental dataset, which was comprised of averages of multiple 4D flow MRI scans and many cardiac cycles over approximately four hours of total scan time. 
There were sources of variation during the experiment that could produce different flow conditions cycle to cycle, including drift in the kinematics of the centrifugal and ventricular pumps in the flow loop and variations in the air pressure pneumatically driving the flow. 
The flow loop was observed during the MRI sequence and small manual adjustments were made to maintain the flow rate of 3.5 L/min. 
We observed that the cycle-to-cycle variations in the volumetric flow rate were relatively small and did not substantially contribute to the overall uncertainty in the velocity.

\subsection{RVOT Geometry Design}
\label{geometrydesign}
The anatomic physical model used in this work was designed to analyze the flow fields local to bioprosthetic valves in a healthy RVOT anatomy. 
The geometry modeled a 11-13 year-old pediatric patient with cardiac output of 3.5 L/min \cite{schiavone2021vitro}. 
The model extended from the outlet of the right ventricle (RV), through the main pulmonary artery (MPA) to the left pulmonary artery (LPA) and right pulmonary artery (RPA) branches after the first bifurcation. The healthy geometry design was based on measurements made from MRI data of the right heart and pulmonary arteries of six healthy subjects between ages 11 and 13. MRI data were collected for clinical purposes and used for modeling under an IRB-approved protocol. For each patient, we measured the diameters of the MPA, LPA, and RPA in the sagittal and axial planes. The measurement location was at the valve annulus while the locations for the LPA and RPA were immediately downstream of the bifurcation.  We measured the turning angle, radius of curvature, and arc length of the RVOT and MPA in both sagittal and axial planes for each subject. The other key parameters were the three angles at the bifurcation: LPA to RPA, MPA to LPA, and MPA to RPA. Measurements were done in ImageJ (Bethesda, MD).

Median values for all subjects of the diameters, lengths, radii of curvature, and bifurcation angles were used to construct the healthy pulmonary geometry. Median values were chosen to reduce the effect of outliers in the patient measurements, though differences between median and mean were generally small. The model design was scaled to fit the 25mm bioprosthetic valve used in this experiment, resulting in an MPA diameter of 25mm, LPA diameter of 12mm, and RPA diameter of 14mm. All scaled values fell within the range of measured values from the patient cohort, resulting in a scaled model that accurately represented the 11-13 year-old patient population.

\begin{figure*}[t!]
    \centering
    \includegraphics[width=0.95\textwidth]{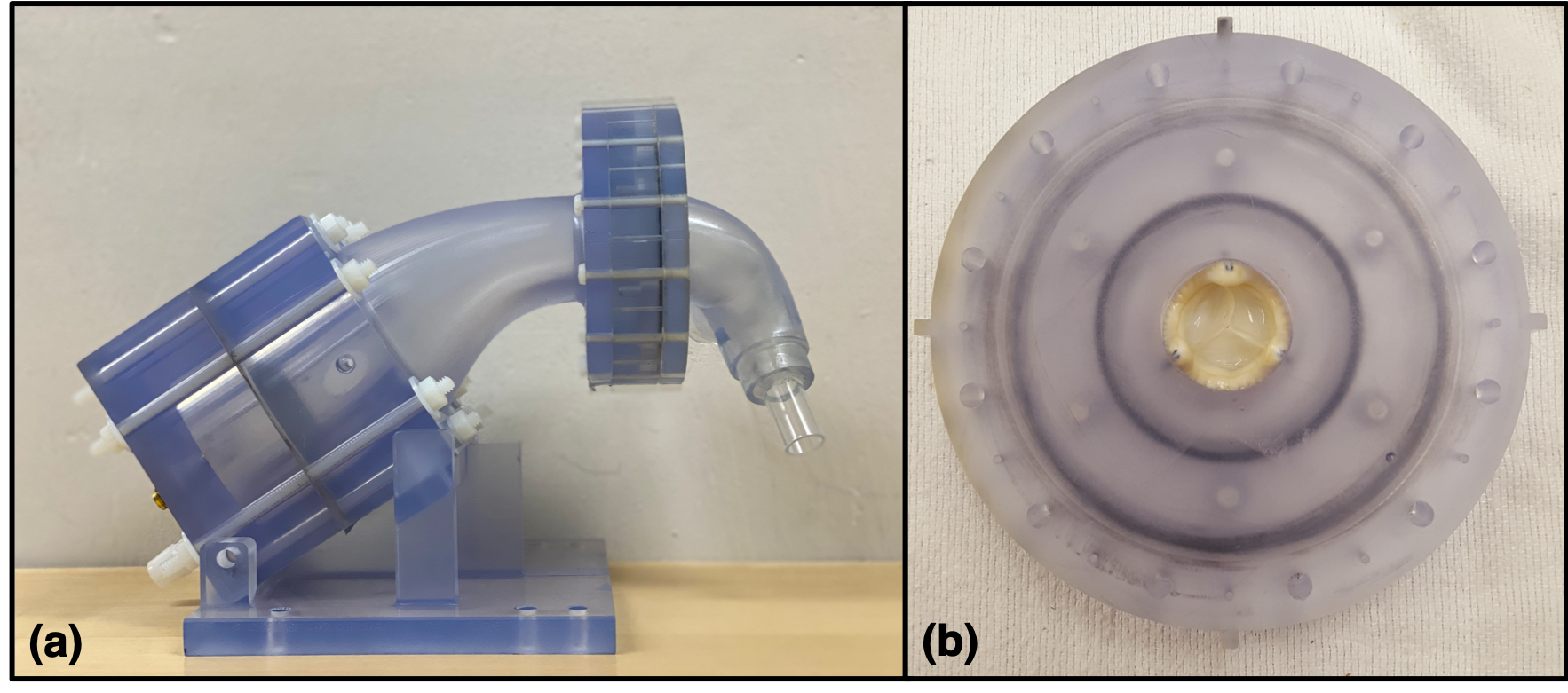}
    \caption{3D printed healthy RVOT model in full assembly (a) and with the valve clamped in the valve-holder component (b).} 
    \label{fig-healthymodelprinted}
\end{figure*}

The model was designed in SolidWorks (Waltham, MA) and was manufactured using stereolithography (SLA) with DSM Somos WaterShed XC 11122 resin at the W.M. Keck Center at the University of Texas, El Paso. All components of the model and the valve itself are fully MR compatible. The 3D printed healthy RVOT model is shown in Figure \ref{fig-healthymodelprinted}a, with all components fully assembled.

The valve used in this experiment was a 25mm Epic valve (Abbott Cardiovascular, Plymouth, MN), which is a porcine bioprosthetic trileaflet aortic surgical valve, which we placed in the pulmonary position. 
The valve includes support scaffolding at the commissures and a sewing ring that is used to suture the valve to the RVOT vessel in the patient. The valve was implanted into the model using a two-piece component valve-holder piece with a groove that matched the dimensions of the sewing ring on the 25mm Epic valve, which has an estimated internal diameter of 20 mm, as measured via a caliper. This groove was clamped around the sewing ring with the two pieces screwed together to secure the valve (Figure \ref{fig-healthymodelprinted}b).

\subsection{Physiological Flow Loop}
\label{flowloop}

\begin{figure*}[t!]
    \centering
    \includegraphics[width=0.95\textwidth]{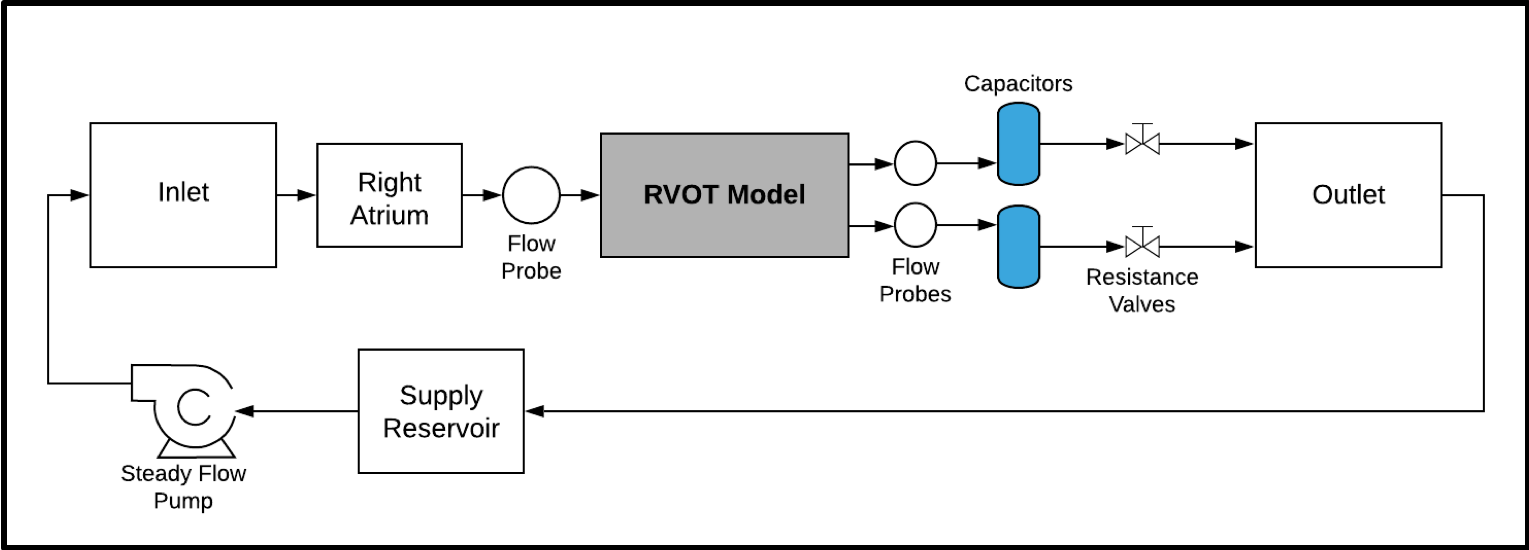}
    \caption{Schematic of the experimental flow loop.} 
    \label{fig-flowschematic}
\end{figure*}

We developed an in vitro flow loop experimental setup that replicates physiological conditions over the cardiac cycle for the pediatric pulmonary system (Figure \ref{fig-flowschematic}). The pulsatile flow loop was driven by a custom designed ventricle box that was inspired by pulsatile ventricular assist devices, such as the Berlin Heart EXCOR (The Woodlands, TX). A thin membrane in the ventricle box was pneumatically driven by a pulsatile air supply to create systole and diastole in the model and the working fluid. The digital trigger signal governing the pneumatic ventricle box was connected to the electrocardiogram (ECG) converter and trigger on the MRI system to signal the start of each cardiac cycle, allowing us to collect gated, phase-locked data.

The ventricle box was the key component of the RVOT model that drives the full flow loop to replicate right heart circulation and pulmonary physiology (Figure \ref{fig-flowschematic}). 
The flow loop included an upstream inlet tank fed by a steady centrifugal pump from a supply reservoir that connected to a flexible bag that acted as a right atrium. 
This atrium proxy fed the ventricle box through a 27mm bioprosthetic tricuspid valve. 
The ventricle box connected to the main printed components of the healthy RVOT geometry, including the component that housed the bioprosthetic valve. 
The outlets of the 3D printed model connected to flexible tubing, which in turn connected to capacitor elements. 
Each capacitor element exited into another segment of flexible tubing which ran to the outlet tank of the flow loop.

All of the experimental flow loop components were tuned so that the pressure and flow rate waveforms reasonably matched physiological conditions. The pressures were measured with Millar (Houston, Texas) SPR-350 pressure transducer catheters inserted in the model through sealed ports. The RV pressure was measured directly upstream of the valve while the MPA pressure was taken immediately downstream of the valve; both catheters were centered in the vessel. The branch flow measurement was collected using a Transonic Systems (Ithaca, NY) ultrasonic flow probe placed around each branch PA. The locations of these flow probes, and the one placed upstream of the model to monitor the cardiac output are seen in Figure \ref{fig-flowschematic}. 
The pressure waveform was averaged over approximately ten cardiac cycles. 

The working fluid for all experiments was a blood analog consisting of 60\% water and 40\% glycerin, with a density of 1.09 g/cm\textsuperscript{3} and a viscosity of 3.9 centipoise at room temperature in the magnet room. Gadolinium was added to the blood-analog fluid to increase the signal contrast during the 4D flow MRI scan. We conducted the experiment at a heart rate of 72 beats per minute and targeted a 50-50 split for flow in the LPA and RPA. Additional details about the development of the physiological flow loop and experimental model and methods can be found in Schiavone et al. \cite{schiavone2021vitro}.

\section{Simulation methods}

\subsection{Construction of the model pulmonary valve}
\label{construction_valve}

We constructed the model valves using a design-based approach to elasticity, which we introduced in our previous studies  \cite{kaiser2019modeling,kaiser2020designbased}. 
With this method, we specified that the valve leaflets must, via tension in the leaflets, support a pressure. 
From this specification, we derived and solved a nonlinear partial differential equation for mechanical equilibrium of the leaflets under the prescribed load. 
By tuning free parameters and boundary conditions for these differential equations, we designed a closed configuration of the valve, including the tensions in the loaded configuration. 
From this closed configuration, we then derived the reference configuration and material properties of the model valve. 
This model was fiber based, in that the structure was discretized as a system of one-dimensional curves that fill a region in space. 
Alternatively, the model may be viewed as a system of nonlinear springs that includes a continuum limit. 
These methods were found to be highly effective under physiological pressures, opening and closing repeatedly over multiple cardiac cycles and achieving realistic flow rates with physiological driving pressures.

To design the model valve, we represented the leaflet as an unknown parametric surface 
\begin{align}  
\mathbf X(u,v) : \Omega \subset \mathbb{R}^{2} \to \mathbb{R}^{3} . 
\end{align}
We assumed that the parameters conform to the fiber and cross fiber directions of the leaflets, meaning that for curves of constant $v$ on which $u$ varies represent the fibers, and curves of constant $u$ on which $v$ varies represent curves in the cross-fiber direction.  
Let single bars $| \cdot |$ denote the Euclidean norm. 
We denoted the unit tangents to the surface as 
\begin{align}
\frac{\mathbf X_{u}}{ | \mathbf X_{u} |} \quad \text{ and } \quad \frac{\mathbf X_{v}}{ | \mathbf X_{v} |}, 
\end{align}
which represented the local fiber and cross-fiber directions, respectively. 
These directions were not prescribed or required to be orthogonal. 
We denoted the magnitude of local membrane tensions in the circumferential and radial directions as $S$ and $T$, respectively.

We then prescribed a uniform pressure load of $p = 30$ mmHg, the approximate diastolic pressure load in the experimental data immediately following the closing transient. 
We then considered the mechanical equilibrium on an arbitrary patch of leaflet $[u_{1}, u_{2}] \times [v_{1}, v_{2}]$. 
This requirement gives the following integral balance of pressure and leaflet tensions: 
\begin{align} 
&0 = \int_{v_{1}}^{v_{2}}   \int_{u_{1}}^{u_{2}}    p  \left(  \mathbf X_{u}(u,v) \times \mathbf X_{v}(u,v) \right)    du dv  \\ 
&\hspace{-3pt}+ \int_{v_{1}}^{v_{2}}  \left(  S(u_{2}, v) \frac{ \mathbf X_{u} (u_{2}, v) }{|  \mathbf X_{u} (u_{2}, v) |} - S(u_{1},v)   \frac{ \mathbf X_{u} (u_{1}, v) }{|  \mathbf X_{u} (u_{1}, v) |}   \right)    dv \nonumber  \\ 
&\hspace{-3pt}+ \int_{u_{1}}^{u_{2}} \left(  T(u, v_{2})  \frac{\mathbf X_{v} (u, v_{2})}{|\mathbf X_{v} (u, v_{2})|} - T(u,v_{1})  \frac{\mathbf X_{v} (u, v_{1}) }{|\mathbf X_{v} (u, v_{1})|} \right)  du . \nonumber
\end{align}
Then, we applied the fundamental theorem of calculus to convert all single integrals to double integrals over the patch and swapped the order of integration formally as needed to obtain
\begin{align} 
0 = &\int_{v_{1}}^{v_{2}}  \int_{u_{1}}^{u_{2}} \bigg(   p  (  \mathbf X_{u} \times \mathbf X_{v} )    
+  \frac{\partial}{\partial u}  \left( S \frac{ \mathbf X_{u} }{ |\mathbf X_{u}| } \right) +  \frac{\partial}{\partial v}  \left( T \frac{ \mathbf X_{v} }{|\mathbf X_{v}|} \right)    \bigg)  \;  du dv . 
\end{align}
This equation represents the integrated form of the equations of equilibrium. 
The patch is arbitrary, and so, formally assuming sufficient smoothness, the integrand must be zero. 
This gave the following partial differential equation form of the equations of equilibrium: 
\begin{align} 
0 = p  (  \mathbf X_{u} \times \mathbf X_{v} )  +   \frac{\partial}{\partial u}  \left( S \frac{ \mathbf X_{u} }{ |\mathbf X_{u}| } \right)  +  \frac{\partial}{\partial v}  \left( T \frac{ \mathbf X_{v} }{|\mathbf X_{v}|} \right).    
\label{eq_eqns}
\end{align}
As written, this system of differential equations is not closed, as it has five unknowns, the three components of $\mathbf X$ and the two tensions $S,T$ and only three equations, one for each component. 
To close the system, we temporarily specified that
\begin{align}
S(u,v)  = \alpha \left( 1 - \frac{1}{1 + |\mathbf X_{u}|^{2} / a^{2} } \right), \quad T(u,v)  = \beta \left( 1 - \frac{1}{1 + |\mathbf X_{v}|^{2} / b^{2} } \right). \label{dec_tension} \end{align}
This functional form allowed the solution of the differential equation to find heterogeneous tensions to support the prescribed pressure load, while preventing extreme local tensions from occurring. 
The parameters $\alpha,\beta,a,b$ are tunable free parameters that we set to match the gross morphology of the prosthetic valve. 
The parameters $\alpha,\beta$ represent the maximum allowable tension in the circumferential and radial directions. 
The parameters $a,b$, which have units of cm, were tuned to control the gross morphology of the leaflets.

To tune these parameters and design a valve that corresponded to the prosthetic valve, we measured its gross morphology with a ruler and caliper. 
The free edge length of each leaflet was approximately 2.2 cm. 
We estimate the leaflet height at 1.3 cm, but this was highly dependent on loading; even pulling the leaflets flat along the ruler changed lengths substantially. 
The internal radius of leaflet attachment to the annulus was 1.0 cm and annular height was 1.1 cm.  
To match these values, we set the parameters $\alpha,\beta$ to $1.3 \cdot 10^7$ and $4.6 \cdot 10^5$ dynes, respectively.  
The value of $a$ varied linearly from $42$ cm at the annulus to $113$ cm at the free edge. 
Taking a variable value of $a$ was necessary to avoid excessive billowing of the leaflet near the annulus while maintaining adequate free edge length. 
The value of $b$ was set to $140$ cm.

We discretized the system of equations \eqref{eq_eqns} including the tension formulas \eqref{dec_tension} with centered finite differences. 
The position of the valve at the annulus and commissures was prescribed as a Dirichlet boundary condition. 
At the free edge, we prescribe zero-tension (homogeneous Neumann) boundary conditions. 
The resulting nonlinear system of equations was solved with Newton's method with line search. 
Additionally, we applied the method of continuations on pressure, in which pressure is adaptively increased from an initial value of zero to the prescribed value of $p = 30$ mmHg, and the solution with the previous pressure is used as the initial guess for Newton's method with the subsequent pressure.

We then used the solution to equations \eqref{eq_eqns}, which represents the predicted loaded configuration of the valve, to derive a reference configuration and constitutive law.  
For each link in the discretized model, the solution included the loaded length and the tension needed to support the prescribed pressure load. 
Based on the experiments of Yap et al. \cite{yap2009dynamic}, we prescribed uniform stretch ratios of $\lambda_{c} = 1.15$ circumferentially and $\lambda_{r}$ = 1.54 radially, then used this information to compute the reference length for each link from its loaded length. 
For a link with length $L$, we then solved $\lambda = L/R $ for the rest length $R$ at the stretch $\lambda$ corresponding to the direction of the link.  
We took the tension/stretch relationship to be exponential with a zero at $\lambda = 1$, and based the shape, but not the local stiffness, on the strip biaxial tests of May-Newman et al. \cite{may2009hyperelastic}.
Based on a nonlinear least squares fit to their data, we took the exponential rate to be $\eta_{c} = 57.46$ circumferentially and $\eta_{r} = 22.40$ radially.  
For each link in the discretized, predicted loaded configuration with tension $\tau$ and the appropriate exponential rate and stretch for the circumferential or radial direction, we solved 
\begin{align}
\tau = \kappa (e^{\eta (\lambda - 1)} - 1)
\end{align}
for the local stiffness coefficient $\kappa$. 
Since the solution to equations \eqref{eq_eqns} includes heterogeneous tensions, this resulted in heterogeneous material properties in the leaflets.

To generate an initial configuration suitable for FSI simulations, we sought a configuration of the leaflets that is open and free from external loading. 
Using the constitutive law we just set to determine tensions, we again solved the equilibrium equations \eqref{eq_eqns} with $p = 0$ mmHg. 
On the physical valve, a small portion of the leaflets point was attached radially inward from the commissures. 
To model this geometric feature, we force one eighth of the free edge points starting at each commissure of each leaflet to coincide (approximately 1/4 of each leaflet total), thus pinching the leaflets together. 
The remainder of the free edge was fixed as a Dirichlet boundary condition to ensure the leaflets do not self-intersect. 
This model has pre- or residual stretch and tension in this configuration, though substantially less than the predicted loaded configuration. 
Further, a configuration with zero tension, given the reference lengths that are computed locally for each link in the discretized model, may not exist.

Until this point, we used a membrane formulation of the leaflets. 
Next, we moved to a thickened formulation, while still using a fiber-based material model. 
To achieve a realistic thickness for the valve, we extruded the membrane for the valve normally in each direction, to form three adjacent membranes with a total thickness of $0.44$ mm, as reported in Sahasakul et al. \cite{sahasakul1988age}. 
This also served to mitigate the ``grid aligned artifact'' associated with large pressure differences across zero-thickness membranes in the IB method \cite{kaiser2019modeling}.
The stiffness of coefficients for each layer were set to be one third of the membrane stiffness coefficients computed previously. 
Linear springs of rest length $0.22$ mm were placed to keep the layers together through the simulation. 

The gross morphology of the model valve that emerged from this process is shown in Figure \ref{vessel_valve}. 
The free edge was 2.87 cm, corresponding to 3.3 cm in the predicted loaded configuration. 
After the pinching the leaflets together at the commissures, this left approximately 2.1 cm of free edge rest length per leaflet free to move independently of the other leaflets, within measurement error of \textpm 0.1 cm from the free edge length of 2.2 cm measured on the prostheses. 
The leaflet rest height was 0.94 cm corresponding to a predicted loaded height of 1.44 cm. 
The measured leaflet height of 1.3 cm is nearly the predicted loaded height of 1.44 cm, which may be because the leaflets are so compliant in the radial direction, that pulling them flat to measure them achieved substantial stretches. 
  The fiber orientation of the model runs from commissure to commissure and qualitatively matches experimental observations \cite{SAUREN198097}, though direct quantitative comparison is beyond the scope of this work.
One minor limitation is that we do not add bending rigidity to the leaflets, beyond what emerges from the thickening process described above, and thus may not accurately capture leaflet flutter or other similar behaviors. 
Based on the thickness of $0.44$ mm, we estimated the mean tangent modulus at the predicted loaded stretches as $6.7 \cdot 10^{7}$ dynes/cm$^{2}$ circumferentially and $8.5 \cdot 10^{4}$ dynes/cm$^{2}$ radially.
The prosthetic valve tissue is fixed in glutaraldehyde, and literature values for the fully-recruited circumferential tangent modulus of fixed porcine aortic valve tissue vary widely. 
Based on the experimental measurements of Billiar and Sacks and their constitutive law for valves fixed under 4 mmHg of pressure, 
we evaluated their constitutive law at the relevant stretches $\lambda_{r}$ and $\lambda_{c}$ and estimated the circumferential tangent modulus to be $4.0 \cdot 10^{7}$ dynes/cm$^{2}$ \cite{billiar2000biaxial2}. 
Rousseau et al. reported moduli ranging from $5.1 \cdot 10^{7}$ to $2.3 \cdot 10^{8}$ dynes/cm$^{2}$, depending on the applied preload during fixation \cite{rousseau1983elastic}. 
Sung et al. reported moduli ranging from $3.0 \cdot 10^{8}$ to $5.2 \cdot 10^{8}$ dynes/cm$^{2}$, depending on fixation pressure   \cite{sung1999mechanical}. 
Thus our estimated tangent modulus falls within the range of existing studies, so we considered our resultant modulus in good agreement given the complexity of the steps involved, phenomenological nature of the constitutive law and uncertainties in experiments. 
We do not have access to the precise material properties of the prosthetic valve, and further, the only literature we could find on the material properties of a similar prostheses reported the tangent modulus at one particular loading, which did not appear to be at a relevant stretch for comparisons with our model \cite{kalejs2009st}. 
Thus, our model has material properties in a reasonable range for a fixed aortic valve prostheses (placed in the pulmonary position in our simulations), but it does not directly model the material properties of the prostheses.

These steps completed the derivation and design of the model valve. 
We then used the model valve and constitutive law described throughout this section as initial conditions for fluid-structure interaction simulations.

\subsection{Construction of the vessel and scaffold}

\begin{figure*}[t!]
\centering
\setlength{\tabcolsep}{2pt}
\begin{tabular}[t]{c c}
\includegraphics[width=.55\textwidth]{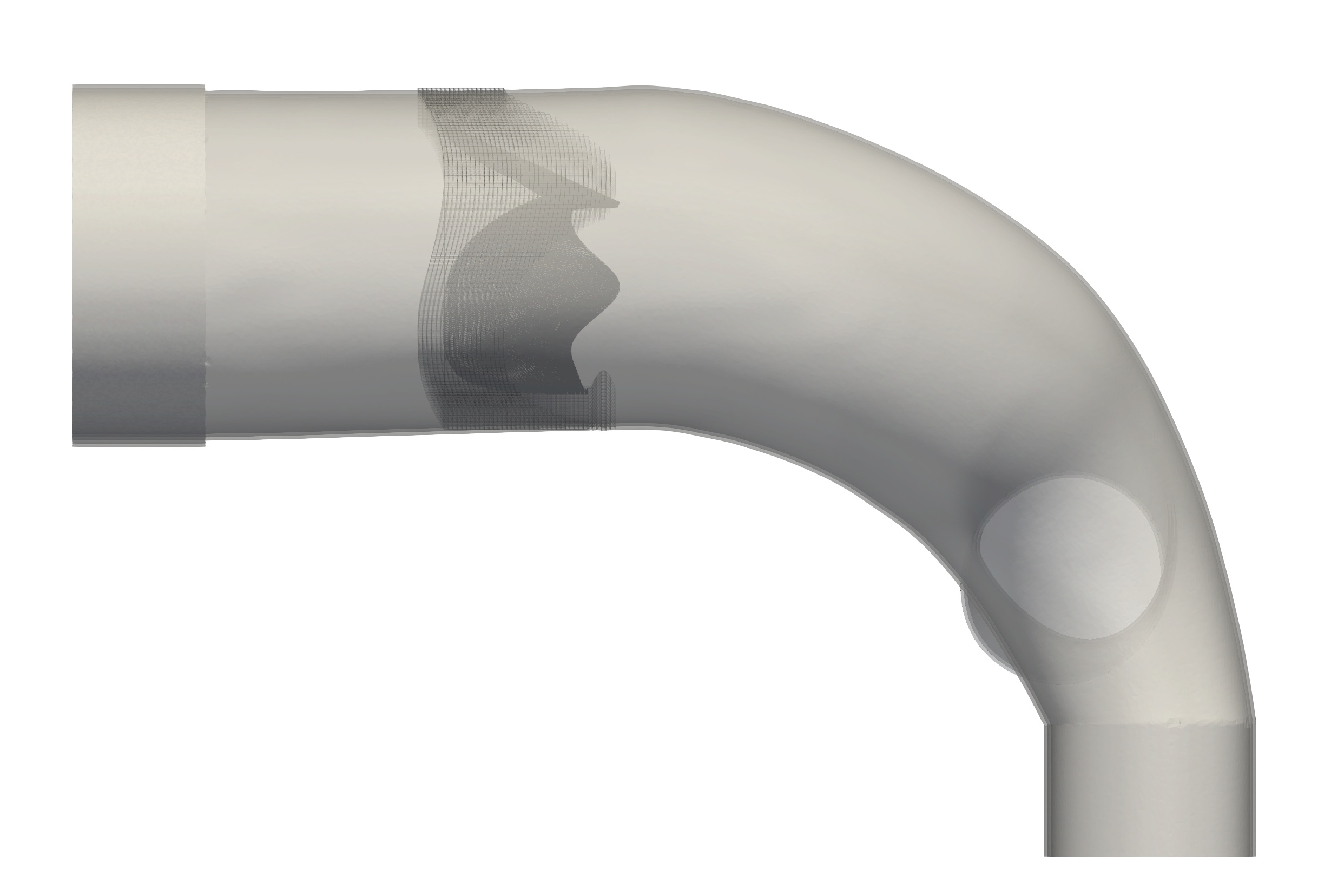} & 
\includegraphics[width=.3\textwidth]{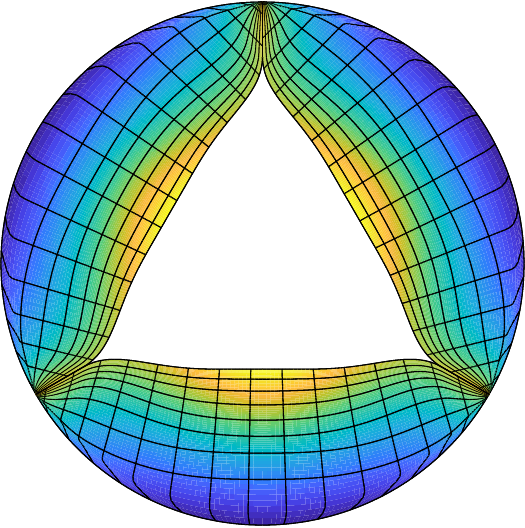}  
\end{tabular}
\caption{ The model vessel, with the scaffold and valve in place (left).
The model pulmonary valve under zero pressure, as used to construct initial conditions for FSI simulations (right). } 
\label{vessel_valve}
\end{figure*}

We constructed the model vessel for FSI simulations from data from the MRI scans (Figure \ref{vessel_valve}).  
The signal magnitude of 3D printed model material is distinct from the signal of the fluid in the scans, and we applied a thresholding operation to generate a three-dimensional model of the printed vessel surface. 
Using the MRI data ensured that the MRI and simulation coordinates were consistent in space and there were no alignment or registration errors. 
While using the files that generated the 3D printed model would have offered more spatial fidelity, the potential error in flow fields due to any mis-registration would have likely been much more substantial. 
Using Meshmixer (San Rafael, CA), we smoothed the mesh to remove stair-step effects and removed artifacts from the valve scaffold. 
We then remeshed to the desired edge length of 0.25 mm and extruded the model 0.25 mm and 0.5 mm to create a three-layer structure. 
As in the valve, this serves to eliminate the ``grid aligned artifact'' that can occur with pressure differences across thin membranes in the IB method \cite{kaiser2019modeling}.
Flow extenders of length 1 cm were added to the vessel at the inlet and both outlets to ensure that the normal to the vessel was aligned with the normal of the fluid box at the inlets and outlets. 
In FSI simulations, the vessel was held in place using target points, stiff springs of zero rest length that connect the current position of each model node to its desired position (Section \ref{fsi}). 
Additional linear springs are placed on each edge in the triangulated model. 
These springs are not meant to model a particular material and only serve to keep the vessel rigid and stationary throughout the simulation. 

The valve leaflets are mounted to a scaffold, which we measured with a ruler and caliper as with the leaflets. 
The bottom of the scaffold has variable height with amplitude 0.2 cm, with the highest points aligned with the commissures, and we prescribed the bottom curve to take the functional form $0.2 \left( \frac{1}{2}(\cos(3\theta)+1 \right)$.
The top of the scaffold has minimum height $h_{min} = 0.65$ cm and maximum height $h_{max} = $1.5 cm aligned with the commissures, measured with the valve resting on a table. 
We prescribed the top of the scaffold to take the form $h_{min} + h_{max} | \cos \left(\frac{3}{2} \theta \right) |^p$, where $p$ is a power to be determined.
We measured the scaffold height at two additional locations, then estimated  $p = 3.1$ with a least squares fit. 
From its minimum radius of 1 cm, the radius of leaflet attachment, we extended the scaffold 0.3 cm radially to form a subset of a cylinder. 
This thickness, slightly larger than the real scaffold thickness of 0.25 cm, ensured the scaffold overlapped with the vessel and prevented leaks between the vessel and scaffold. 
Finally, a small nub of thickness 0.15 cm and height 0.2 cm was placed inward radially at the commissure tip to model the bulging of the scaffold at that location. 
As with the vessel, the position was held nearly constant with target points and stiff linear springs connecting nodes, here placed with a cylindrical topology. 
We do not model the texture or irregularities of the real scaffold structure, but believe these measurements are accurate to approximately $\pm$1 mm throughout the model.

\subsection{Fluid-structure interaction}
\label{fsi}

We used the immersed boundary method for fluid-structure interaction simulations \cite{ib_acta_numerica}. 
The immersed boundary method uses two reference frames, an Eulerian or laboratory frame for fluid quantities and a Lagrangian or material frame for structure quantities. 
In one distinctive feature of the IB method, the fluid interacts with the structure through an Eulerian-frame body force, which in turn is computed from the Lagrangian frame structure force. 
Let $\rho,\mu$ denote the fluid density and viscosity, respectively. 
Let $\mathbf x$ represent a fixed location in space, and $t$ time. 
The fields $\mathbf u, \mathbf p$ represent fluid density and pressure, and $\mathbf f$ denotes the IB body force in the Eulerian frame.
Let $\mathbf s$ label a material point of the structure, which we previously denoted $u,v$ in Section \ref{construction_valve} but switch to avoid confusion with fluid velocity, and $\mathbf X (\mathbf s,t)$ denote the position of the structure point with label $\mathbf s$. 
The field $\mathbf F$ represents the structure force in the Lagrangian frame. 
Let $\delta$ denote the Dirac $\delta$-function.

The governing equations of the IB method are 
{\allowdisplaybreaks
\begin{align}
\rho \left( \frac{ \partial \mathbf u (\mathbf x, t)}{\partial t}  + \mathbf u (\mathbf x, t) \cdot \nabla \mathbf u (\mathbf x, t) \right) &= - \nabla p (\mathbf x, t) + \mu \Delta \mathbf u (\mathbf x, t) + \mathbf f (\mathbf x, t)  \label{momentum} \\
\nabla \cdot \mathbf u(\mathbf x, t)  &= 0   		\label{mass}		\\
\mathbf F( \, \cdot \, , t) &=  \mathcal F(\mathbf X( \, \cdot \, ,t))  \label{nonlinear_force}  \\
\frac{ \partial \mathbf X(\mathbf s,t)}{\partial t}&=  \mathbf u(\mathbf X(\mathbf s,t), t) 	 \label{interpolate} 	 \\
		&= \int \mathbf u(\mathbf x, t)   \delta (  \mathbf x  - \mathbf X(\mathbf s,t) )  \;  d  \mathbf x   \nonumber  \\ 
\mathbf f(\mathbf x, t)   &= \int  \mathbf F(\mathbf s,t)  \delta(  \mathbf x - \mathbf X(\mathbf s,t)  )   \; d\mathbf s   .       \label{spreading} 
\end{align}  
}

The equations \eqref{momentum} and \eqref{mass} are the Navier Stokes equations governing the dynamics of a viscous, incompressible fluid, plus the extra IB body force $\mathbf f$. 
Equation \eqref{nonlinear_force} represents the Lagrangian frame force via the mapping $\mathcal F$, which determines the field $\mathbf F$ as a function of the entire configuration of the structure. 
Equations \eqref{interpolate} and \eqref{spreading} are the interaction equations, which couple the two frames. 
Equation \eqref{interpolate} performs velocity interpolation and specifies that the structure moves with the local fluid velocity. 
The Lagrangian frame force is spread to the Eulerian grid via equation \eqref{spreading}. 

We ran simulations using the open-source solver IBAMR (Immersed Boundary Adaptive Mesh Refinement) using a staggered grid discretization of the fluid domain \cite{IBAMR,griffith2010parallel}. 
Simulations were run on the Shrelock Cluster at Stanford University on four nodes or 96 Intel Xeon Gold 5118 cores with a 2.30GHz clock speed. 
The discrete $\delta$-function used was the five-point kernel, which we selected for its translational invariance \cite{IB5_arxiv} and lack of even-odd condition, which may create artifacts in the flow field when using a staggered-grid flow solver \cite{thesis}.

To drive the simulations, we prescribed pressures at the RV inlet and PA outlets based on the experimental measurements coupled to time-varying resistance boundary conditions.
The motivation for this setup is as follows. 
In the physical experiment, an approximately equal flow split was achieved by tightening a clamp on downstream tubing.
In the simulation, we wished to similarly achieve an approximately equal flow split, but the RPA and LPA portions of the domain had differing resistances due to their geometry, including curvature and length within the domain of interest.
Further, the experimental pressure was measured approximately at the junction of these two vessels and pressures downstream at the simulation outlets were not available to use directly. 
To achieve an approximately even flow split on the two outlets, the total resistance of each outlet must be equal. 
Thus, we added a small resistance during systole to the less resistive outlet to make the sum of the resistance due to geometry and the resistance added to be equal on both outlets. 
During diastole, the valve became heavily loaded in a short duration of time, which created an oscillation in the flow waveform in simulations. 
On preliminary simulations without resistance this ringing persisted for an excessively long duration. 
The experimental pressure waveform includes an oscillation at the corresponding time, but if this oscillation was prescribed directly at the outlets extensive ringing occurred because the frequency of the experimental oscillation was not precisely equal to that of the simulation setup. 
(The time resolution on flow rates was ten frames per cardiac cycle, which was not enough to capture oscillations that may have occurred in the experimental flow rate.)
Thus, we prescribed a resistance to both outlets, which allows the pressure and flow to oscillate in phase, which in turn caused the amplitude of the oscillations in both pressure and flow to decay over time. 
The experiment includes damping downstream of the region of interest that would be very complex to model. 
We selected the value of diastolic resistance by trial and error to introduce numerical damping in the system and reduce ringing in the flow rate. 
We expect that this resistance has minimal influence on the opening dynamics of the valve and almost none on the fully closed state, rather the resistance primarily influences the amplitude, decay rate and duration of the closing transient seen in the flow rate as the valve oscillates and settles into its closed state.  
We applied pressures, rather than prescribed flow rates, because pressures are necessary to load and properly close the valve during diastole. 
If we had applied a flow rate, during diastole the valve would not be loaded by the appropriate pressure difference to close in a physiological manner and would instead remain open under near-zero flow conditions. 
These strategies created reasonable agreement with the experimental pressures and flow rates throughout the cardiac cycle (minus small differences in the pressure oscillation frequency), while at the same time achieving a more even flow split and better damping of the closing transient than prescribing the experimental pressures alone.

To implement these strategies, we smoothed the experimental pressures measured on the RV and PA sides of the valve by taking a weighted average in time via convolution with a normalized cosine bump of radius 0.01 s. 
We then represented each curve as a finite Fourier series with 600 frequencies, which ensured the data were smooth and periodic in time. 
At the outlets, we applied a time-varying resistance boundary condition of the form 
\begin{align}
R(t) Q  = P_{outlet} - P_{exp}
\label{resistance_eqn}
\end{align}
where $R(t)$ is the resistance, $Q$ is the instantaneous flow rate during the simulation, $P_{outlet}$ is the outlet pressure and $P_{exp}$ is the experimental pressures as processed above. 
To achieve an approximately even flow split during systole, we estimated the systolic resistances as follows. 
We ran a steady simulation for 0.2 s with the maximum experimental pressure difference of $p = 15.66$ mmHg prescribed at the RV and zero pressure at both the RPA and LPA.
This resulted in flow rates of $Q_{RPA}$ = 93.4 ml/s and $Q_{LPA}$ = 171.2 ml/s. 
We integrated the simulation pressure field at a slice in the PA junction to estimate a mean pressure of 3.53 mmHg. 
We then estimated that the resistance of the vessels themselves was 27.49 and 47.85 dynes/cm$^{5}$ for the left and right PAs, respectively. 
The resistance at systole is set to $20.36$ s dynes/cm$^{5}$ on the LPA and $0$ on the RPA, which makes the sum of the resistance in the vessel and the added outlet resistance approximately equal for both outlets. 
When the sign of the pressure difference across the vessel changed sign, indicating diastole, $R(t)$ smoothly increased to $1000.0$ s dynes/cm$^{5}$.
The resistance value remained constant through the remainder of diastole, then decreased smoothly to the systolic resistance after the pressure again changed sign, indicating systole.

The cardiac cycle duration was 0.8325 s, and simulations were run for two cycles. 
Results from the second cycle were analyzed and results from the first cycle was discarded due to initialization effects. 
(We subsequently report times in the second cardiac cycle modulo the cycle duration.)
The time step was set to $\Delta t = 7.5 \cdot 10^{-6}$ s, which was the largest value we found to be numerically stable 
given the explicit time discretization of the structure and fluid-structure coupling. 
At peak flow, this time step resulted in an advective Courant Friedrichs Lewy (CFL) number of $\max \| \mathbf u \| \Delta t / \Delta x = 0.036$.
The observed stability limit is thus much less than that expected by the fluid velocity alone, and we do not consider lower time steps further. 
We output 1200 frames per second of the simulation for post processing. 
The fluid resolution was $\Delta x = 0.45$ mm, exactly half that of the MRI resolution in the sagittal direction, and approximately half that of the MRI in the other two directions.  
(A convergence study was performed, see Appendix.)
The structure resolution was targeted to half that of the fluid resolution, or 0.225 mm, but the precise lengths of each link in the model were determined by the steps in Section \ref{construction_valve}. 
We set $\rho = 1.09$ g/cm$^{3}$ and $\mu =$ 3.9 centipoise. 
The vessel was placed into a $9.36 \times 6.48 \times 5.76$ cm box, which corresponds to $208 \times 144 \times 128$ points.
No local mesh refinement is used, which facilitates post-processing the velocity field (Section \ref{post_processing}). 
The valve was placed at the origin with its axis aligned with the $x$ direction. 
The minimum height of the leaflet attachment is placed at -0.2 cm, just below the origin. 
Open-boundary stabilization was applied at the inlets and outlets \cite{CNM:CNM2918}. 
Outside of the inlet and outlets, on faces that include the inlet and outlets, we prescribed no slip boundary conditions. 
On the remaining three faces, we prescribed zero pressure boundary conditions. 
No contact forces were needed, as the IB method automatically prevents contact between structures via the interaction equations and continuity of the velocity field \cite{doi:10.1137/070699780}.

\subsection{Post processing}
\label{post_processing}

We post-processed the simulation data to obtain a phase-averaged, resampled velocity field $\bar{\mathbf{u}}$ in a manner that mimicked experimental data acquisition. 
The scanner reported the phase-averaged velocity field at ten time steps per cardiac cycle, recorded from approximately ten cycles. 
To process simulation data, we computed the mean of the velocity field over the second cycle binned into ten time intervals of duration 0.083 s. 
(We also computed phase averages over three cycles but found no qualitative difference in the resulting flow fields and little quantitative difference between such results, see Appendix.)
We converted phase-averaged velocity field from cell to point data, then linearly interpolated onto the nodes of the MRI mesh using PyVista \cite{sullivan2019pyvista}.

\subsection{Integral metrics}
\label{Integral_metrics}

We analyzed the flow quantitatively with the following metrics. 
We computed the $L^p$ relative error for $p = 1,2$ of phase-averaged, resampled velocity magnitude, 
\begin{align}
\frac{\| \bar{\mathbf{u}} - \mathbf{u}_{exp} \|_{L^p} }{  \| \mathbf{u}_{exp} \|_{L^p} } 
= \frac{  \left( \int | \bar{\mathbf{u}} - \mathbf{u}_{exp} |^{p} \;  d\mathbf x \right)^{1/p} }{  \left( \int | \mathbf{u}_{exp} |^{p} \; d\mathbf x \right)^{1/p}} , 
\end{align}
and the $x$ (axial) component of the phase-averaged, resampled velocity only, which we denote as $\bar u$, 
\begin{align}
 \frac{\| \bar u - u_{exp} \|_{L^p} }{  \| u_{exp} \|_{L^p} } = \frac{ \left( \int | \bar u - u_{exp} |^{p} \;  d\mathbf x \right)^{1/p} }{ \left( \int | u_{exp} |^{p} \; d\mathbf x \right)^{1/p}}.
\end{align}
Both relative errors were evaluated on the entire flow domain interior to the vessel and on three two-dimensional slices, $x = $ 0, 0.625 and 1.25 cm. 
We computed these metrics during the systolic phase only, because during diastole the flow fields are near zero in both simulation and experiment.

We also computed the integral metric $I_1$, which represents the nondimensional streamwise momentum and is computed on two dimensional slices of the domain  \cite{banko2016oscillatory,schiavone2021vitro}. 
The metric is defined as 
\begin{align}
I_{1} &= \left( \frac{ 1}{U_{T}^{2} A}  \iint (\bar{\mathbf{u}} \cdot \mathbf n)^{2} \; dA  \right)^{1/2} =  \left\| \frac{\bar{\mathbf{u}} \cdot \mathbf n} {U_{T} A^{1/2}}  \right\|_{L^{2}} 
\label{I1_def}
\end{align}
where $A$ is the area of the slice, $\mathbf n$ is the unit normal and $U_{T}$ is a velocity scale. 
The velocity scale is set to the maximum of the spatially averaged velocity over time, 
\begin{align}
U_{T} = \max_{t \in [0, 0.8325]}  \left( \frac{1}{A} \iint \left( \bar{\mathbf{u}}(\mathbf x, t) \cdot \mathbf n \right) \; dA \right). 
\label{velocity_scale}
\end{align}
The velocity scale for the simulation and experiment were computed individually. 
This metric is 1 for a flow with a uniform velocity profile equal to $U_{T}$.

\section{Results}
\label{results}

\begin{figure*}[p!]  
\hfill \hfill \includegraphics[width=.25\textwidth]{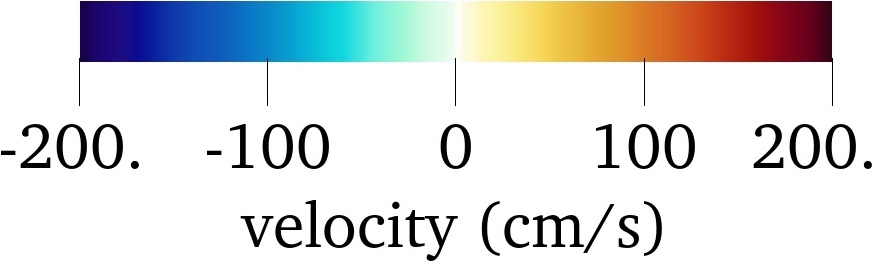}
{
\centering
\setlength{\tabcolsep}{2.0pt}        
\begin{tabular}{ c | c | c}        
& \large Simulation & \large Experiment \\ 
\rotatebox[origin=l]{90}{$t = 0.04 $ s} & 
\includegraphics[width=.42\textwidth]{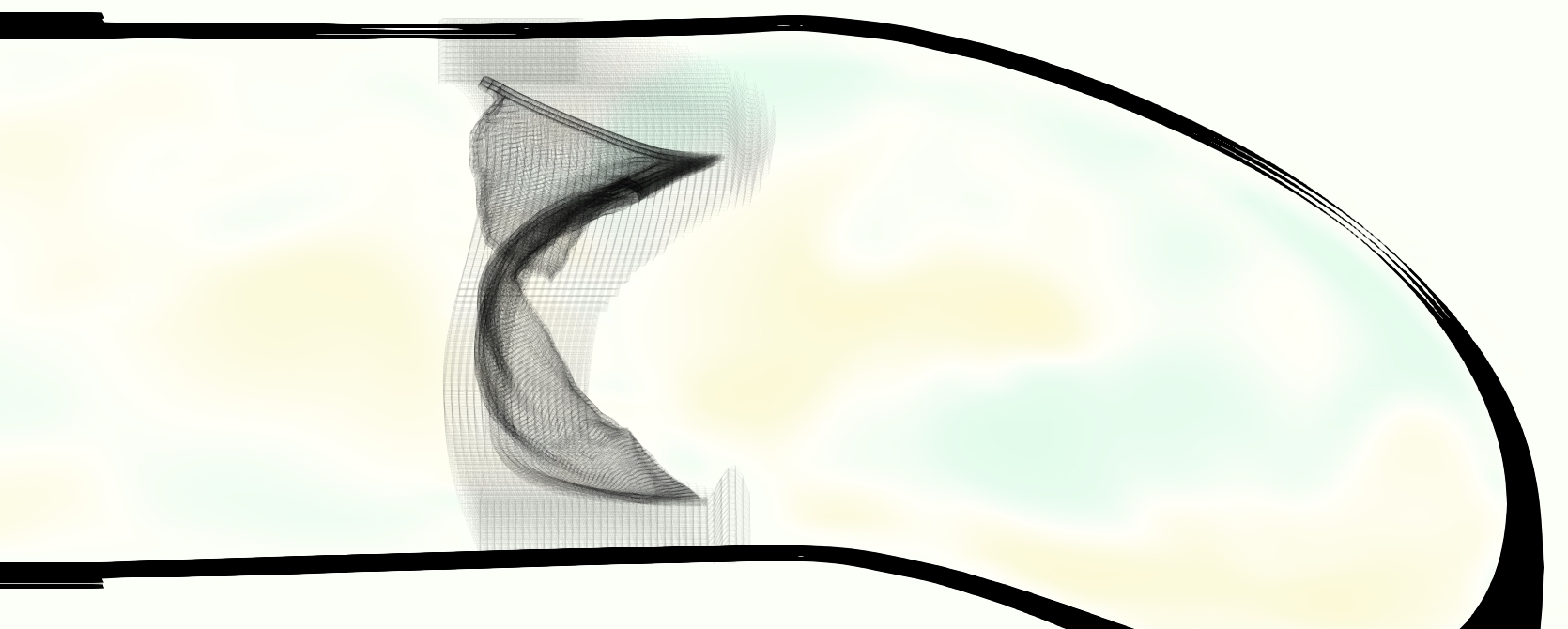} & \includegraphics[width=.42\textwidth]{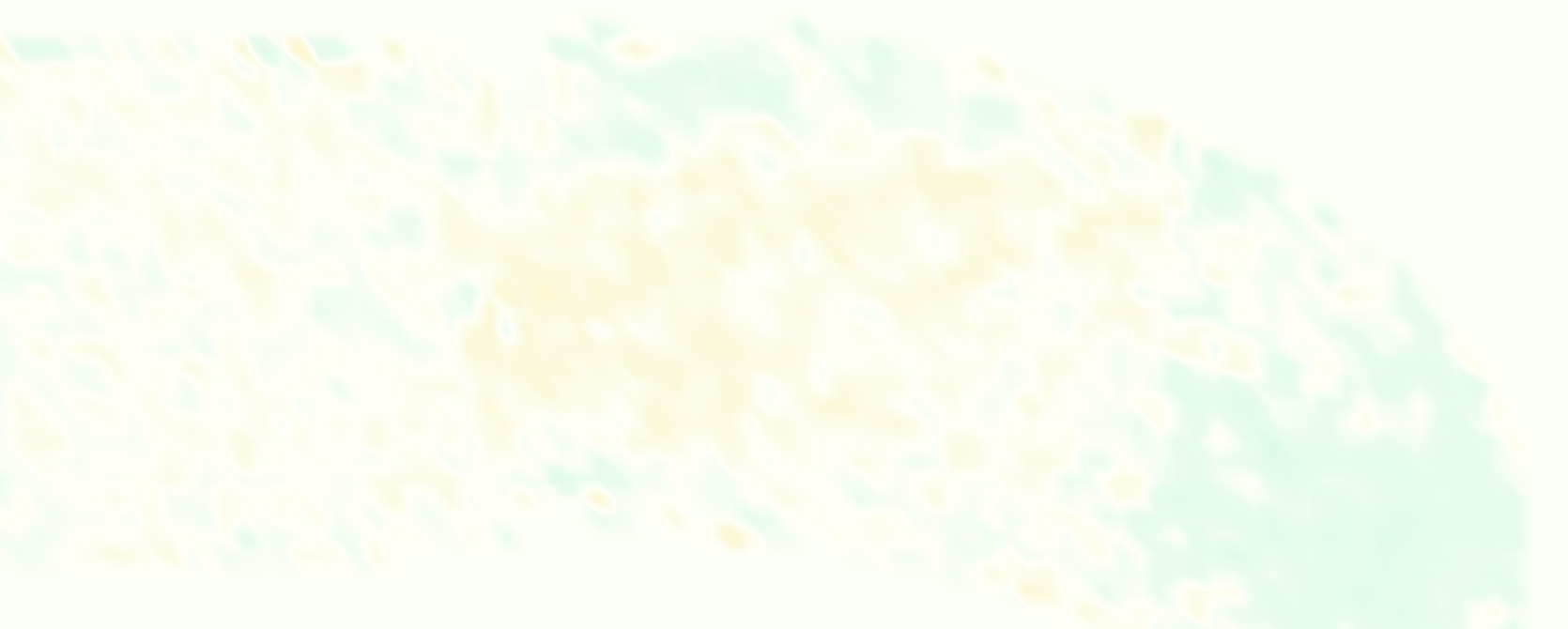} \\  
\rotatebox[origin=l]{90}{$t = 0.12 $ s} & 
\includegraphics[width=.42\textwidth]{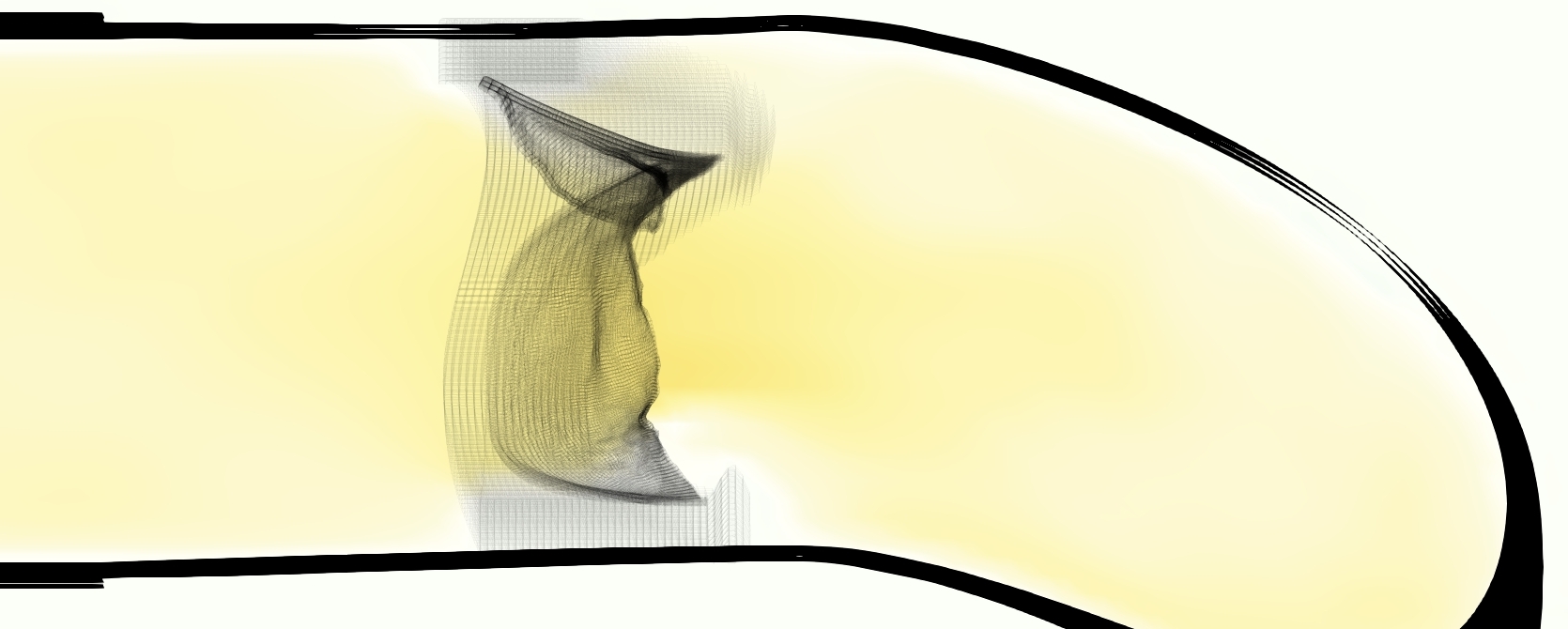} & \includegraphics[width=.42\textwidth]{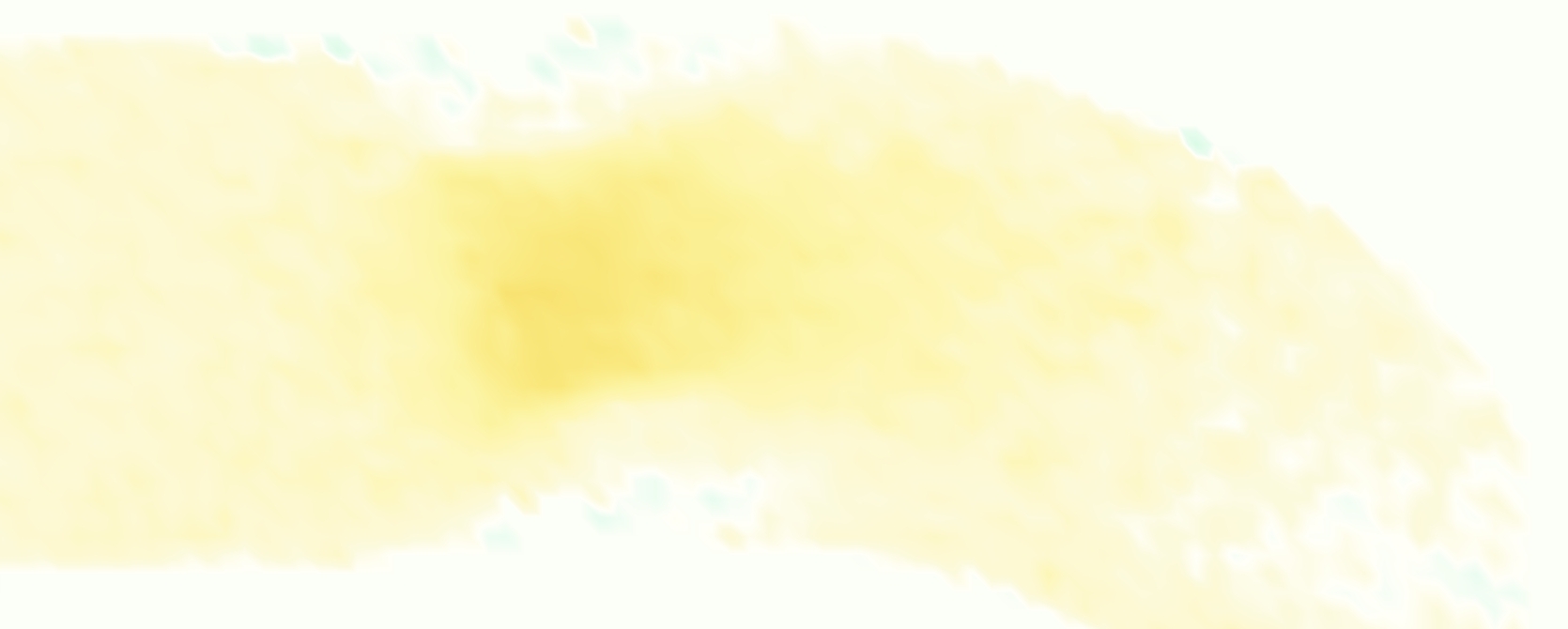} \\  
\rotatebox[origin=l]{90}{$t = 0.21 $ s} & 
\includegraphics[width=.42\textwidth]{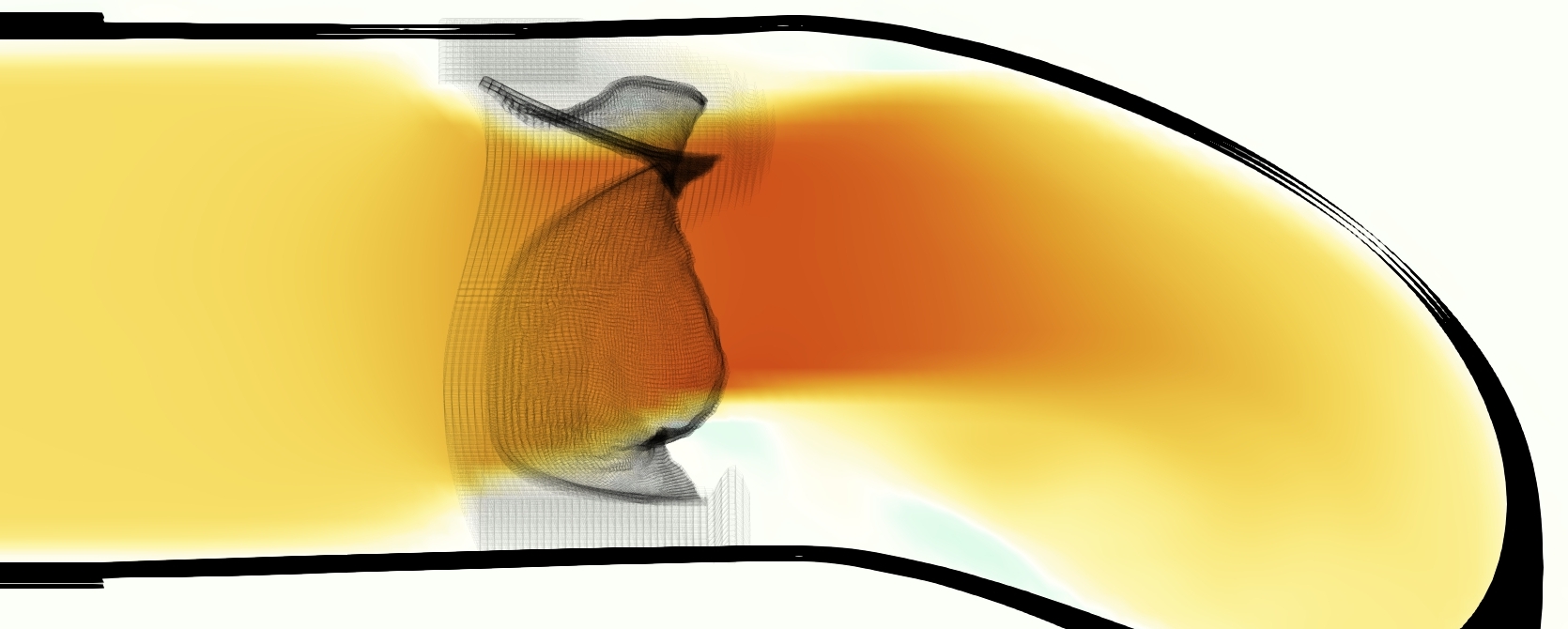} & \includegraphics[width=.42\textwidth]{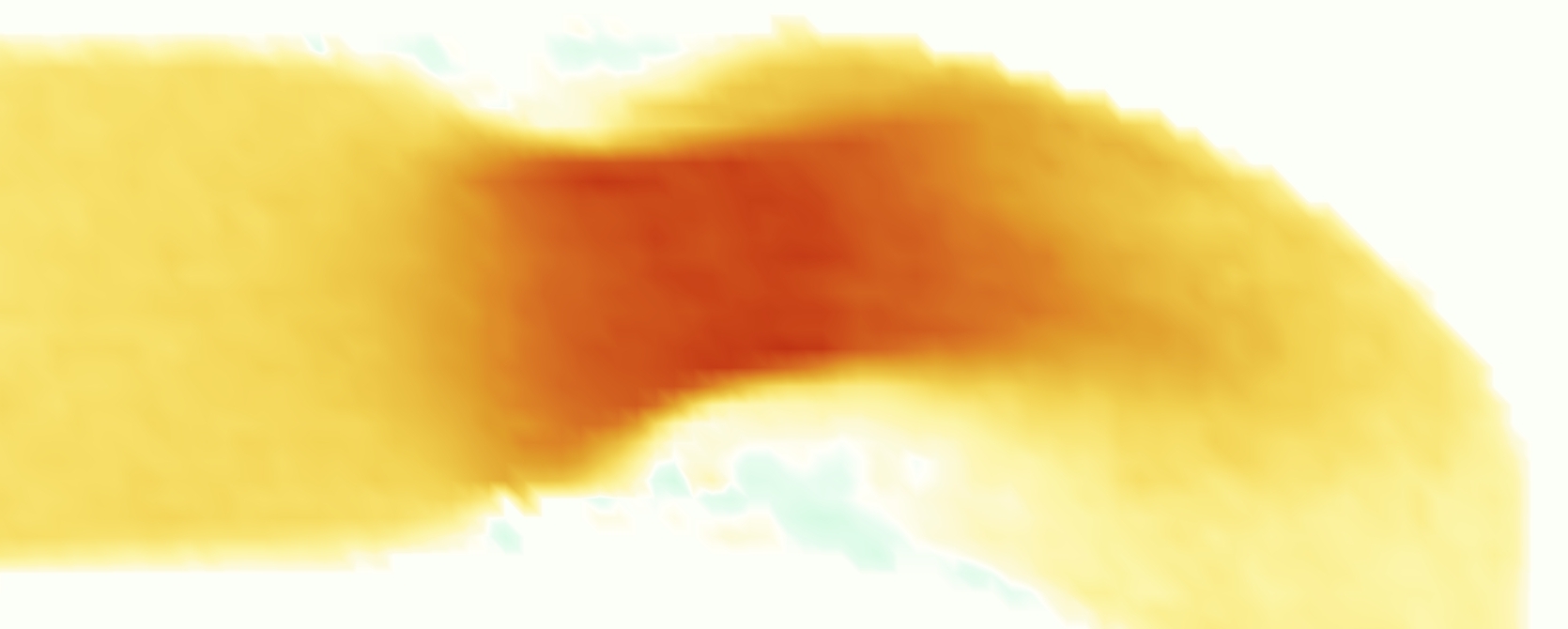} \\  
\rotatebox[origin=l]{90}{$t = 0.29 $ s} & 
\includegraphics[width=.42\textwidth]{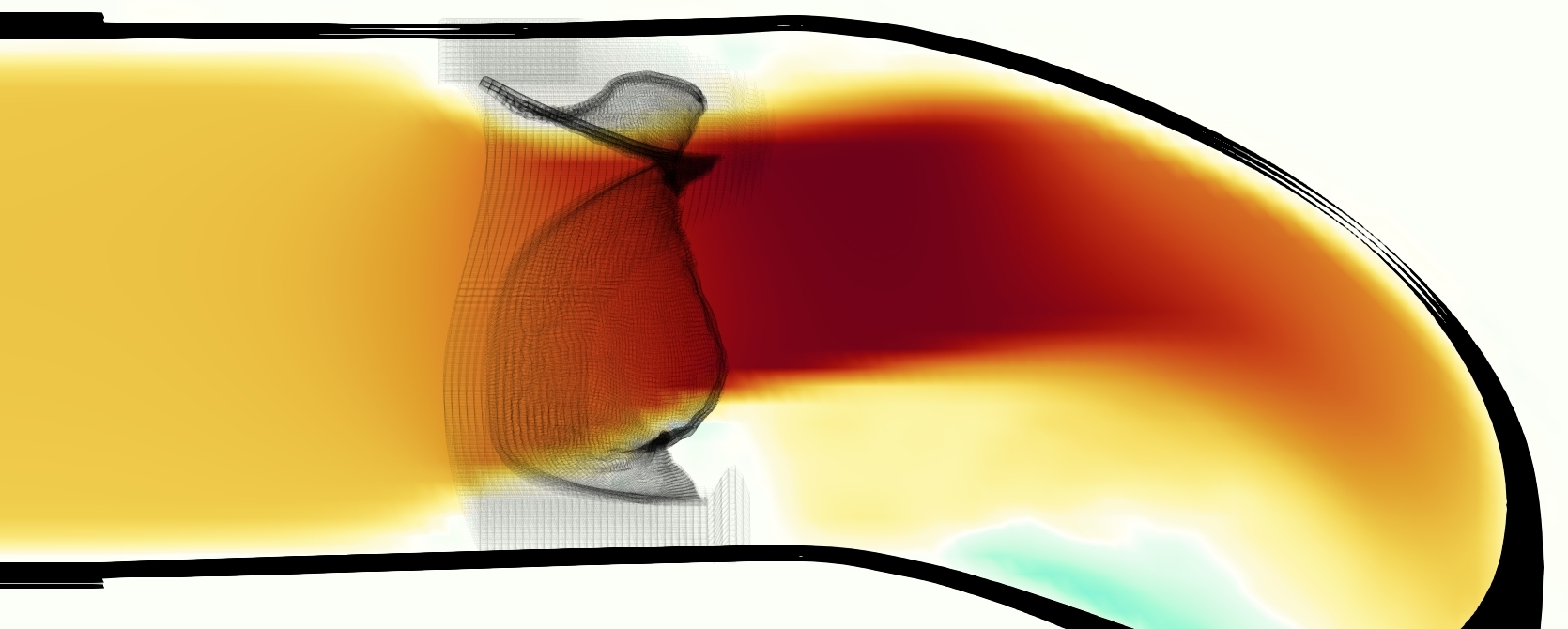} & \includegraphics[width=.42\textwidth]{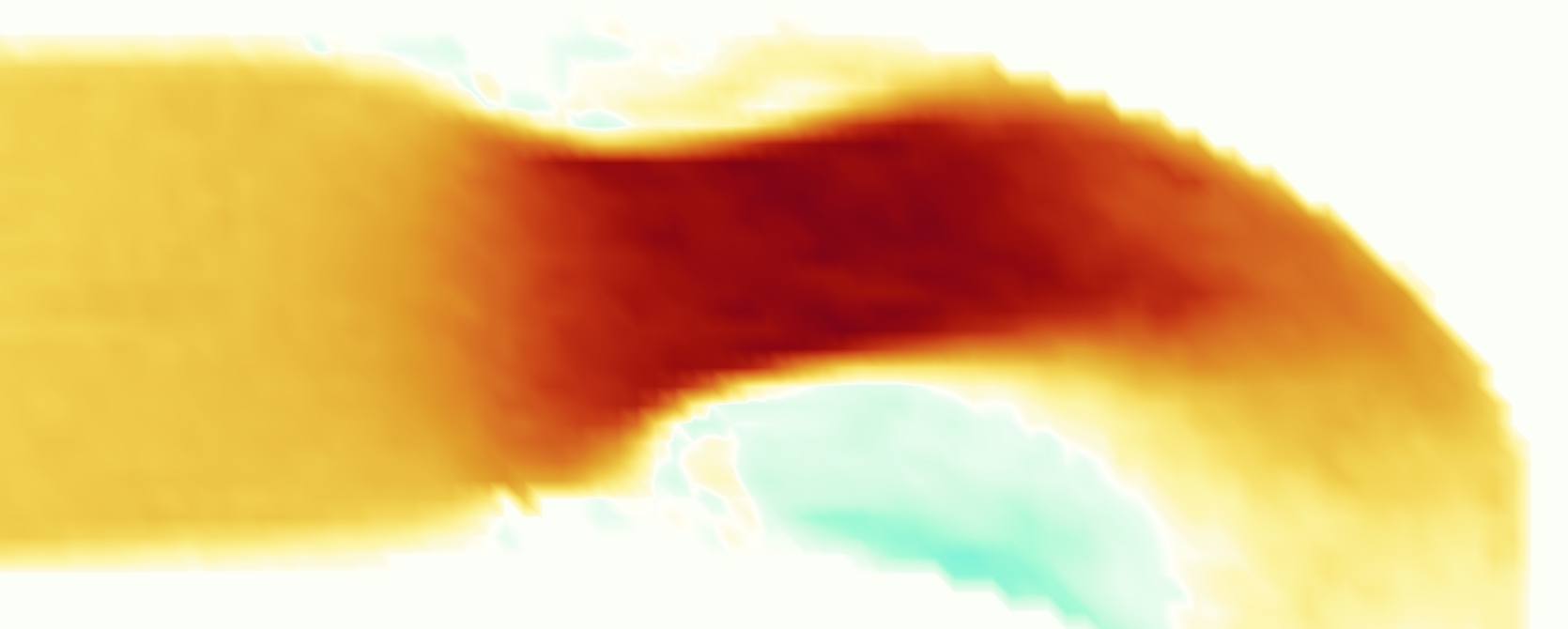} \\  
\rotatebox[origin=l]{90}{$t = 0.37 $ s} & 
\includegraphics[width=.42\textwidth]{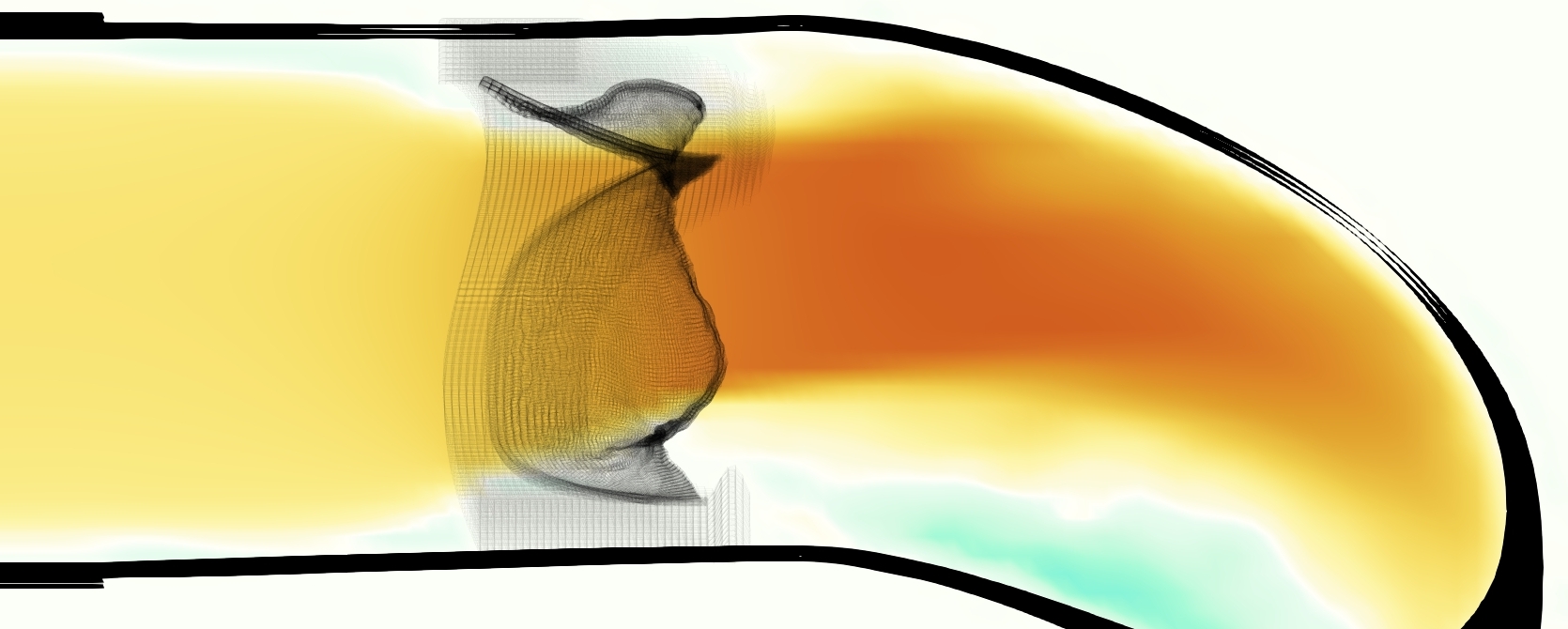} & \includegraphics[width=.42\textwidth]{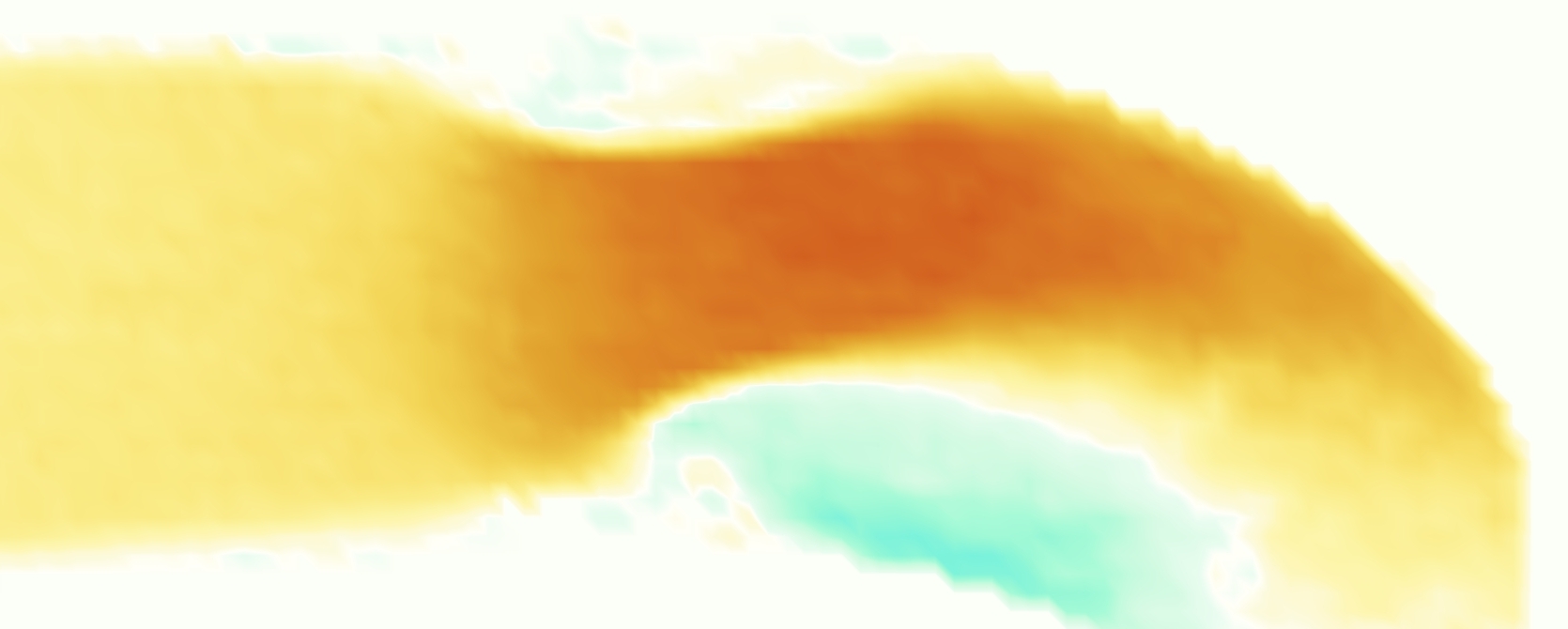} \\  
\rotatebox[origin=l]{90}{$t = 0.46 $ s} & 
\includegraphics[width=.42\textwidth]{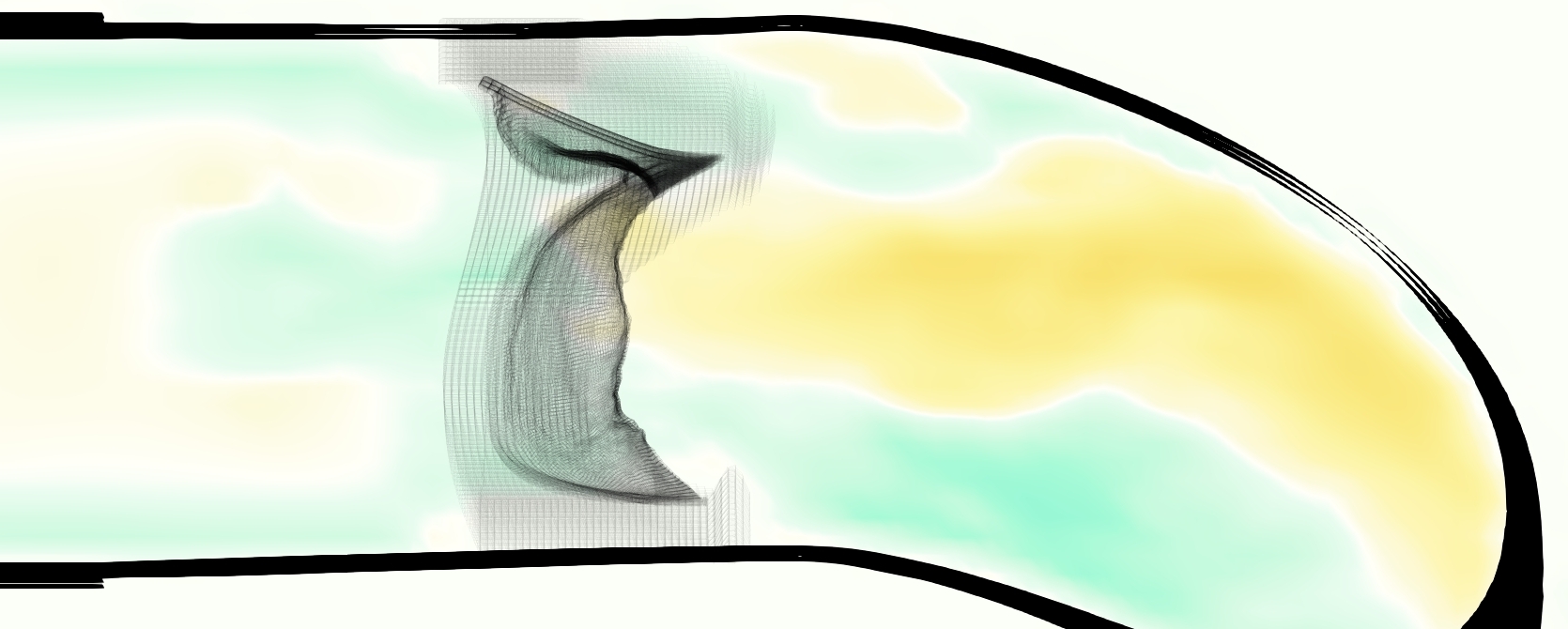} & \includegraphics[width=.42\textwidth]{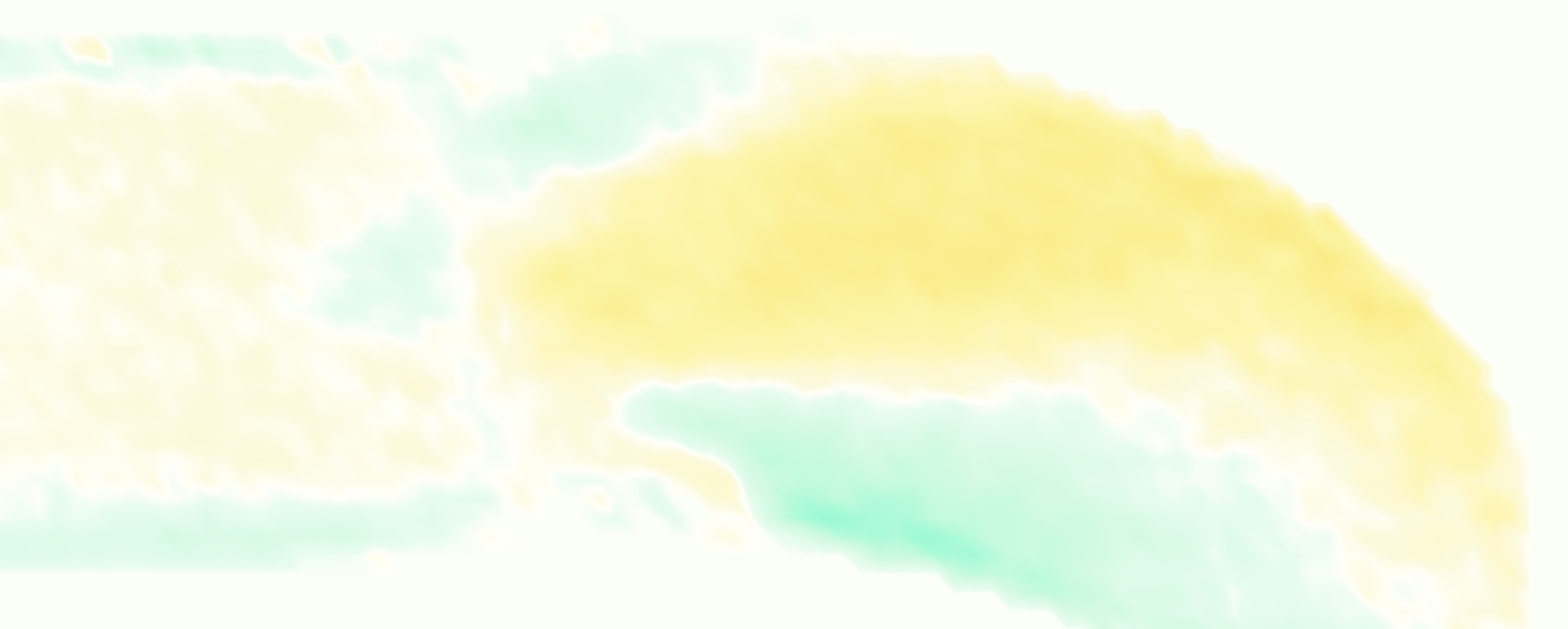} \\  
\rotatebox[origin=l]{90}{$t = 0.54 $ s} & 
\includegraphics[width=.42\textwidth]{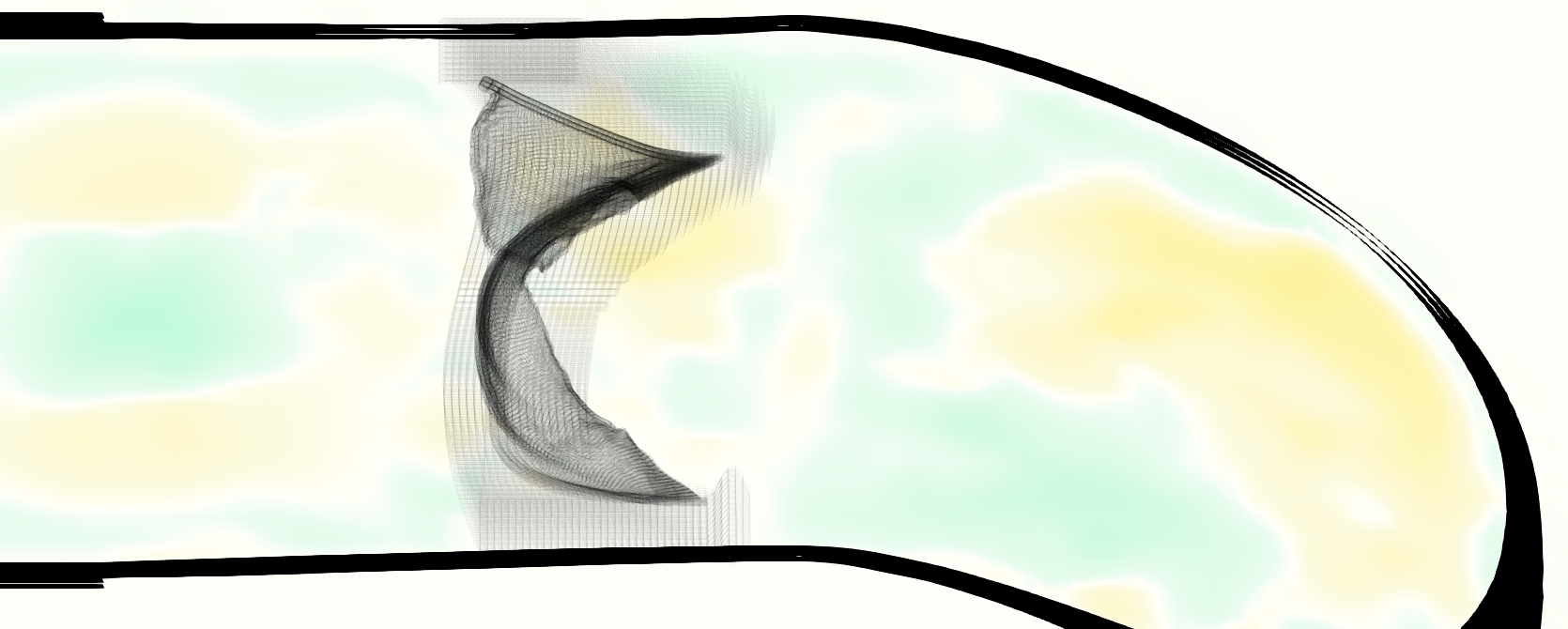} & \includegraphics[width=.42\textwidth]{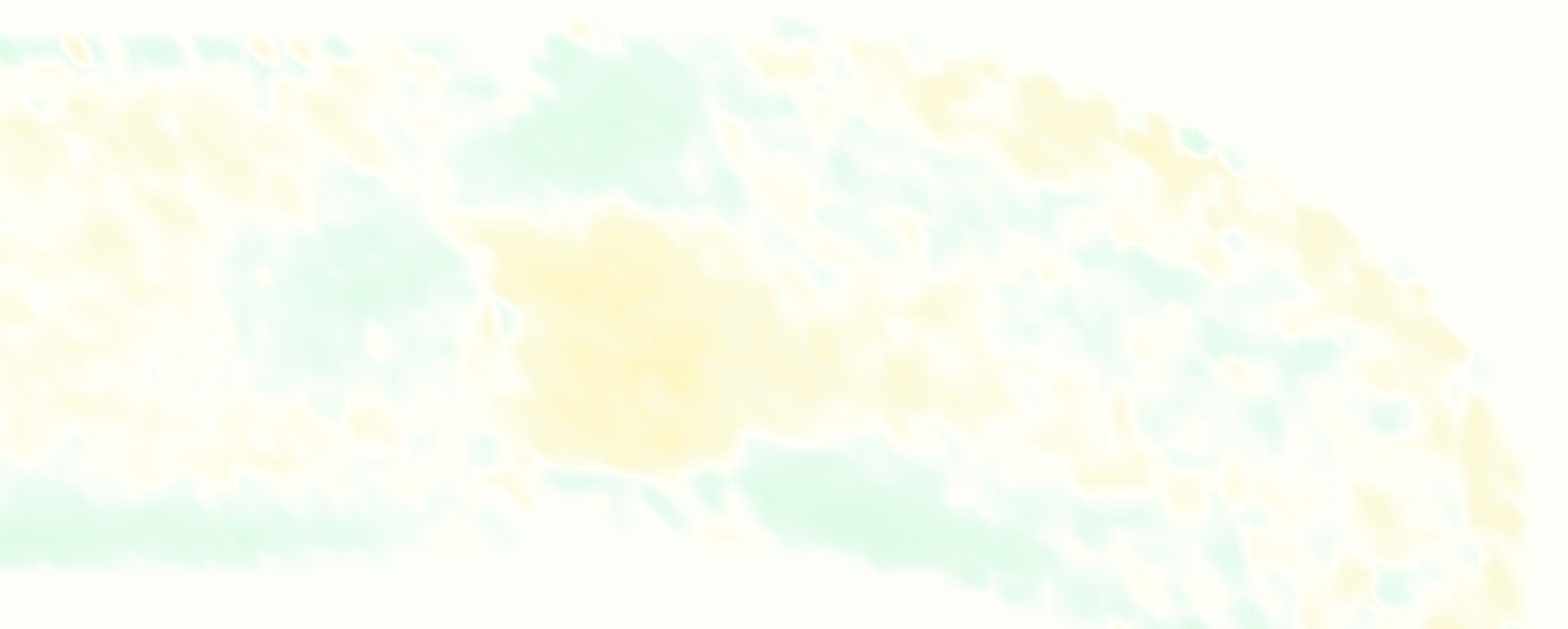} \\  
\end{tabular}
\caption{Sagittal view of the axial component of velocity in simulation (left) and experiment (right).}
\label{sagittal}
}
\end{figure*}

The emergent hemodynamics showed excellent qualitative agreement with experiments. 
Slice views of the axial component of fluid velocity ($x$ component) in the sagittal plane that cuts through the center of the vessel, parallel to the flow direction, are shown in Figure \ref{sagittal}. 
The flow in both the simulation and the experiment began to accelerate at the same time, with a jet forming through the open valve leaflets. 
The velocity, angle, and shape of the jet agreed well between the two cases during systole. 
The simulation captured the slight upward angle of the jet, which was not fully centered in the vessel. 
It also matched the location where the jet impacts the wall and the slower speed of the jet as it turns with the MPA downstream. 
As the flow decelerated, the fluid along the interior curve of the vessel reversed first while forward flow persisted where the jet was strongest, as seen in both the simulation and experimental results.

In both the experiment and the simulation, a separation region of reverse flow developed along the interior curve of the vessel under the core jet through the valve. 
The reverse flow began to develop at the same time in the cardiac cycle and grows throughout systole. 
The simulation lacked reversed flow in the entire region where reversed flow was present in the experiment, but this region developed some flow separation and had much slower flow than elsewhere in the vessel. 
Thus, the simulation captured that this is a distinct region from the core jet through the valve opening. 
There was a smaller amount of slower and slightly reversed flow along the outer curve of the vessel, close to the valve annulus and scaffold support. 
This region was well-matched between the experiment and the simulation.


\begin{figure*}[p!]  
\hfill \hfill \includegraphics[width=.25\textwidth]{colorbar.jpeg}
{
\centering
\setlength{\tabcolsep}{2.0pt}        
\begin{tabular}{c | c | c | c | c | c | c}
 & \large simulation & \large experiment & \large simulation & \large experiment & \large simulation & \large experiment  \\ 
\rotatebox[origin=l]{90}{$t = 0.04 $ s} & 
\includegraphics[width=.14\textwidth]{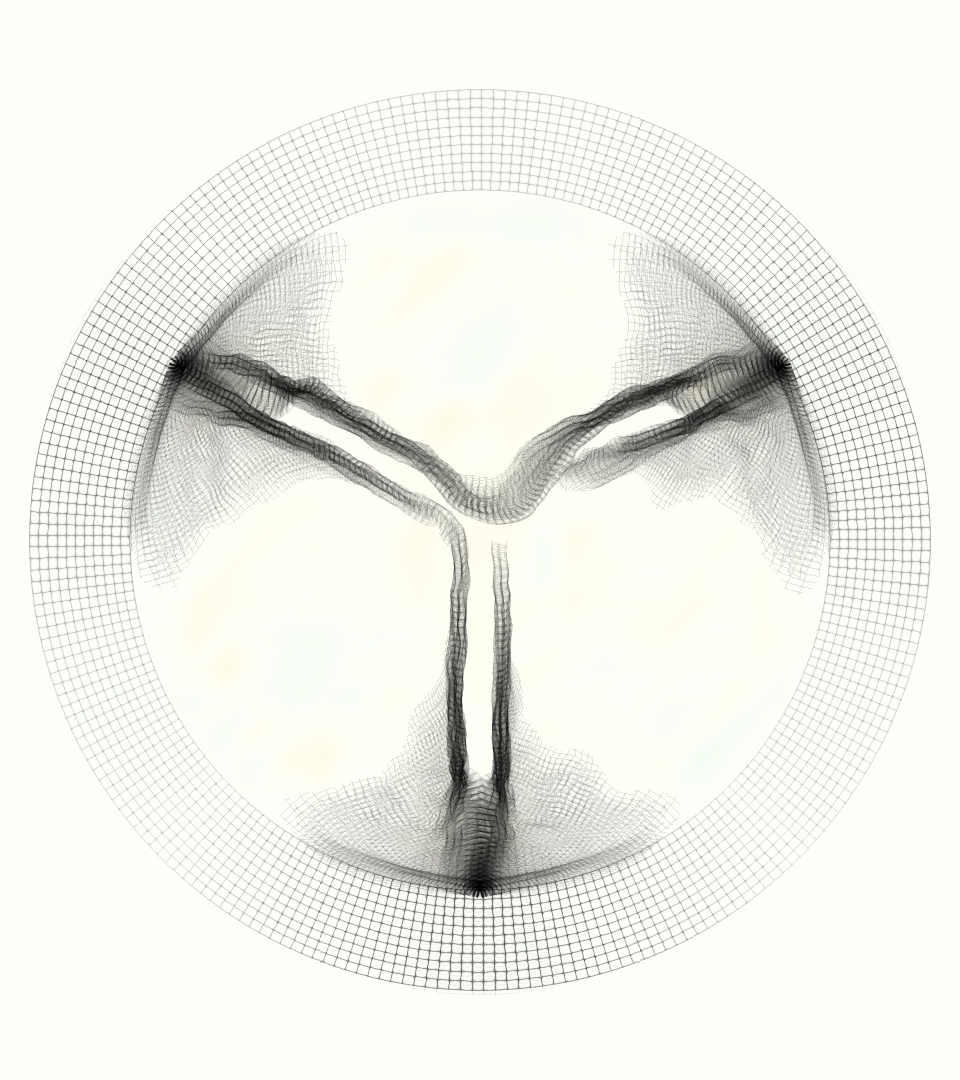} & 
\includegraphics[width=.14\textwidth]{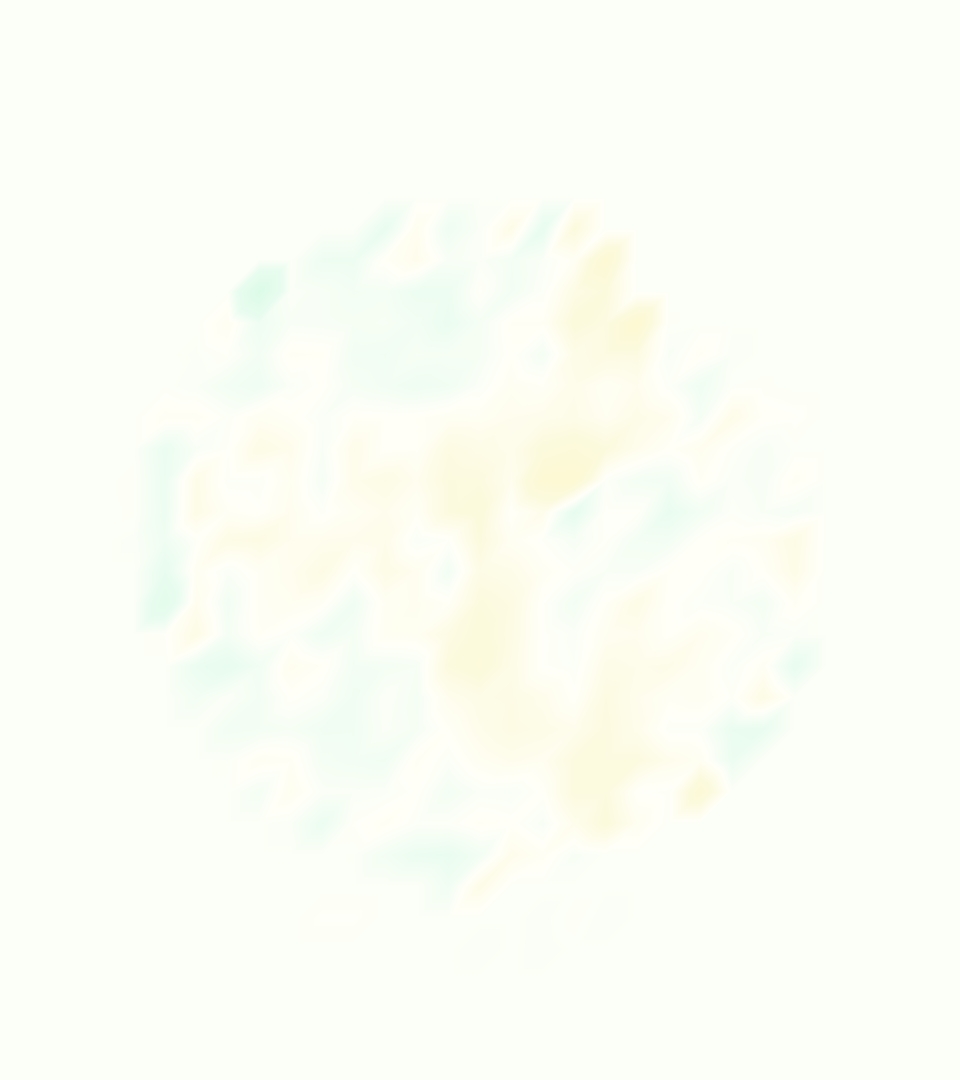} & 
\includegraphics[width=.14\textwidth]{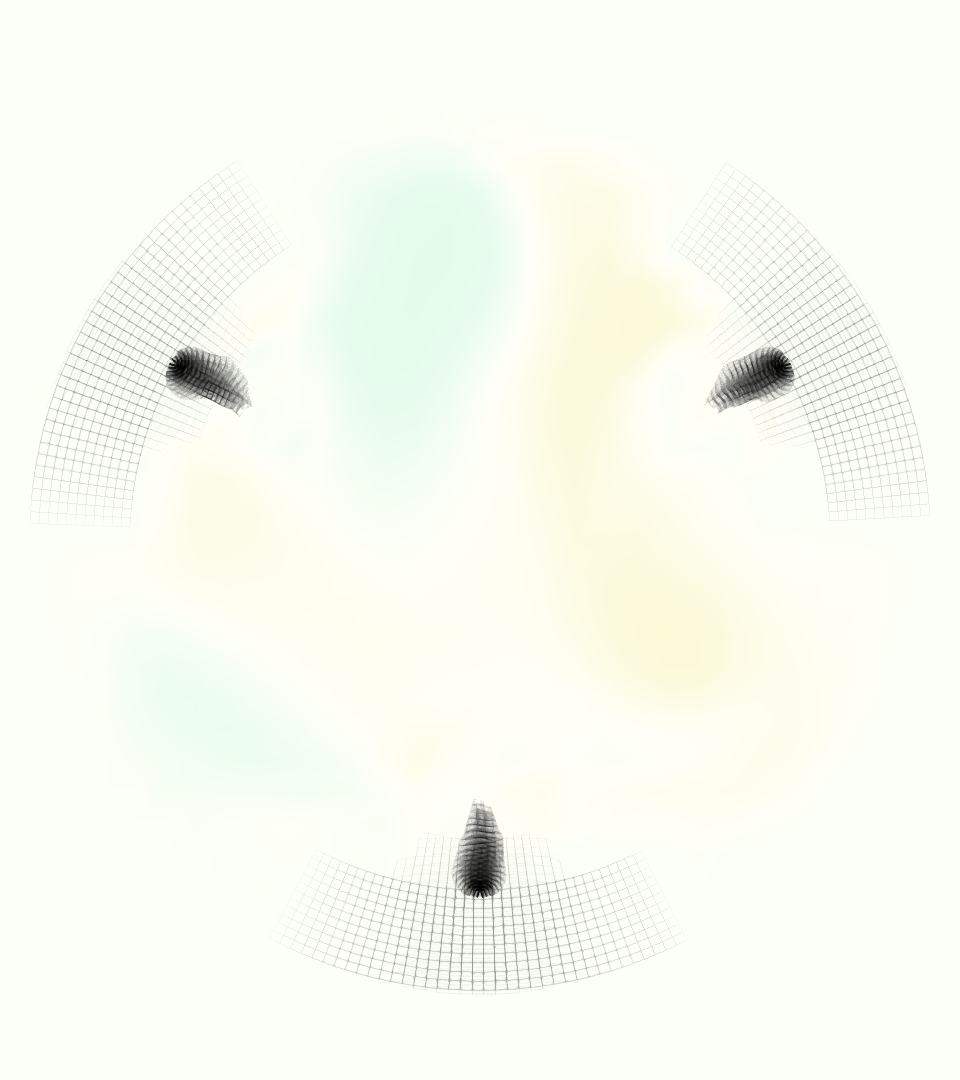} & 
\includegraphics[width=.14\textwidth]{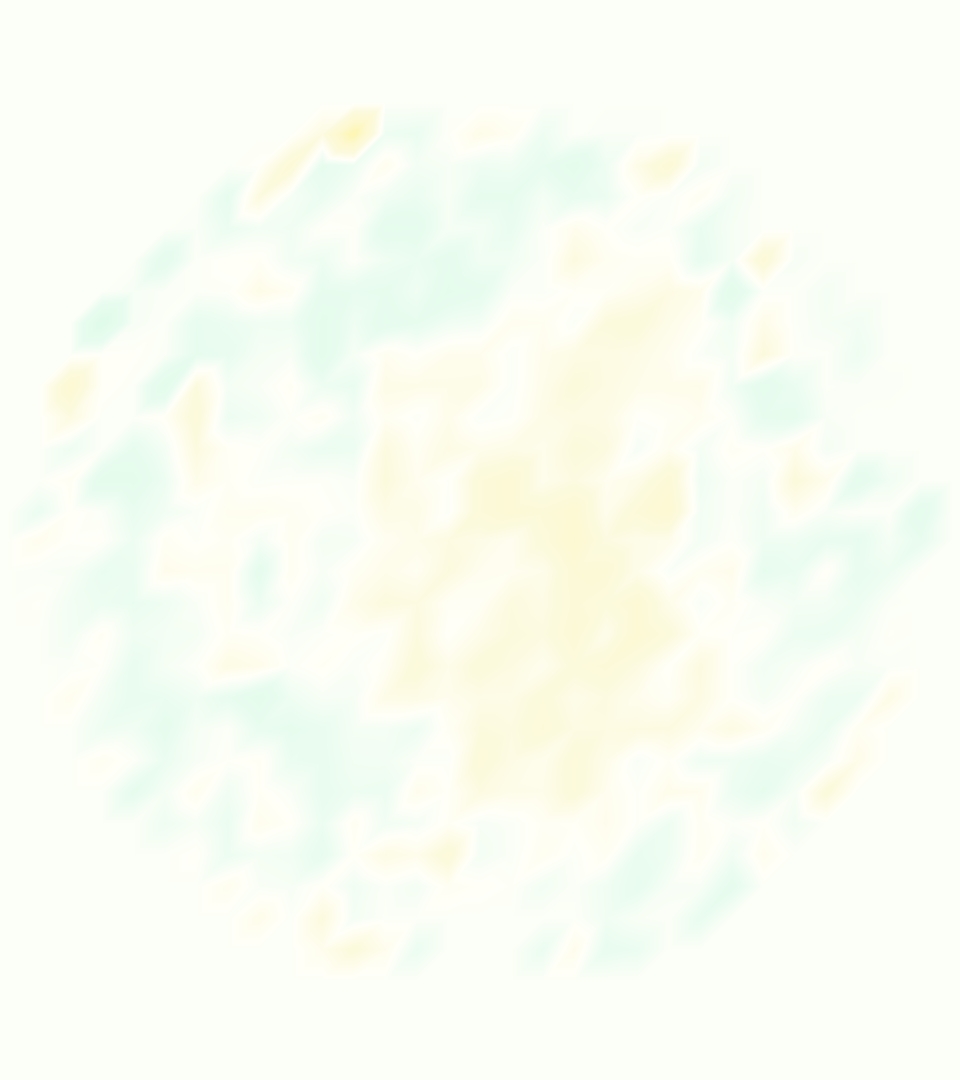} & 
\includegraphics[width=.14\textwidth]{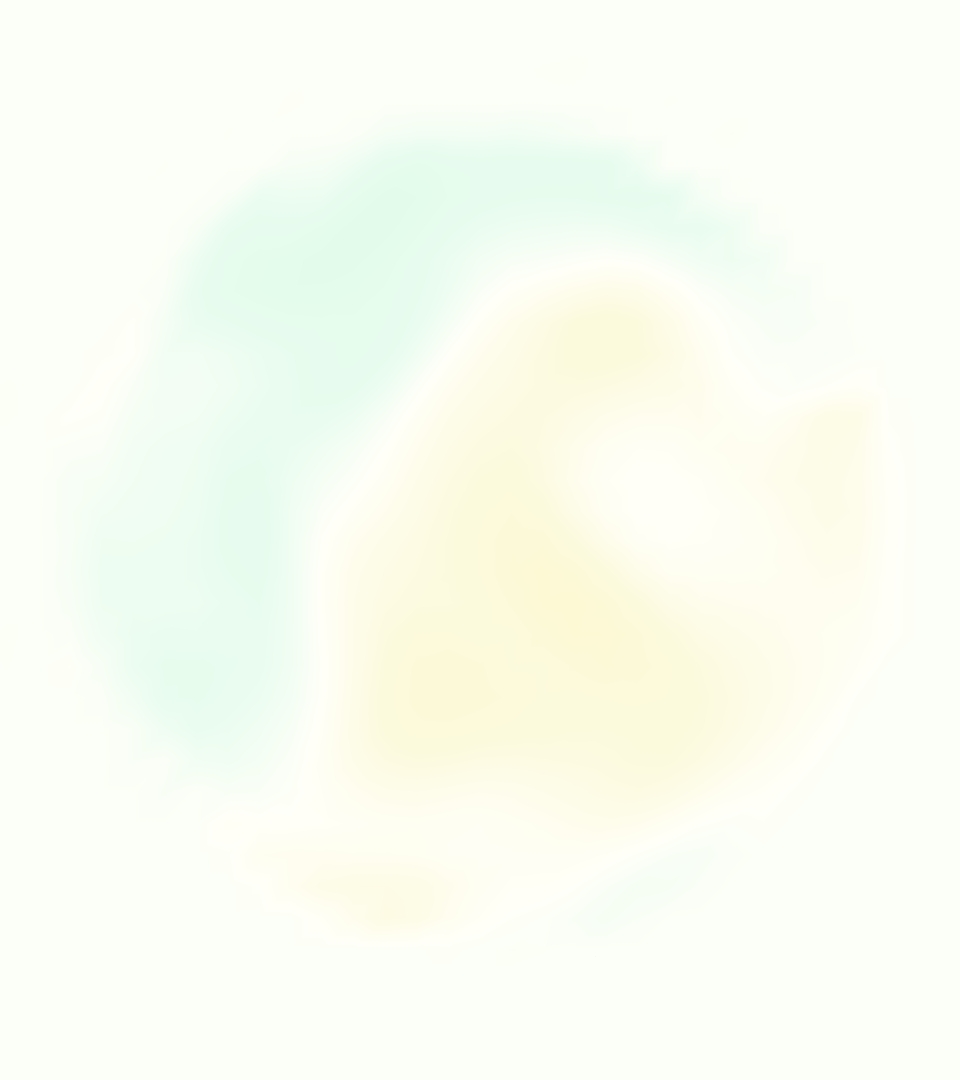} & 
\includegraphics[width=.14\textwidth]{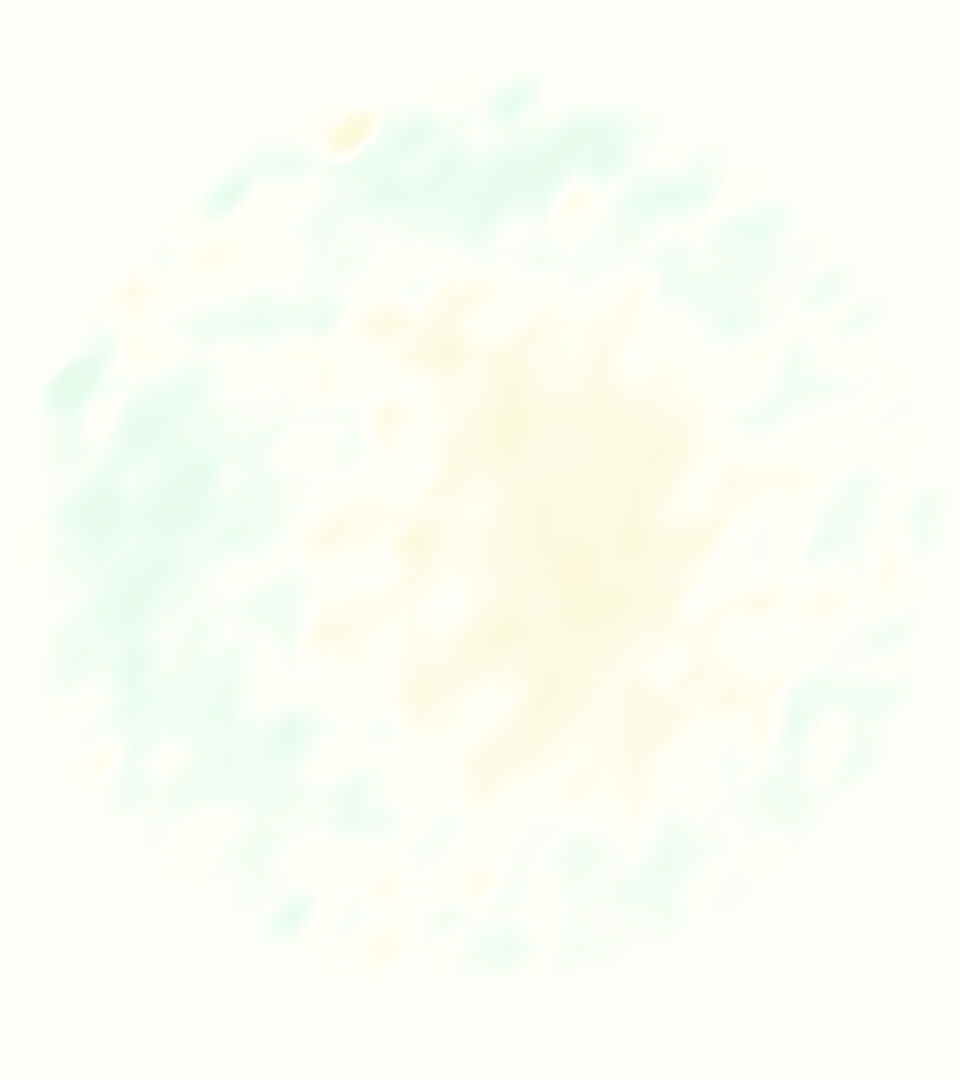}  \\ 
\rotatebox[origin=l]{90}{$t = 0.12 $ s} & 
\includegraphics[width=.14\textwidth]{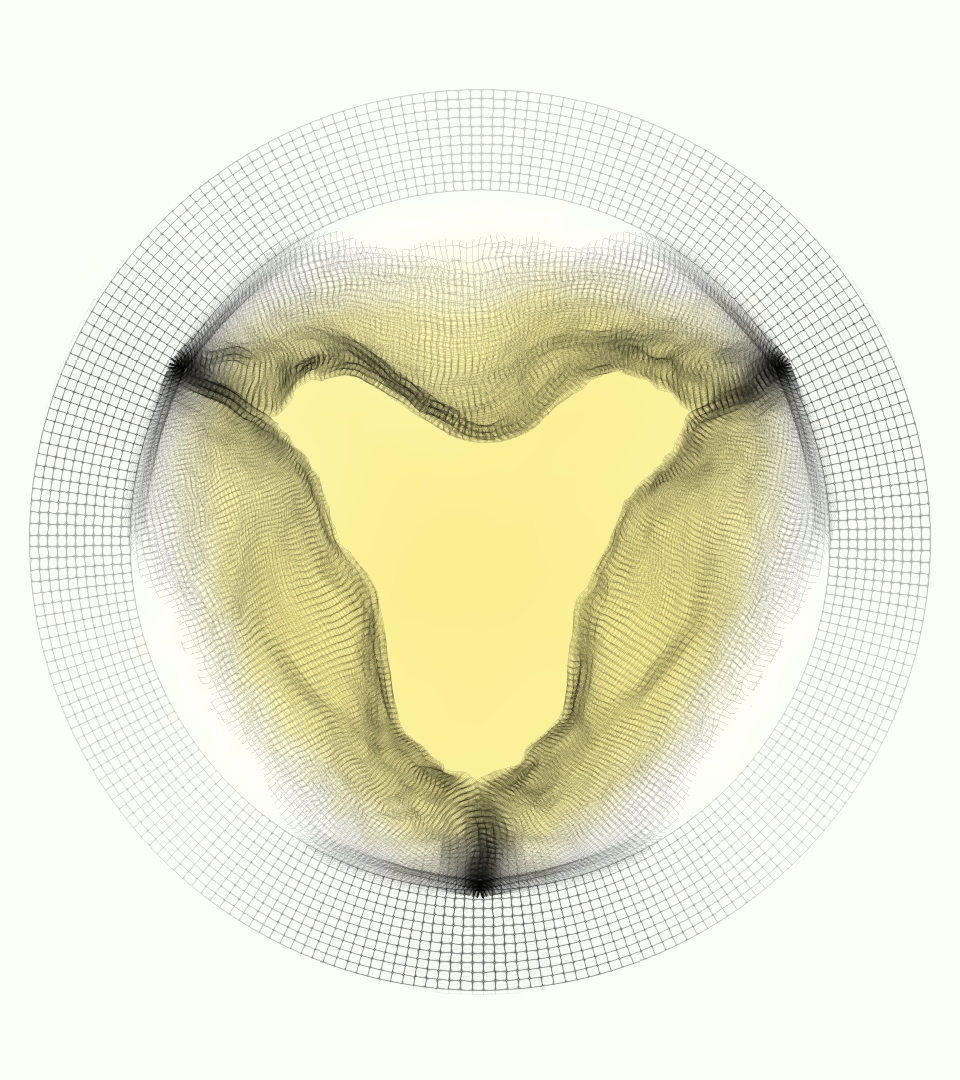} & 
\includegraphics[width=.14\textwidth]{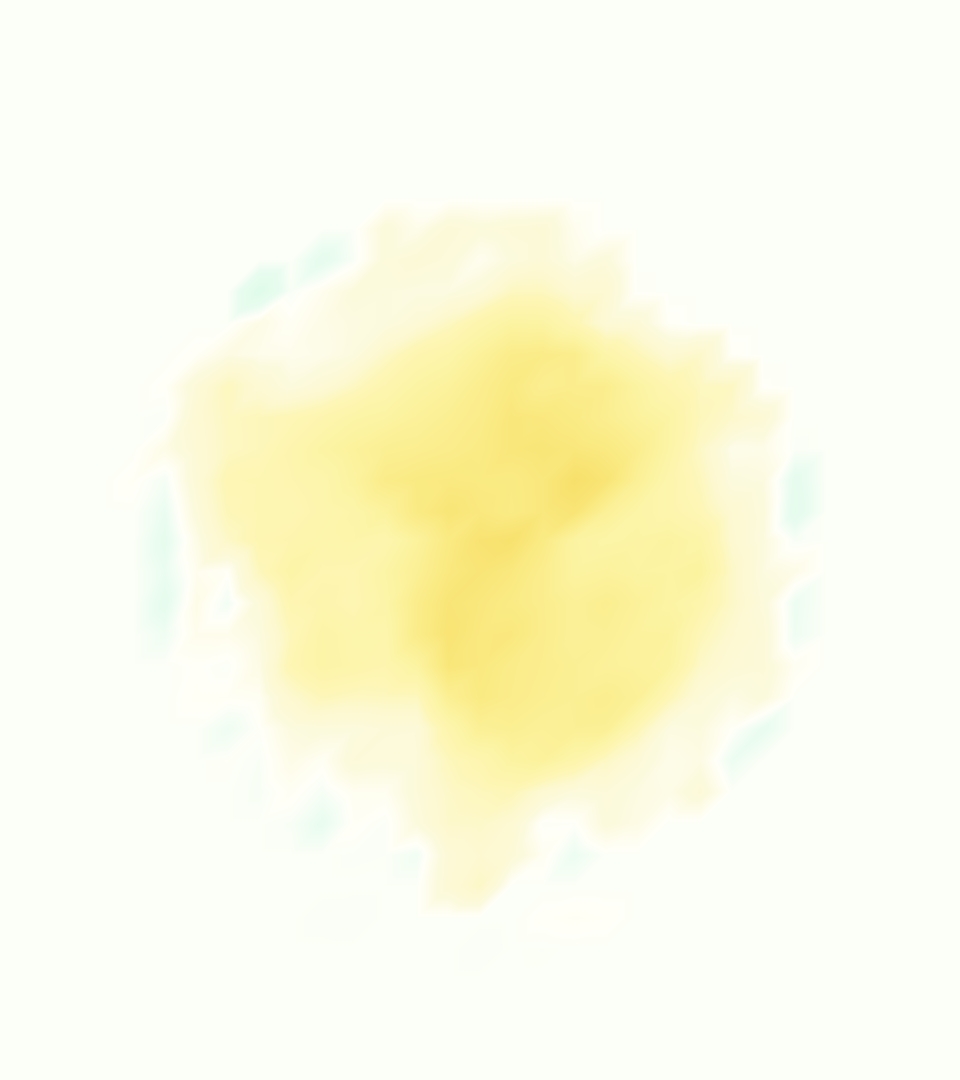} & 
\includegraphics[width=.14\textwidth]{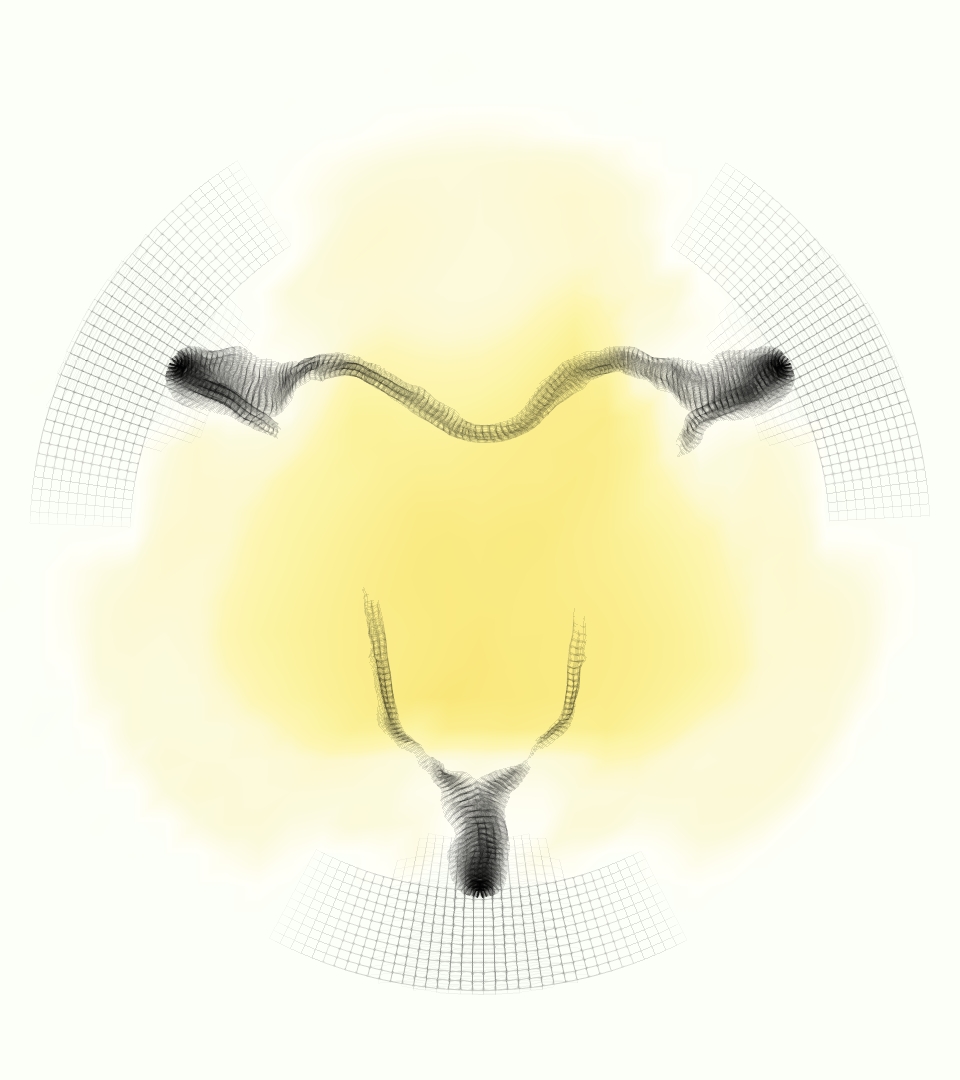} & 
\includegraphics[width=.14\textwidth]{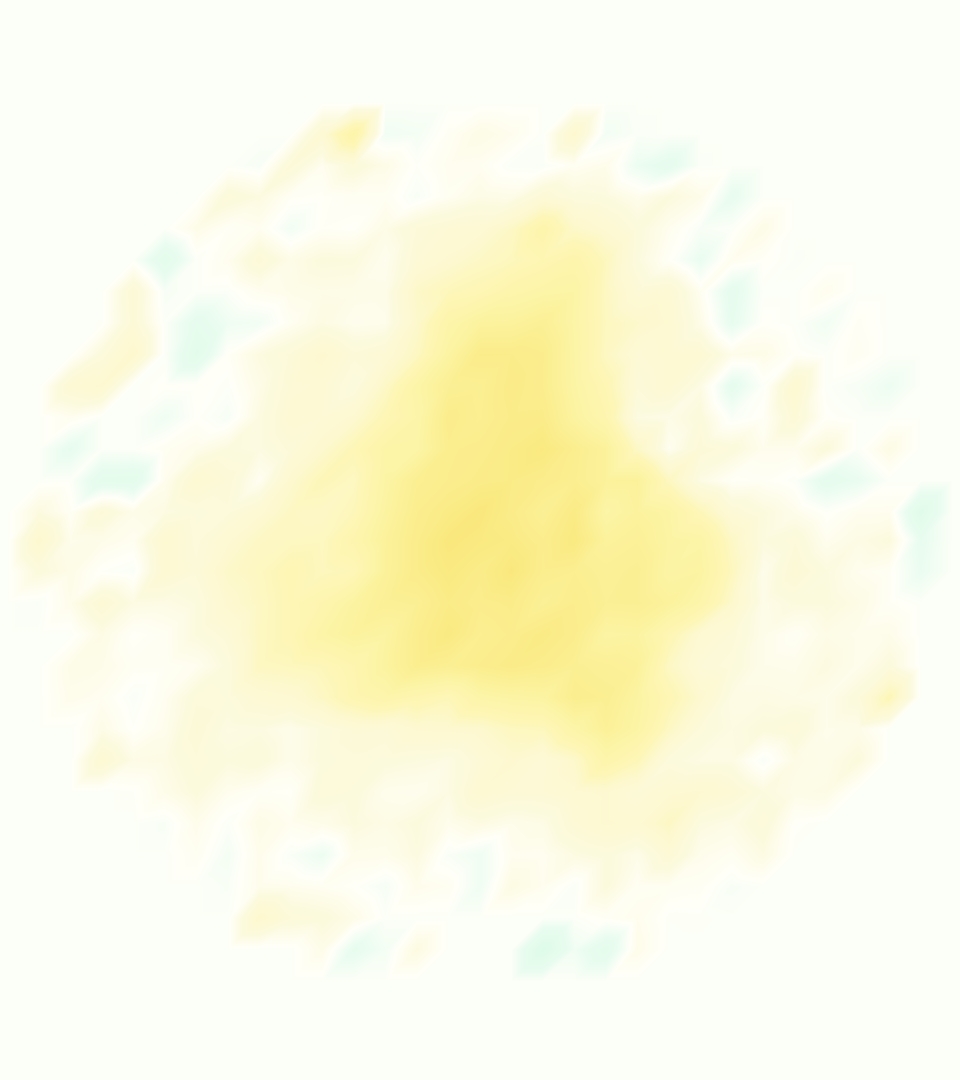} & 
\includegraphics[width=.14\textwidth]{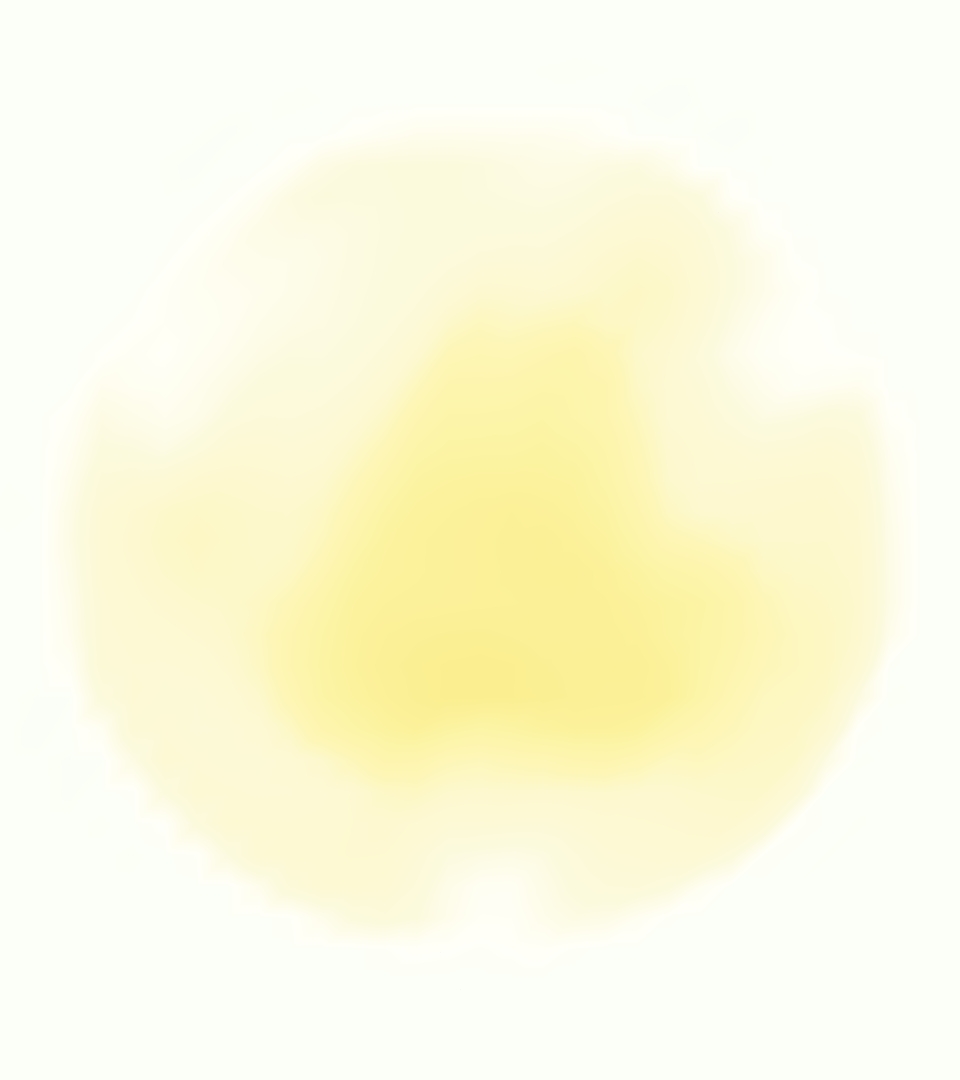} & 
\includegraphics[width=.14\textwidth]{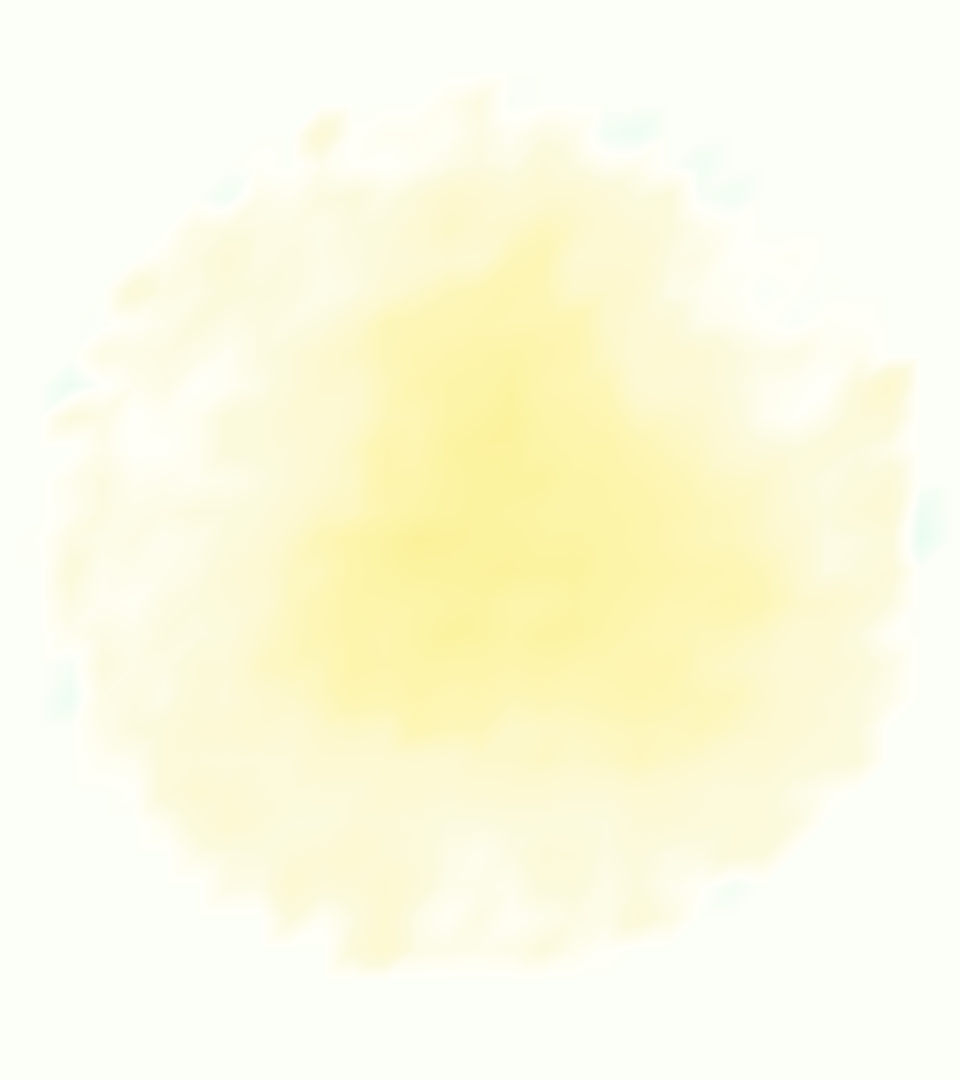}  \\ 
\rotatebox[origin=l]{90}{$t = 0.21 $ s} & 
\includegraphics[width=.14\textwidth]{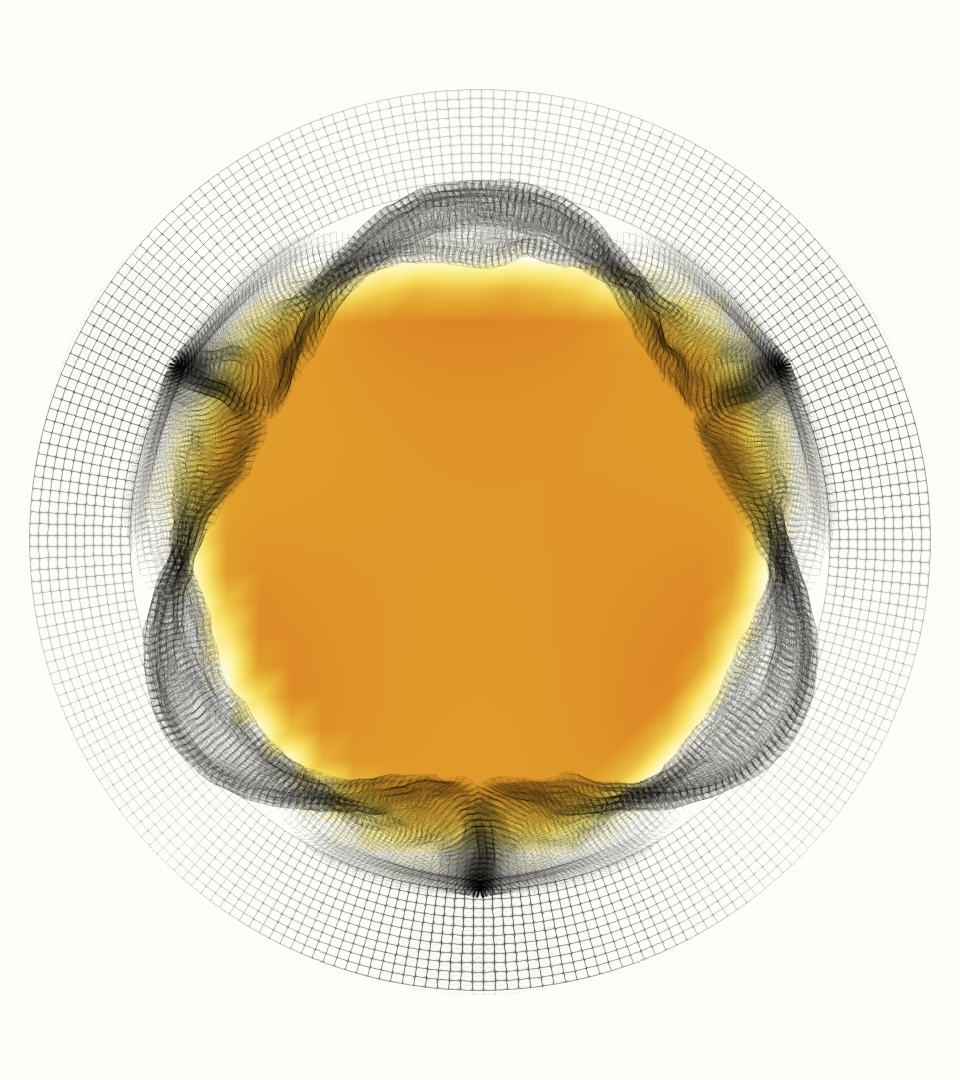} & 
\includegraphics[width=.14\textwidth]{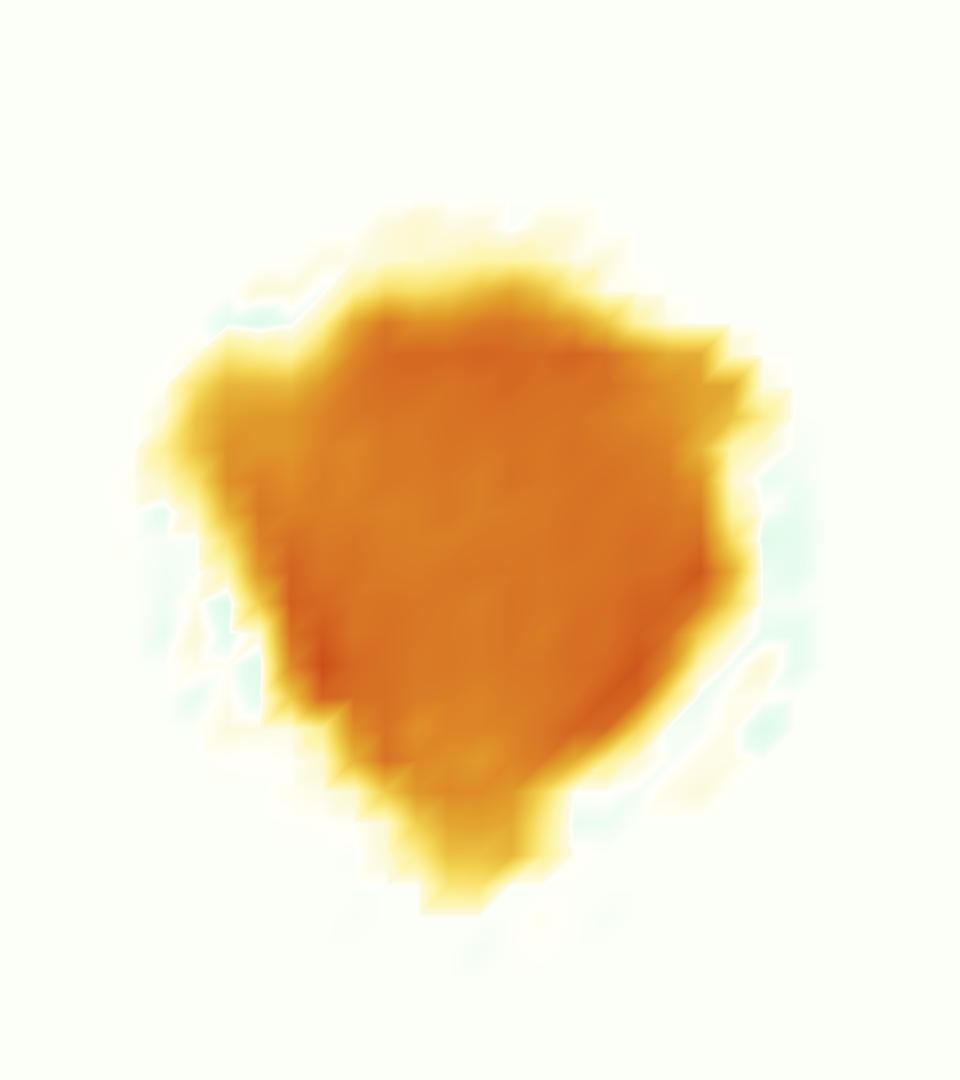} & 
\includegraphics[width=.14\textwidth]{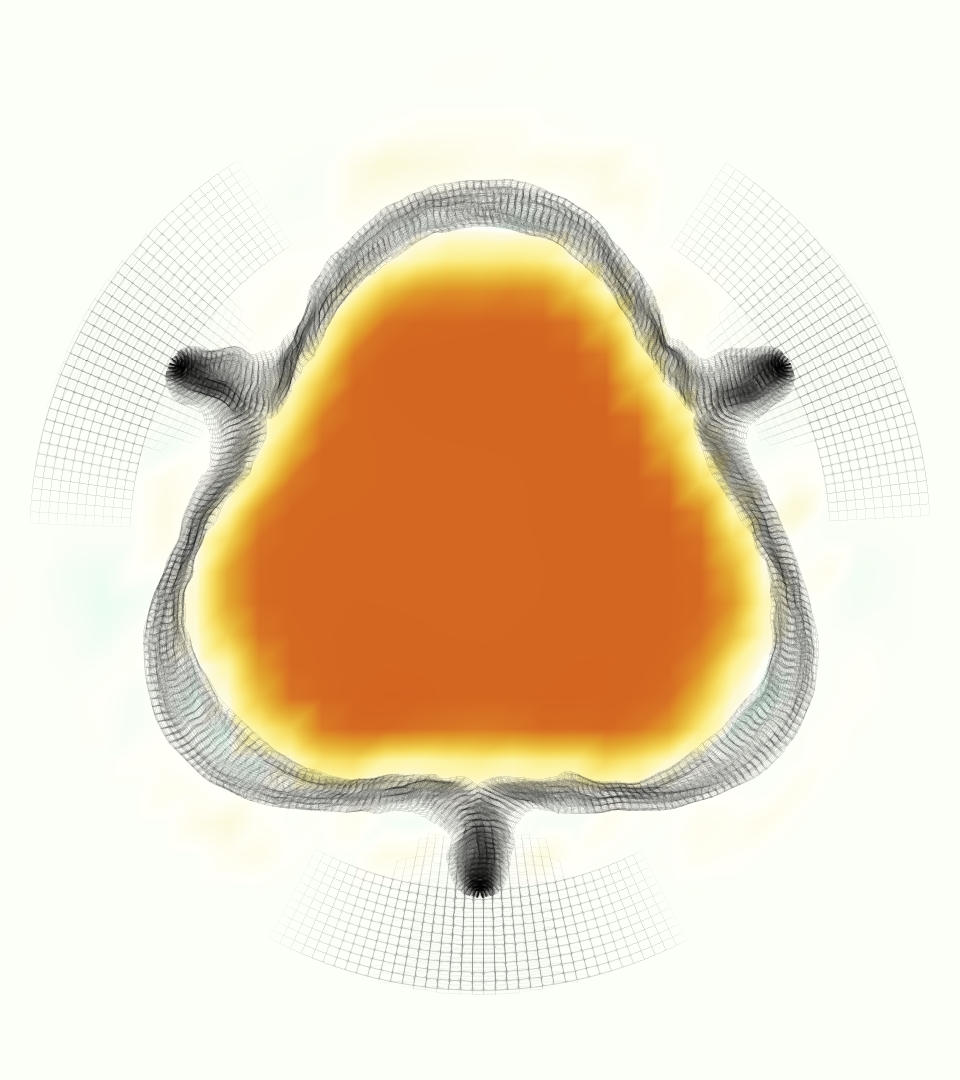} & 
\includegraphics[width=.14\textwidth]{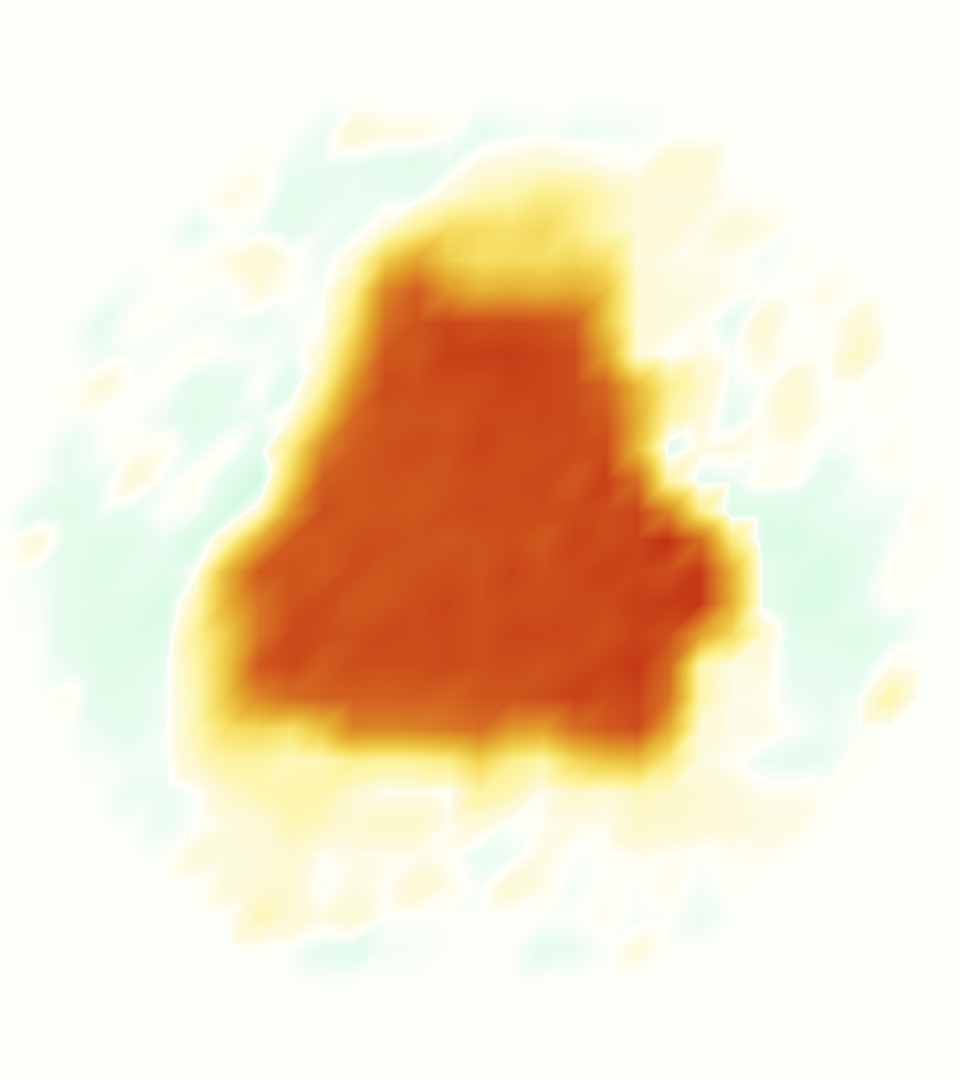} & 
\includegraphics[width=.14\textwidth]{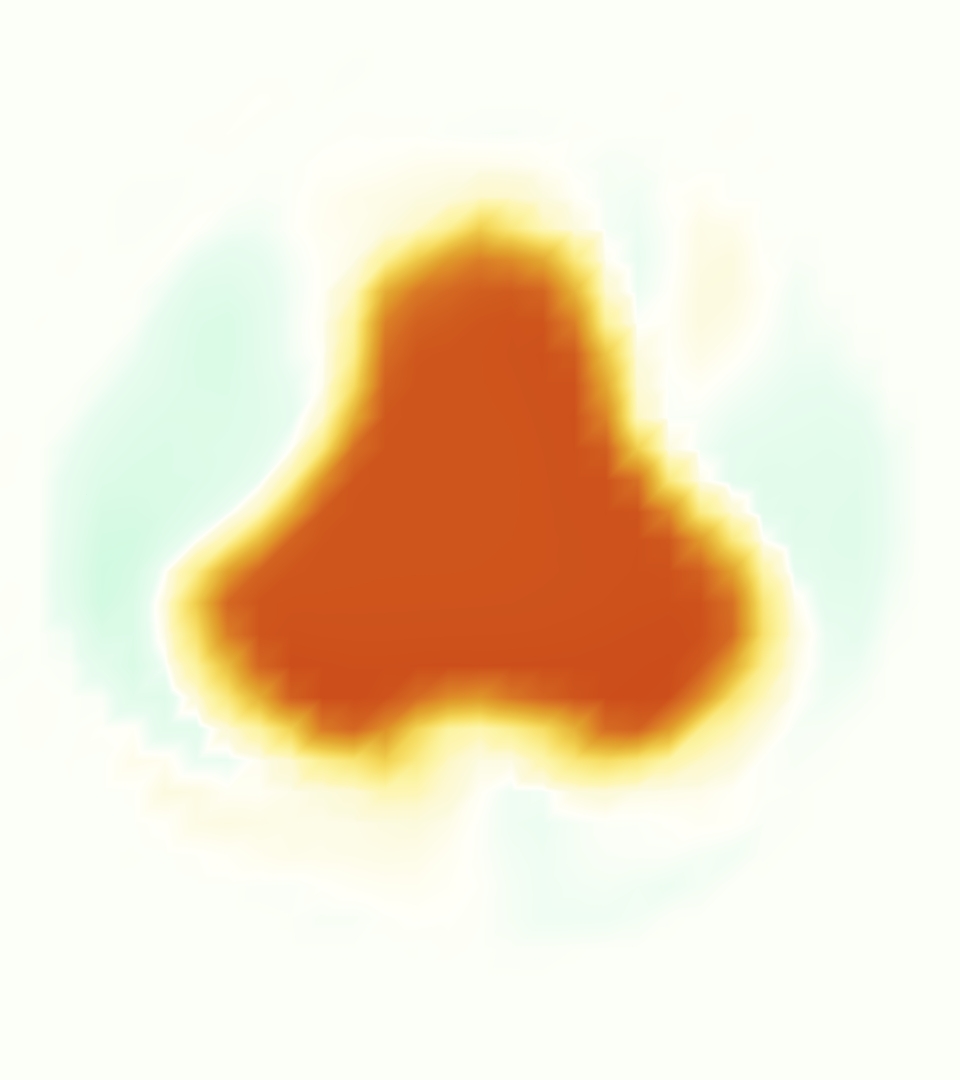} & 
\includegraphics[width=.14\textwidth]{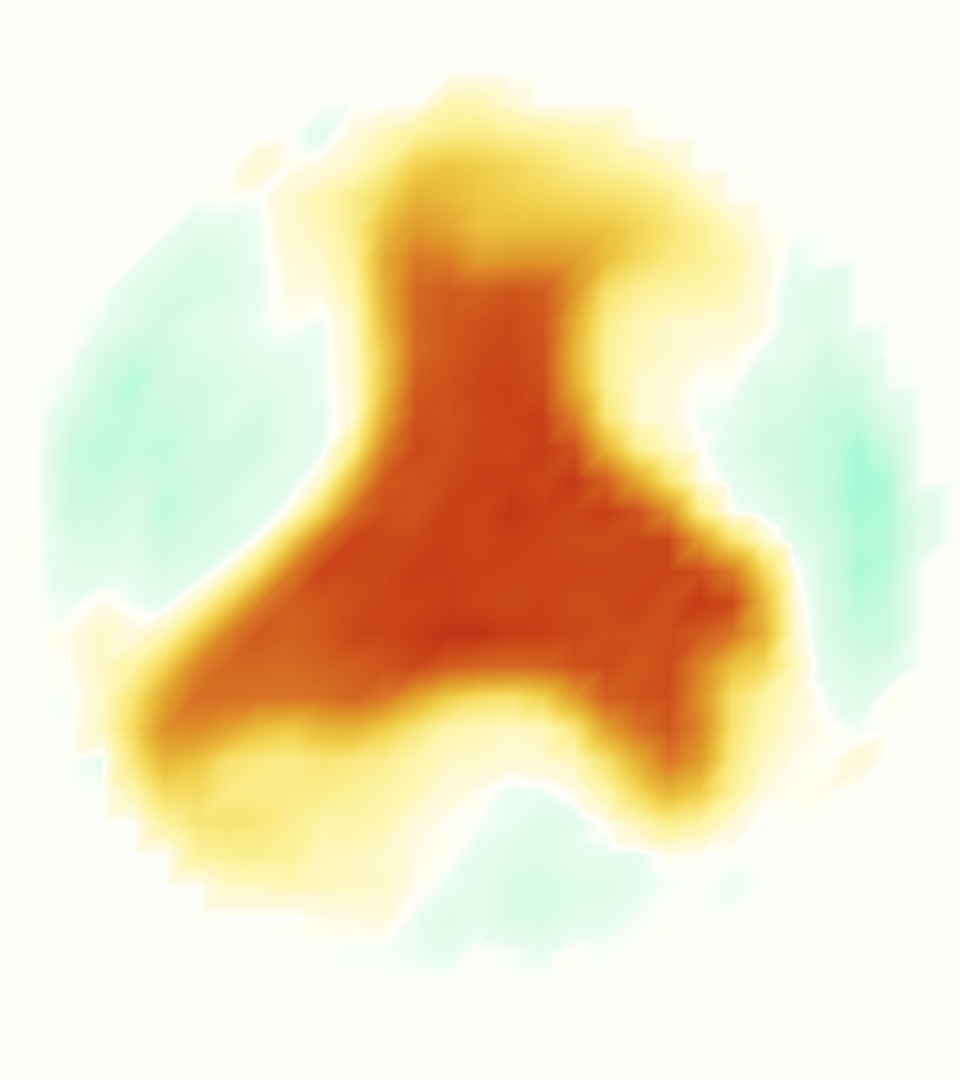}  \\ 
\rotatebox[origin=l]{90}{$t = 0.29 $ s} & 
\includegraphics[width=.14\textwidth]{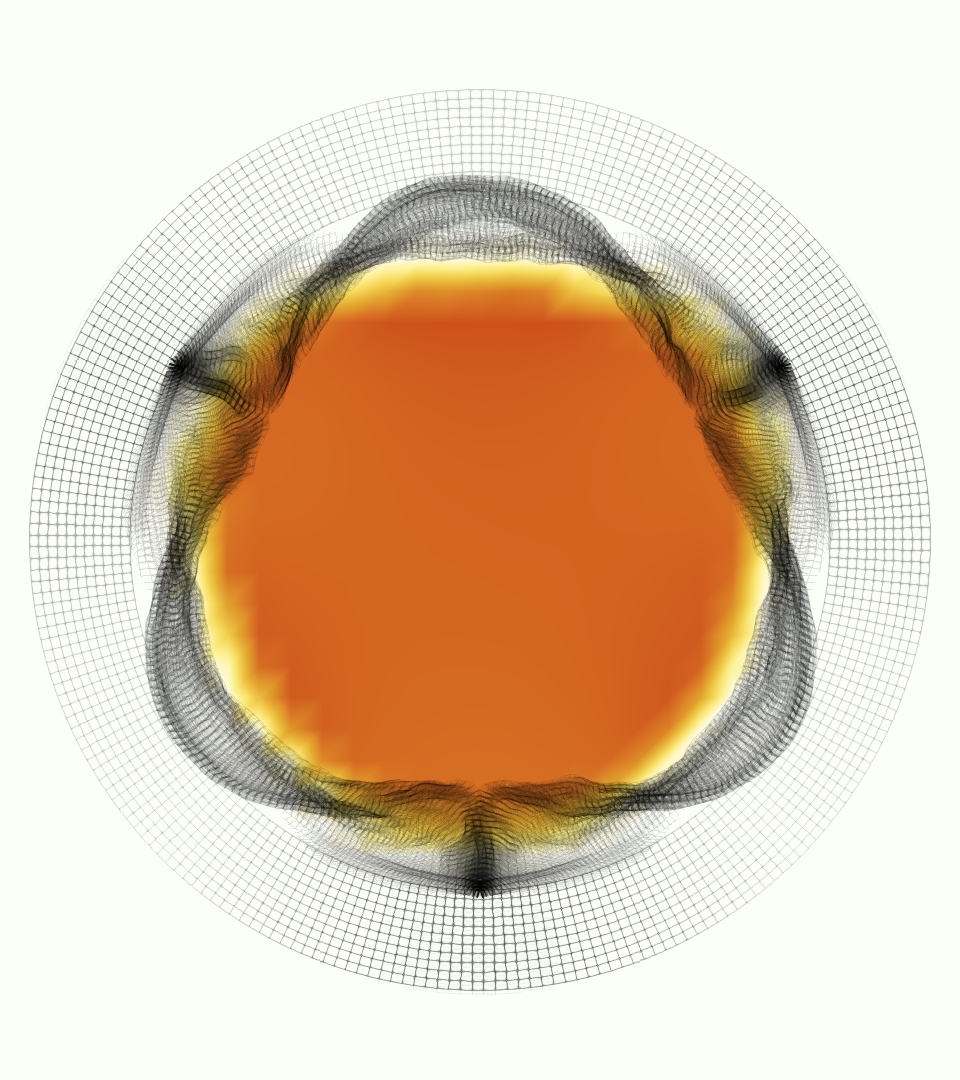} & 
\includegraphics[width=.14\textwidth]{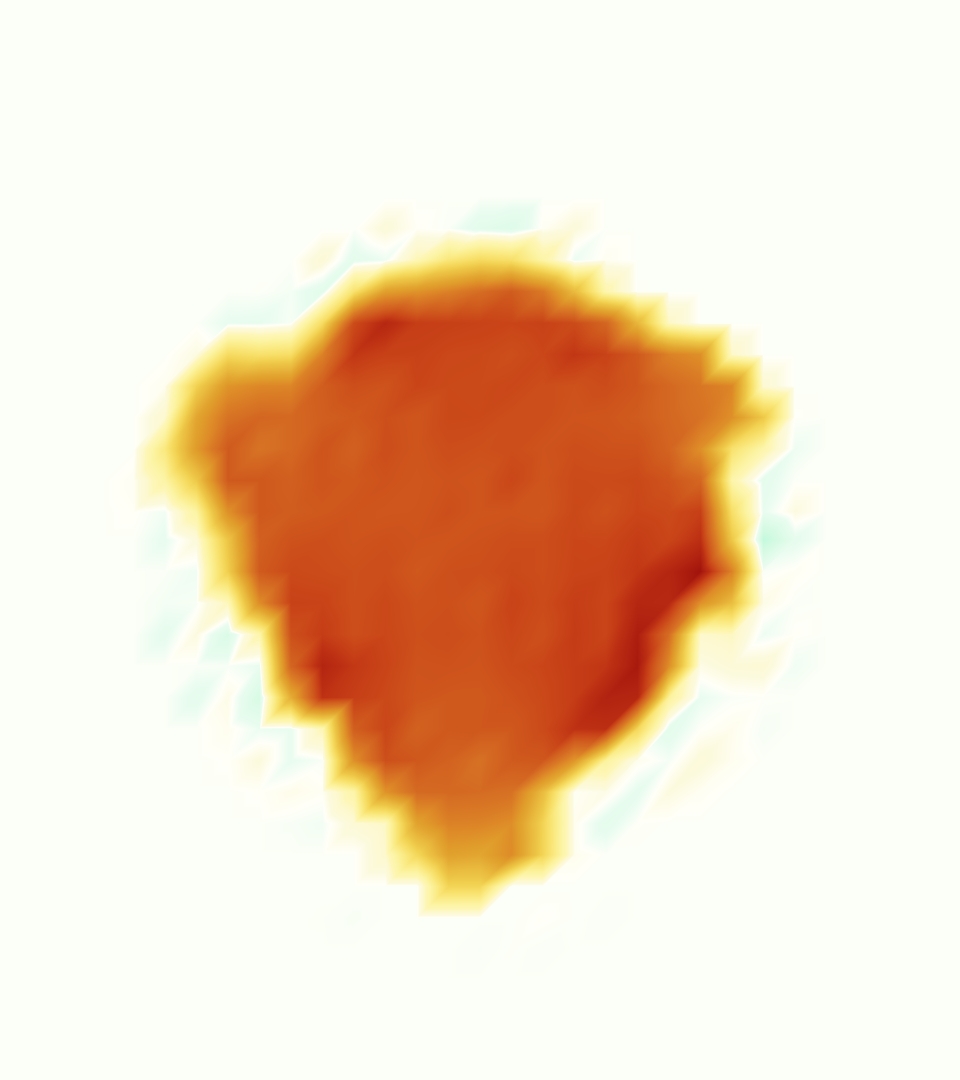} & 
\includegraphics[width=.14\textwidth]{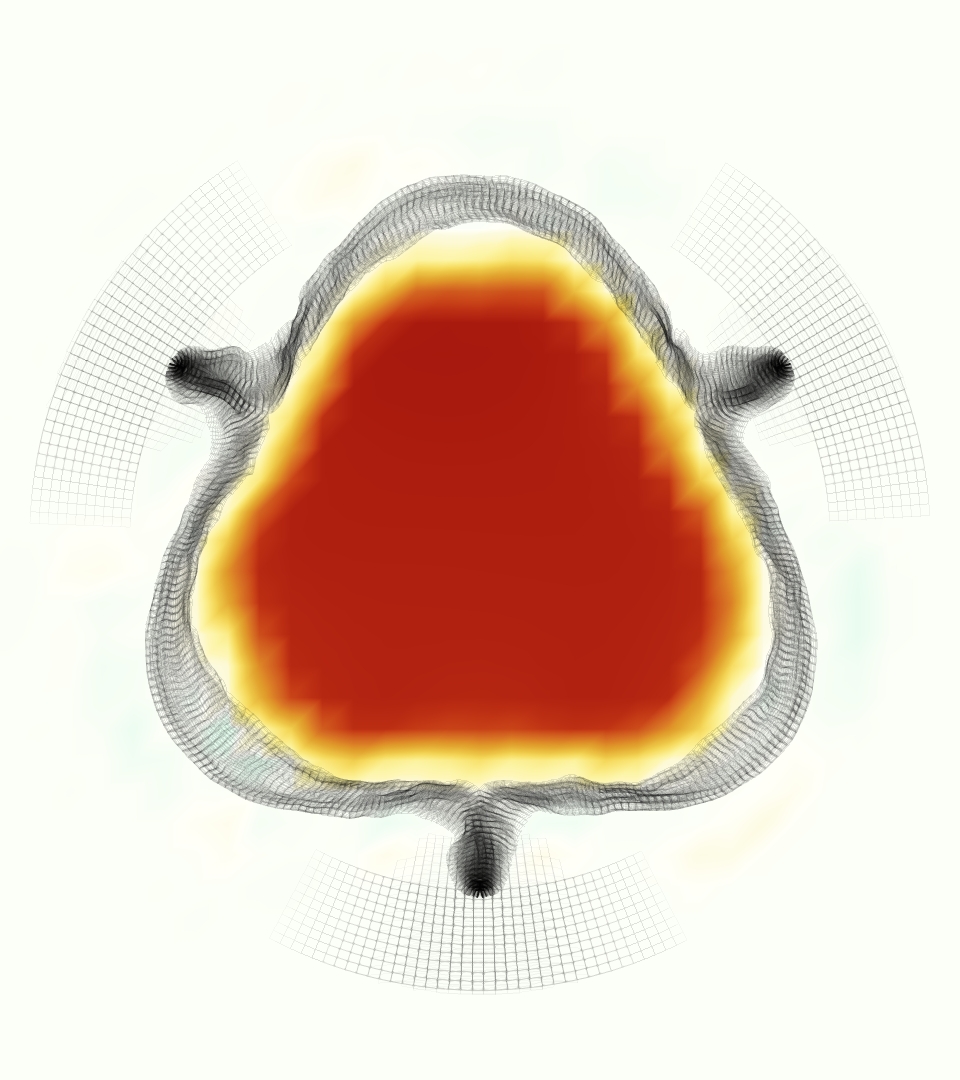} & 
\includegraphics[width=.14\textwidth]{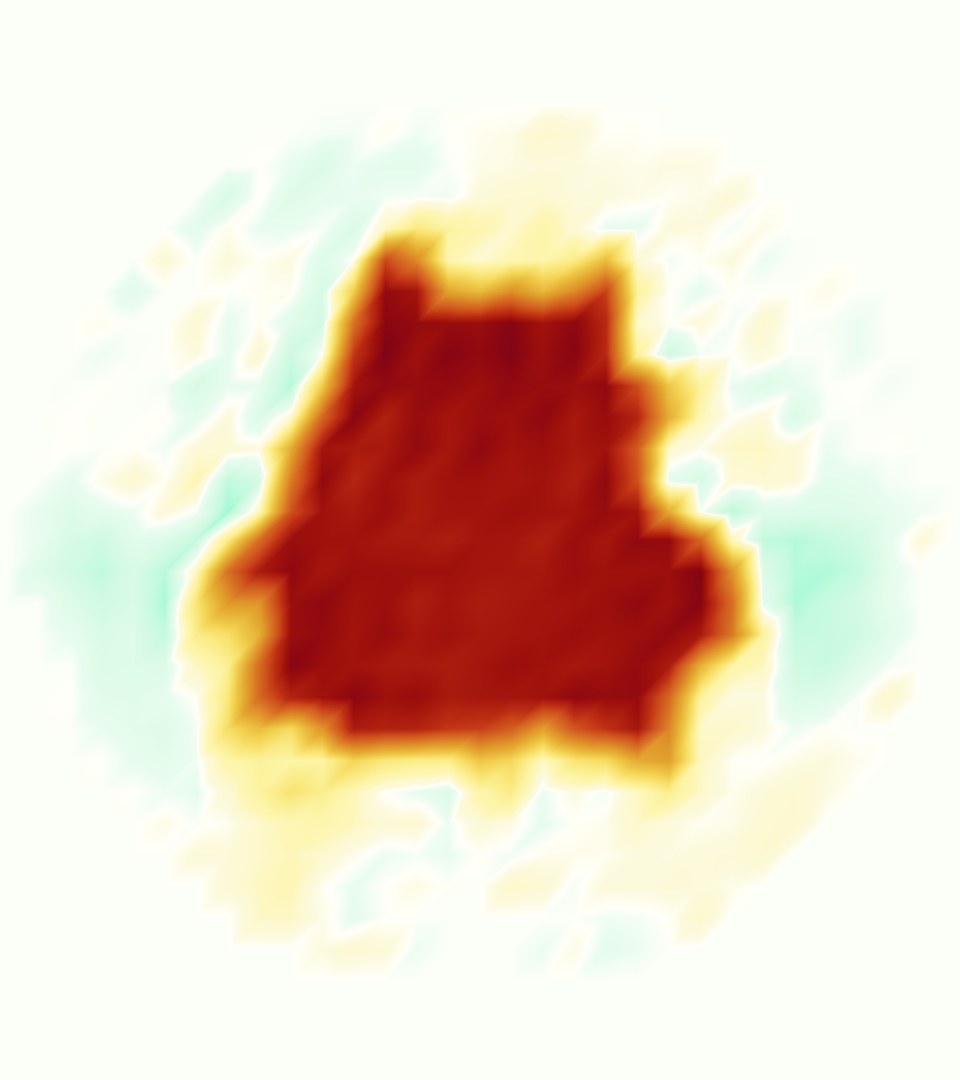} & 
\includegraphics[width=.14\textwidth]{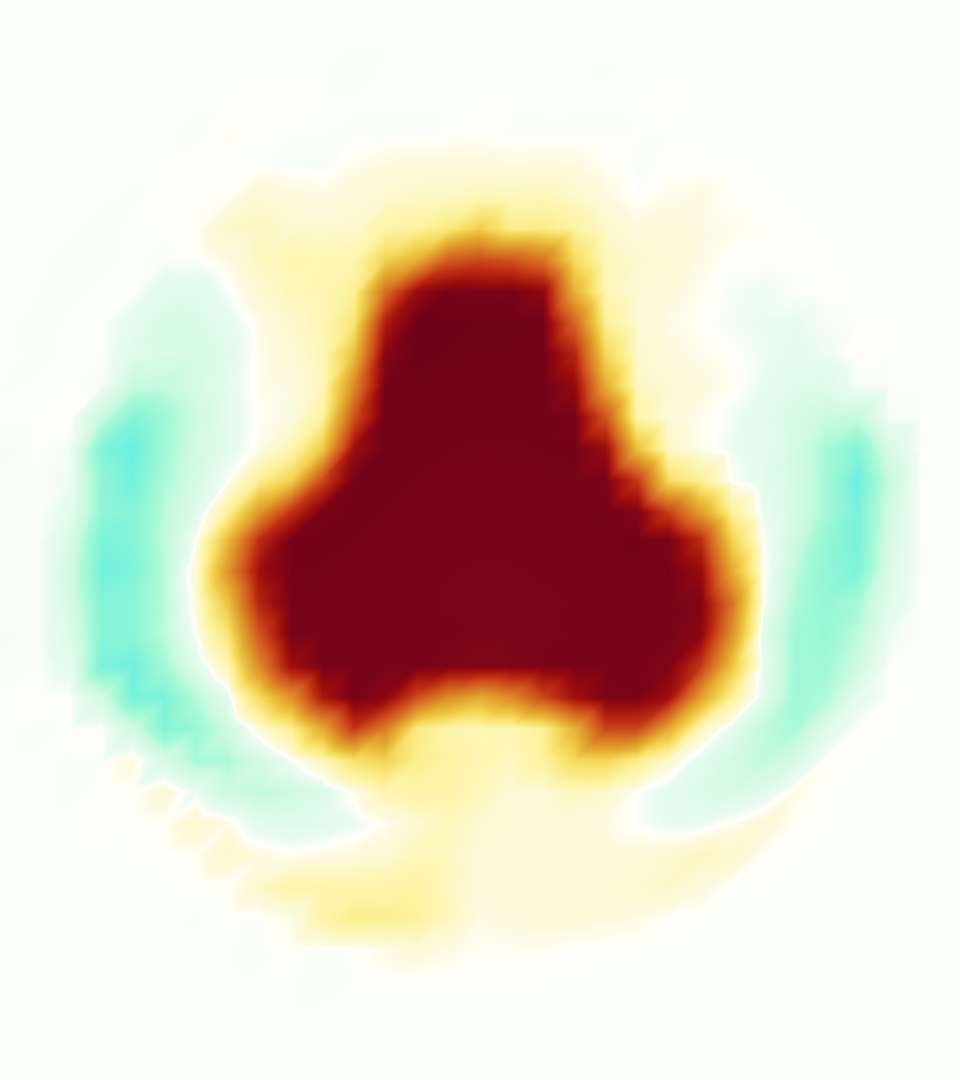} & 
\includegraphics[width=.14\textwidth]{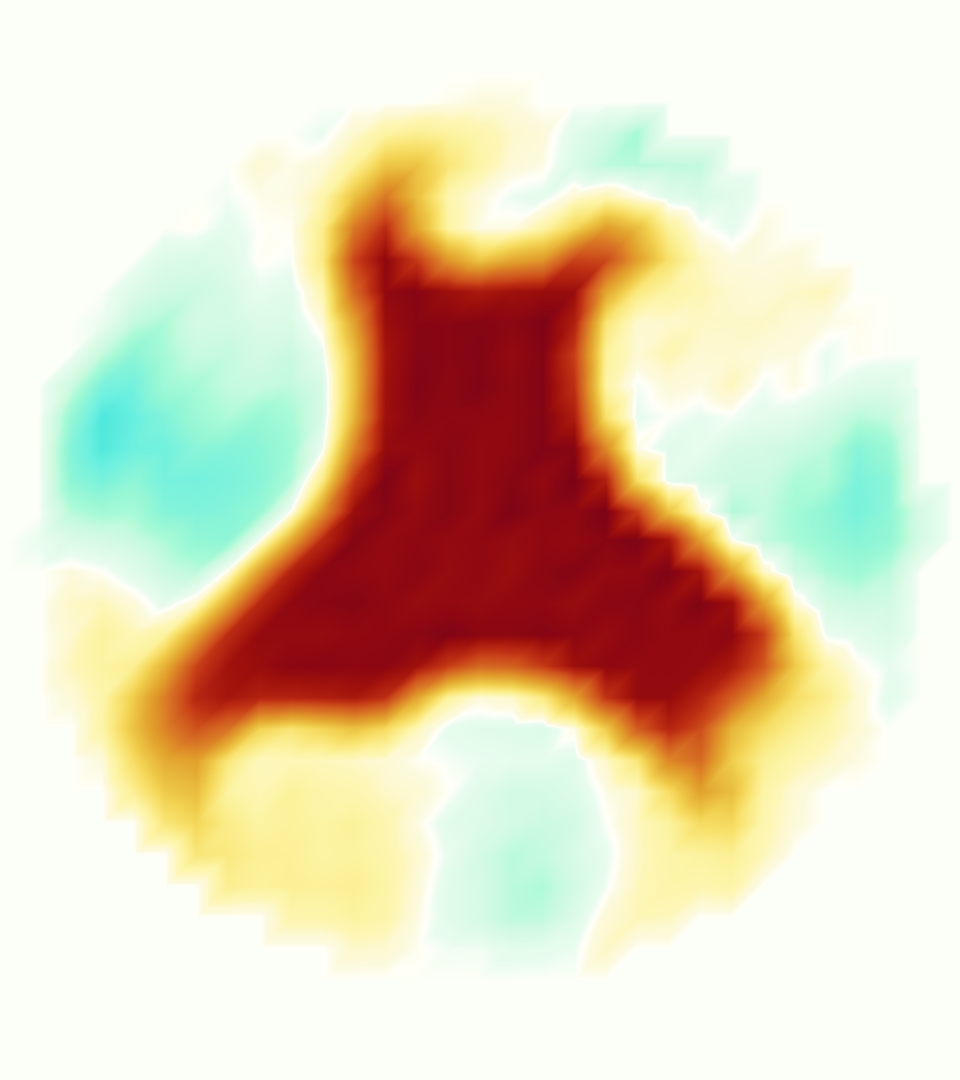}  \\ 
\rotatebox[origin=l]{90}{$t = 0.37 $ s} & 
\includegraphics[width=.14\textwidth]{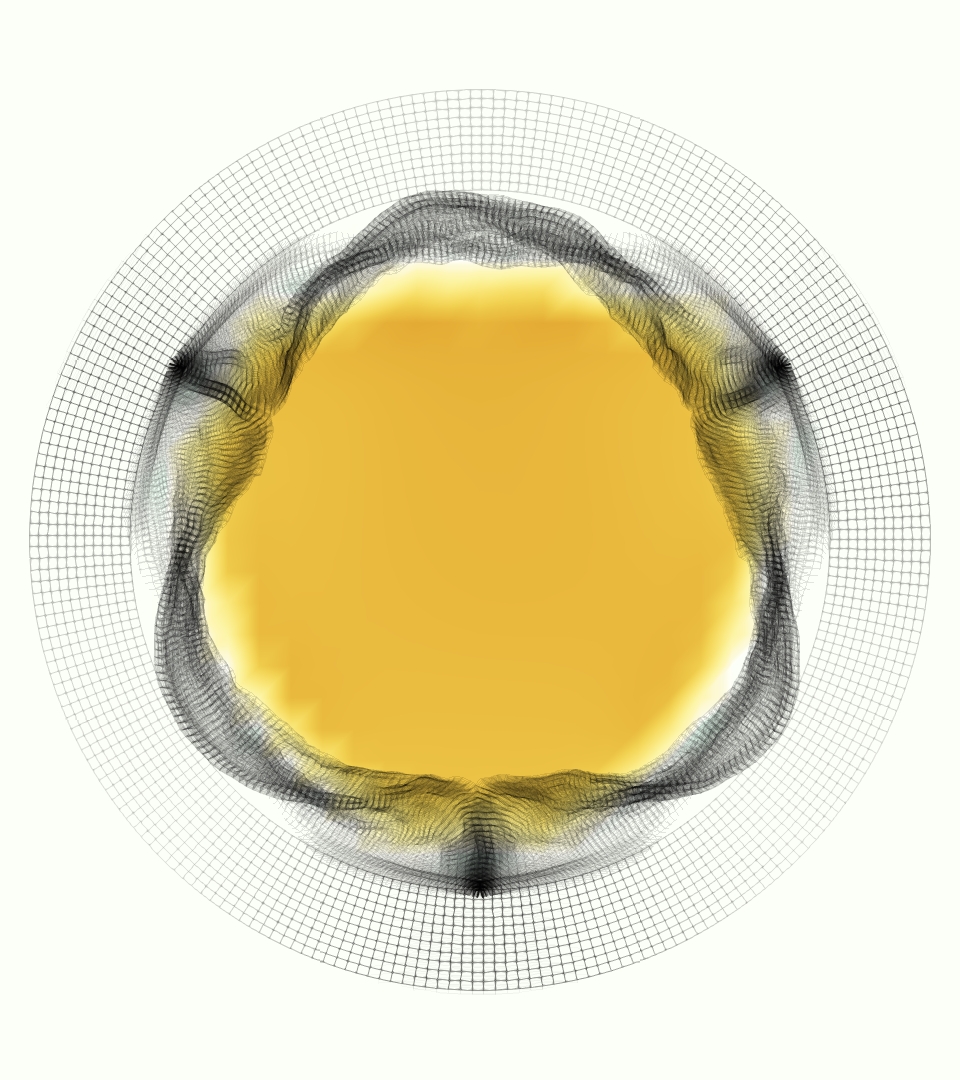} & 
\includegraphics[width=.14\textwidth]{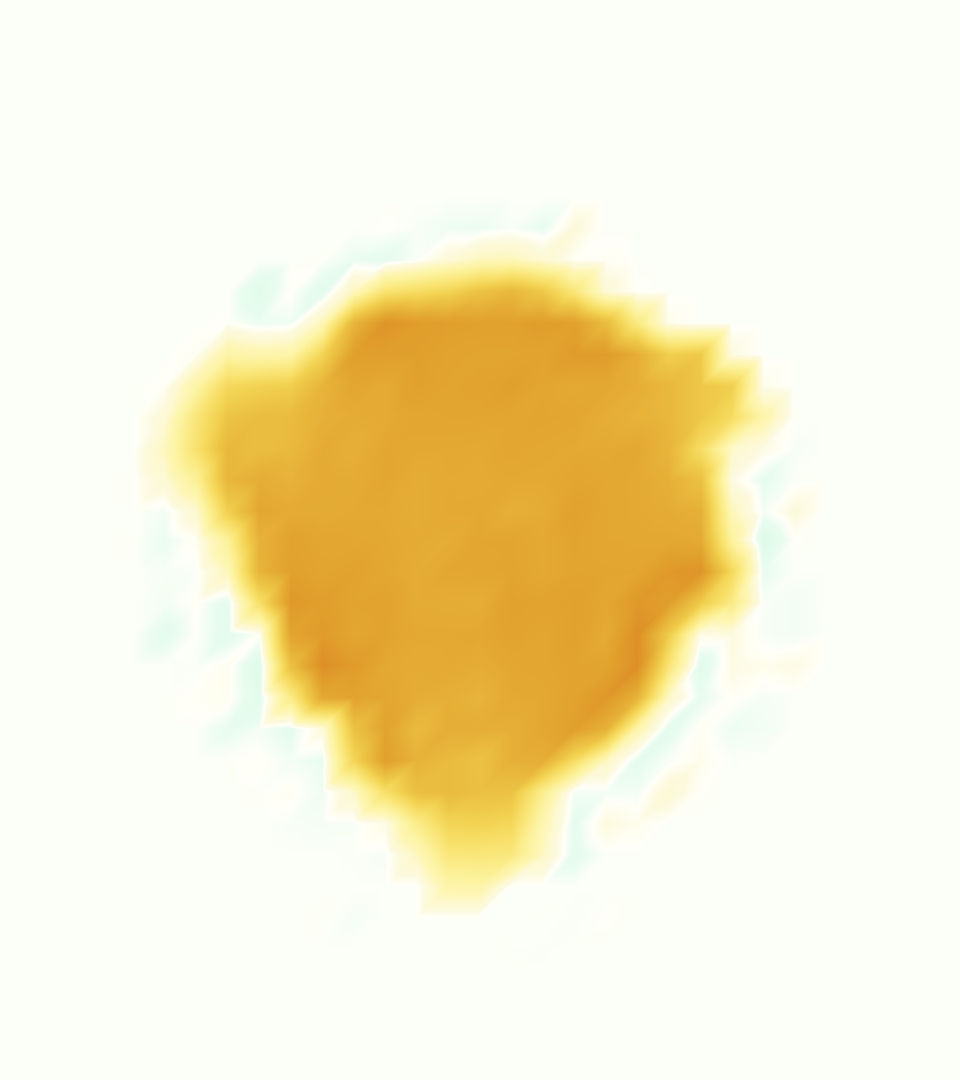} & 
\includegraphics[width=.14\textwidth]{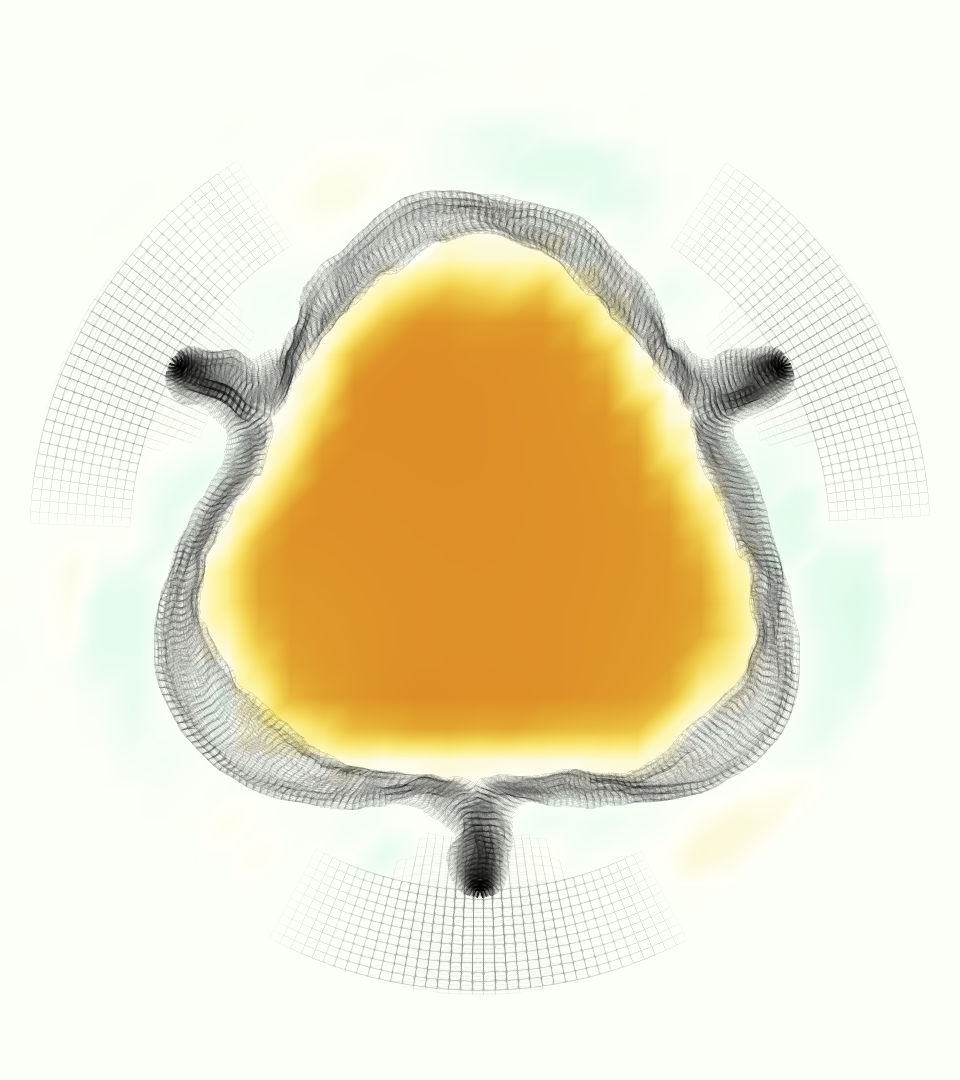} & 
\includegraphics[width=.14\textwidth]{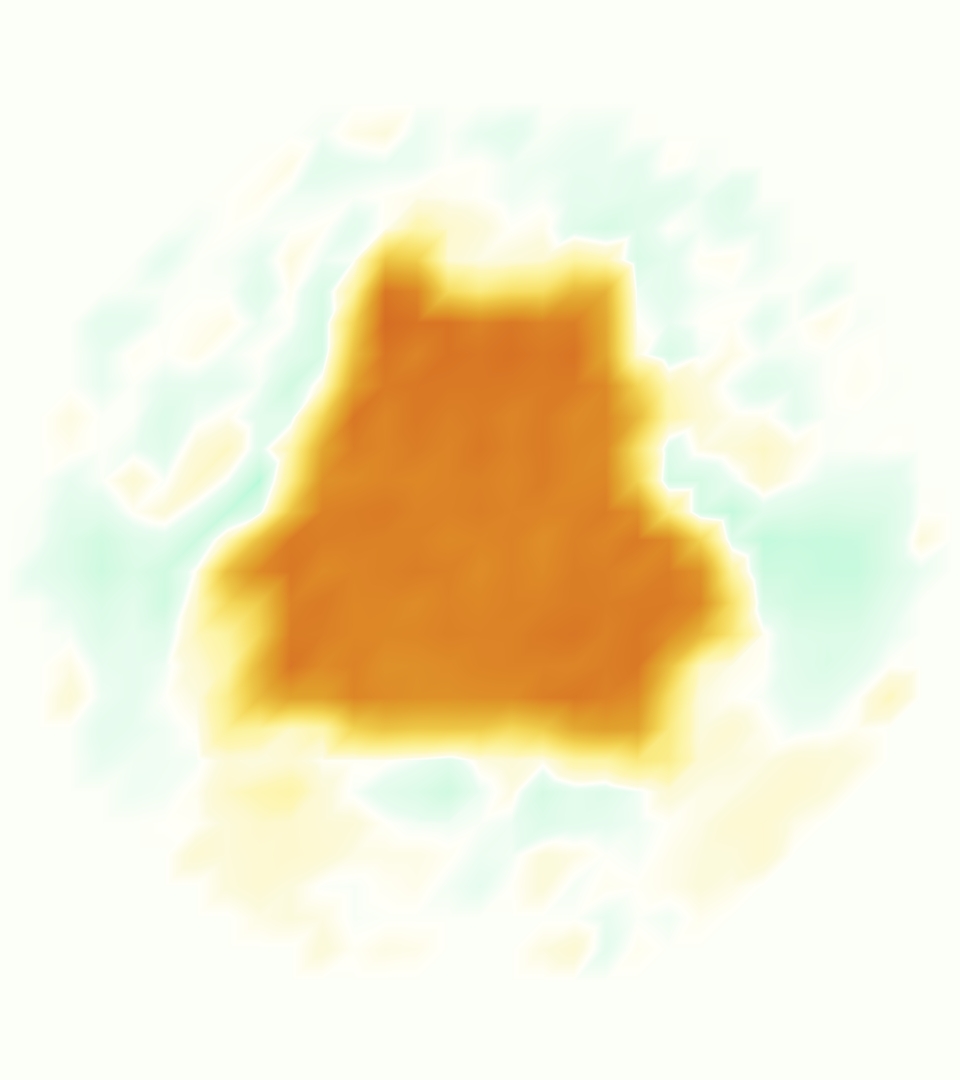} & 
\includegraphics[width=.14\textwidth]{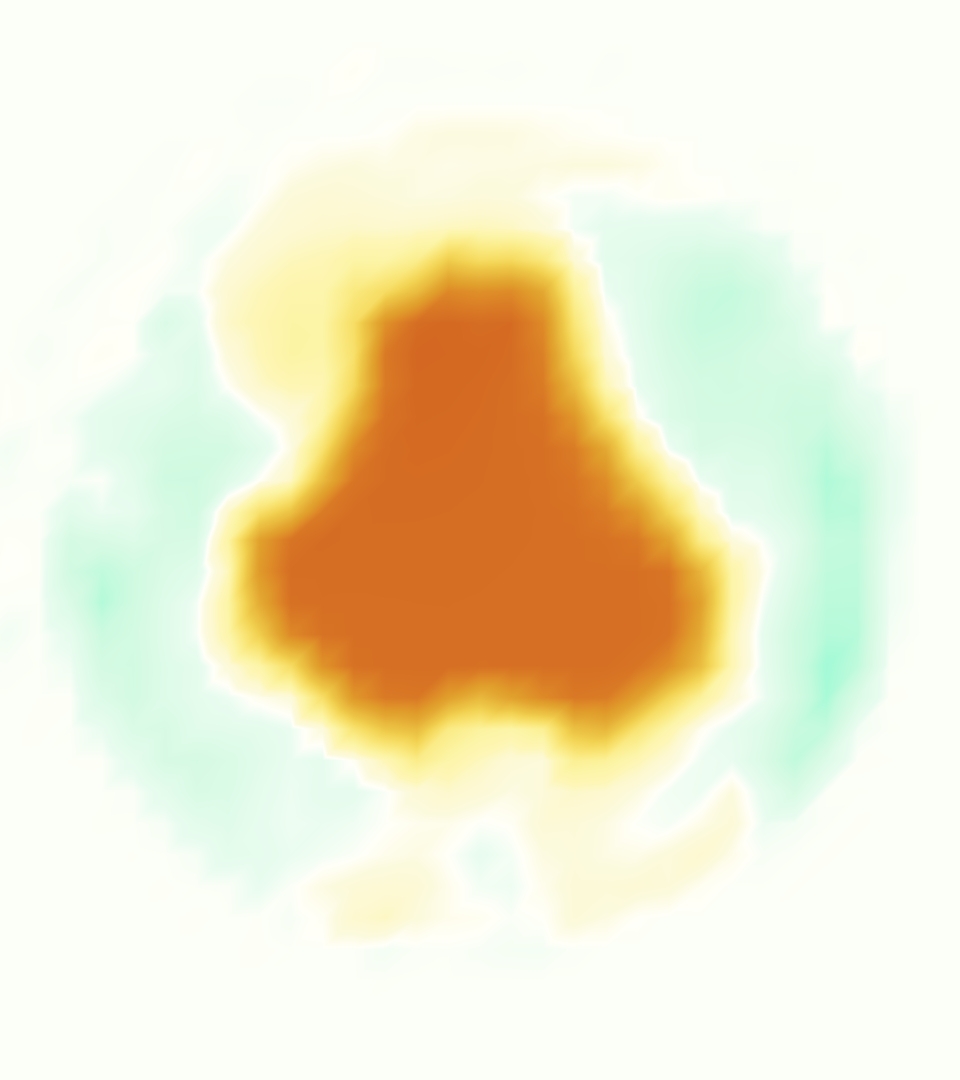} & 
\includegraphics[width=.14\textwidth]{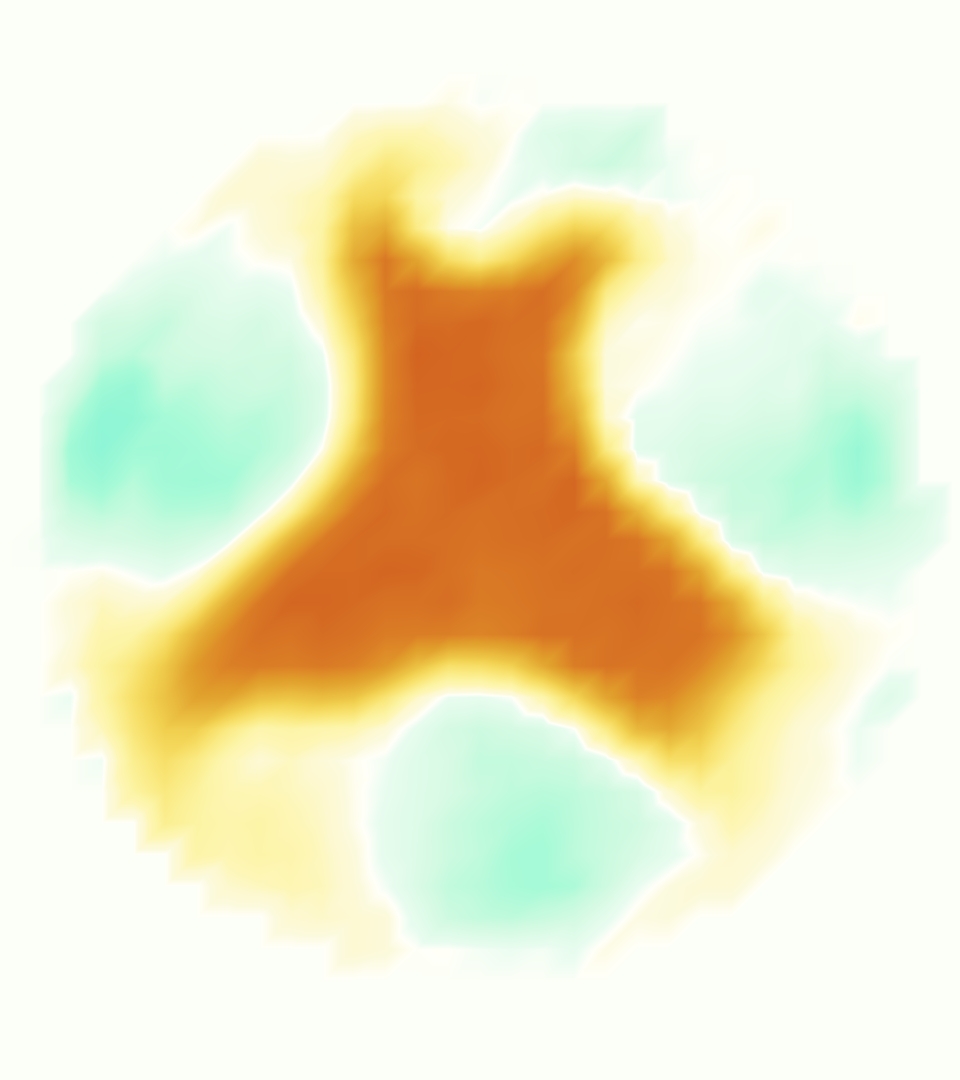}  \\ 
\rotatebox[origin=l]{90}{$t = 0.46 $ s} & 
\includegraphics[width=.14\textwidth]{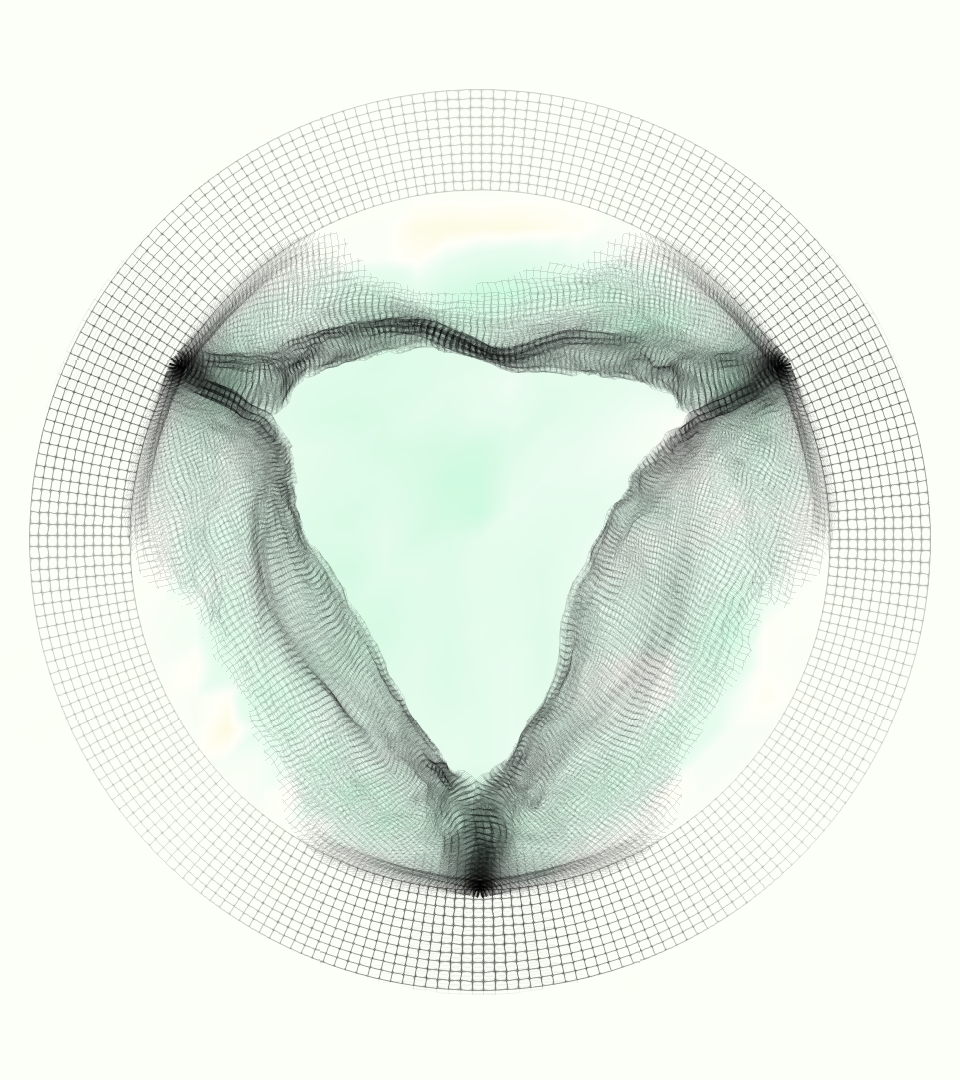} & 
\includegraphics[width=.14\textwidth]{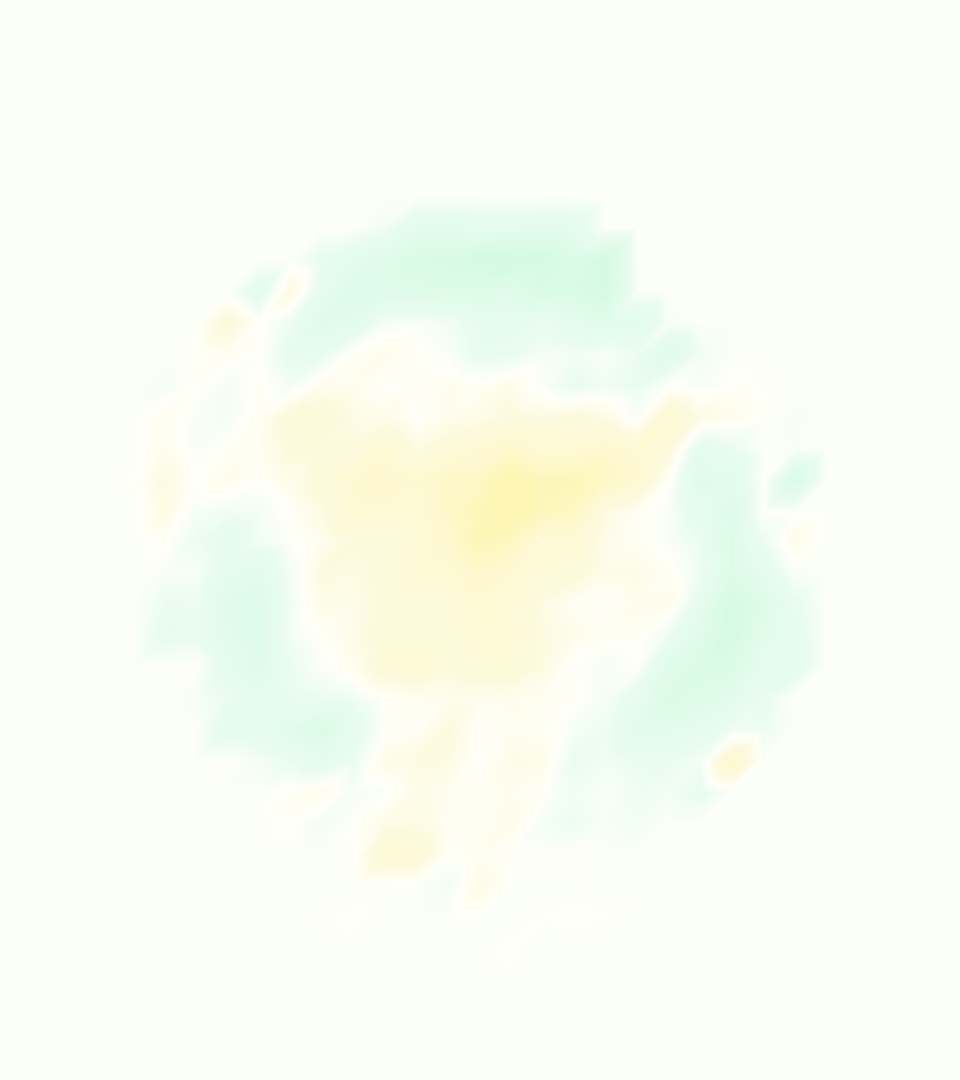} & 
\includegraphics[width=.14\textwidth]{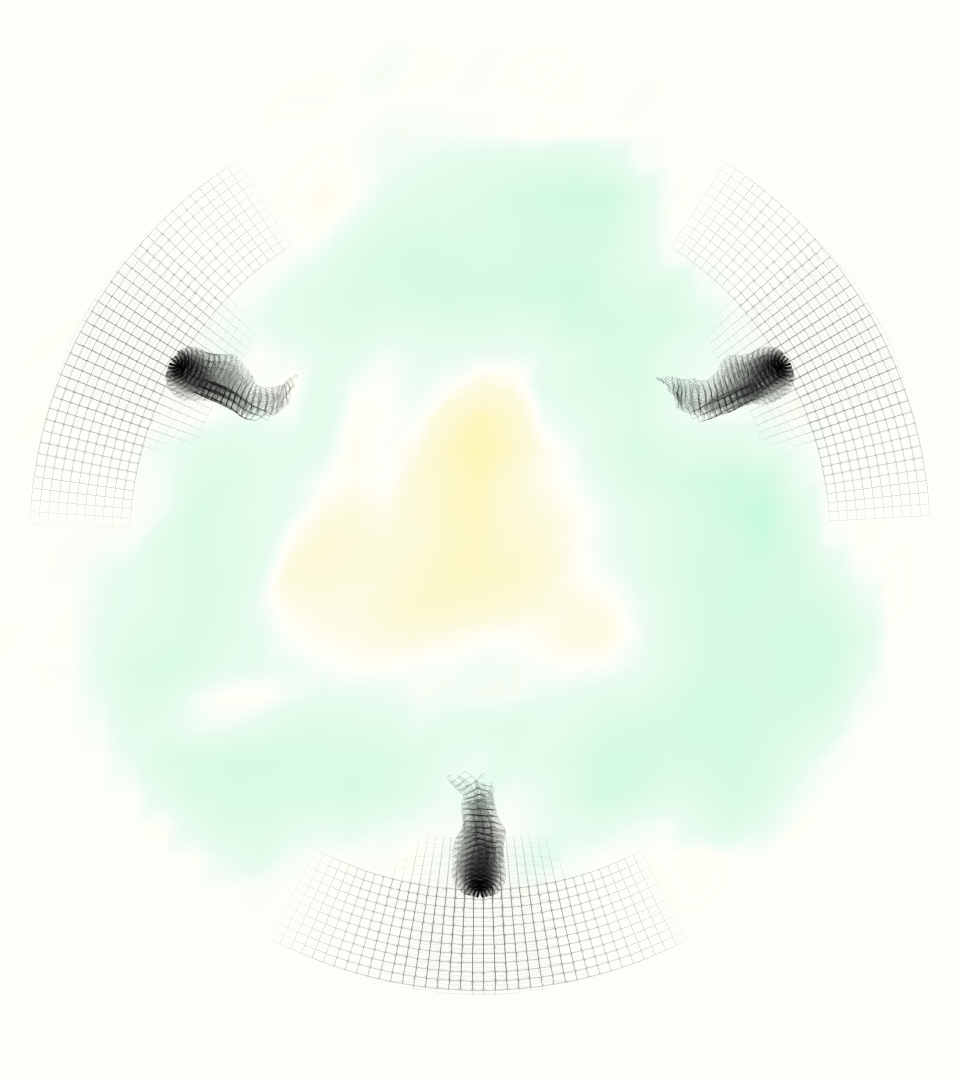} & 
\includegraphics[width=.14\textwidth]{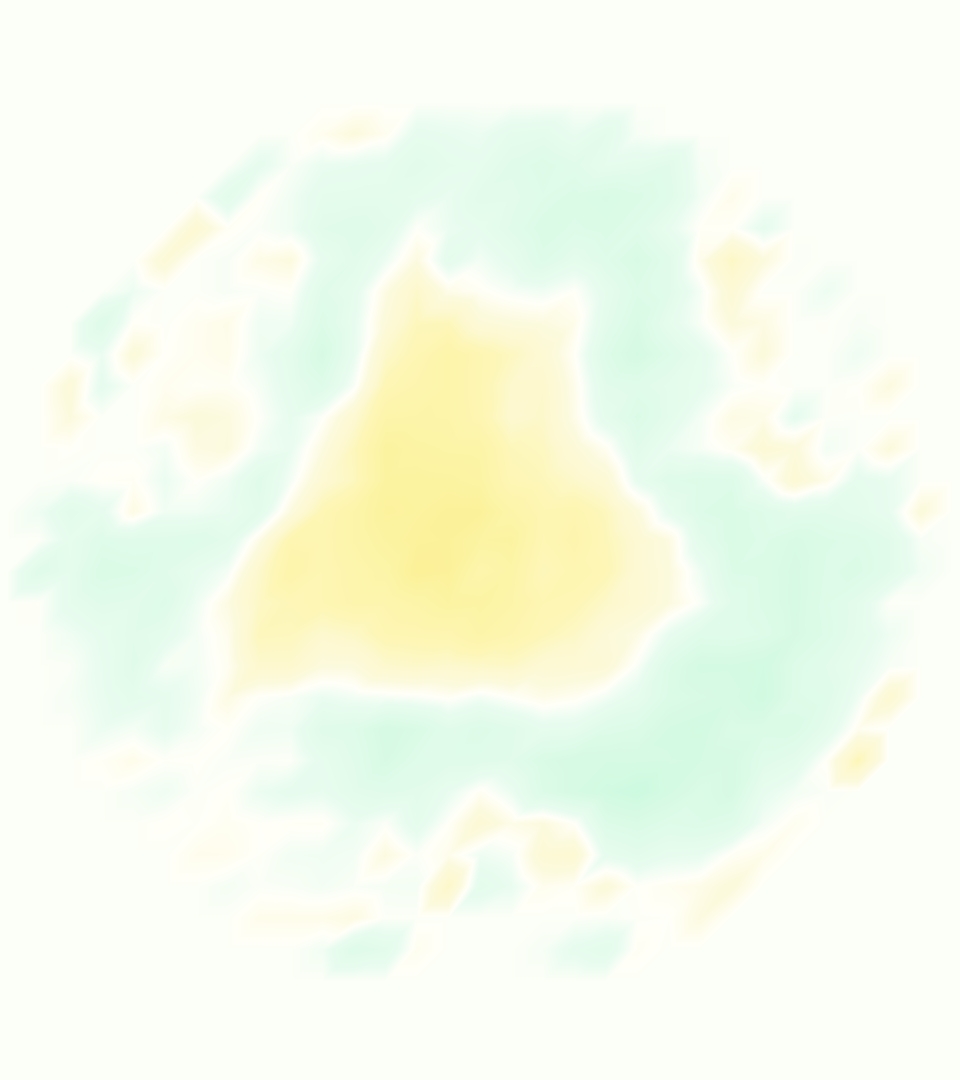} & 
\includegraphics[width=.14\textwidth]{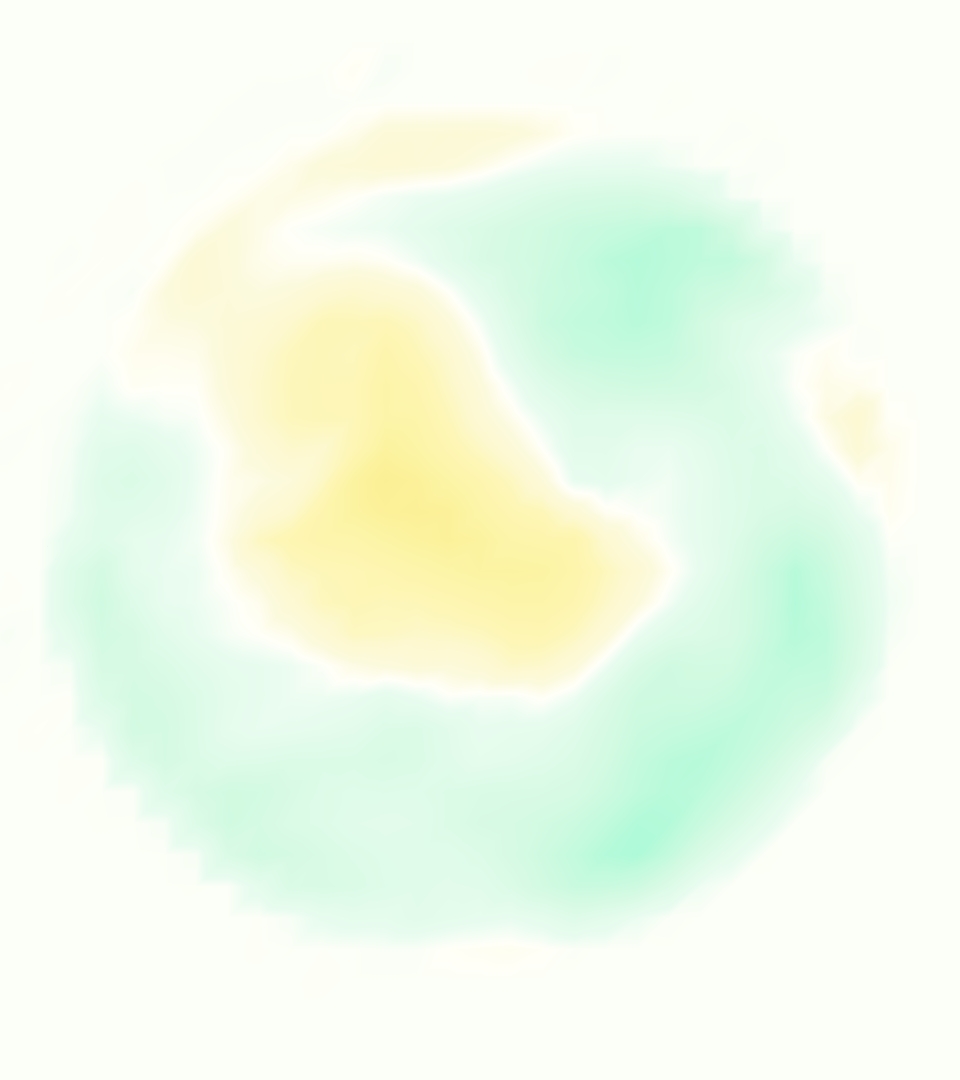} & 
\includegraphics[width=.14\textwidth]{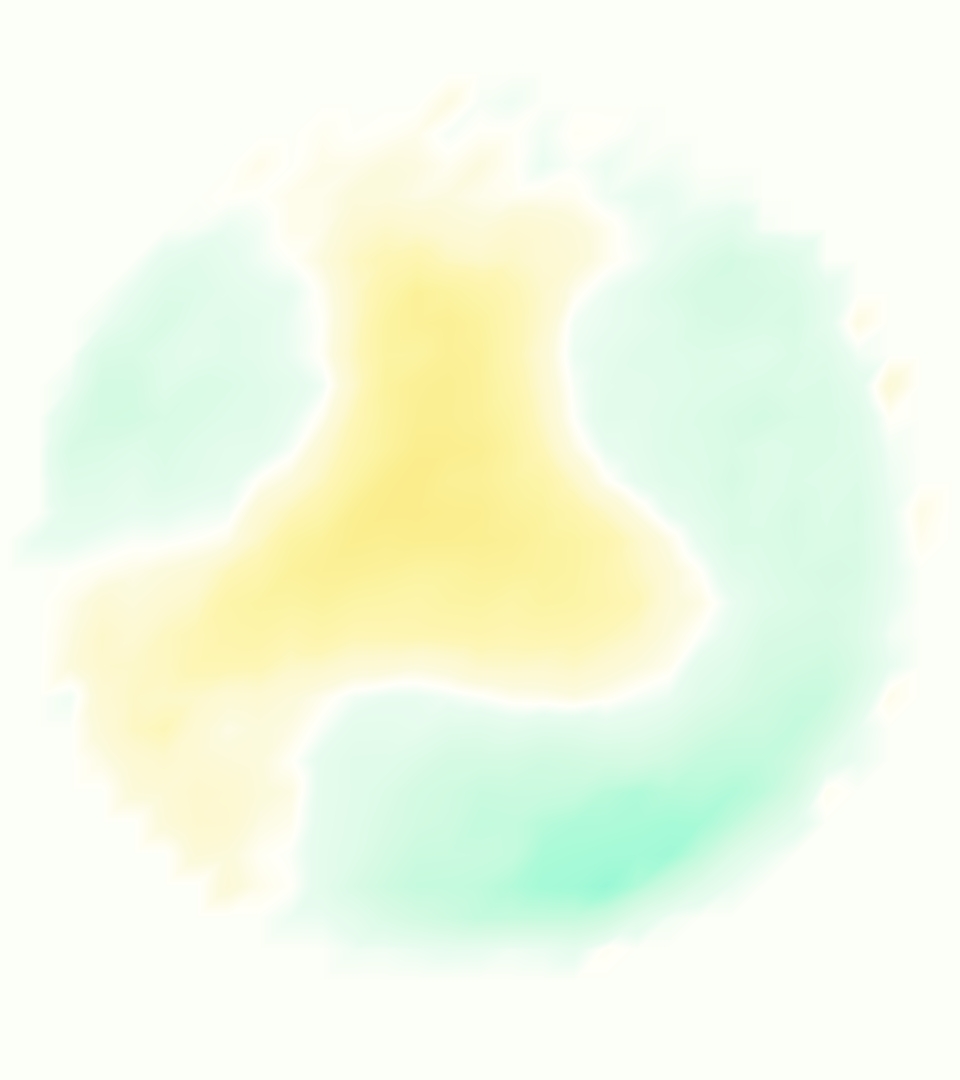}  \\ 
\rotatebox[origin=l]{90}{$t = 0.54 $ s} & 
\includegraphics[width=.14\textwidth]{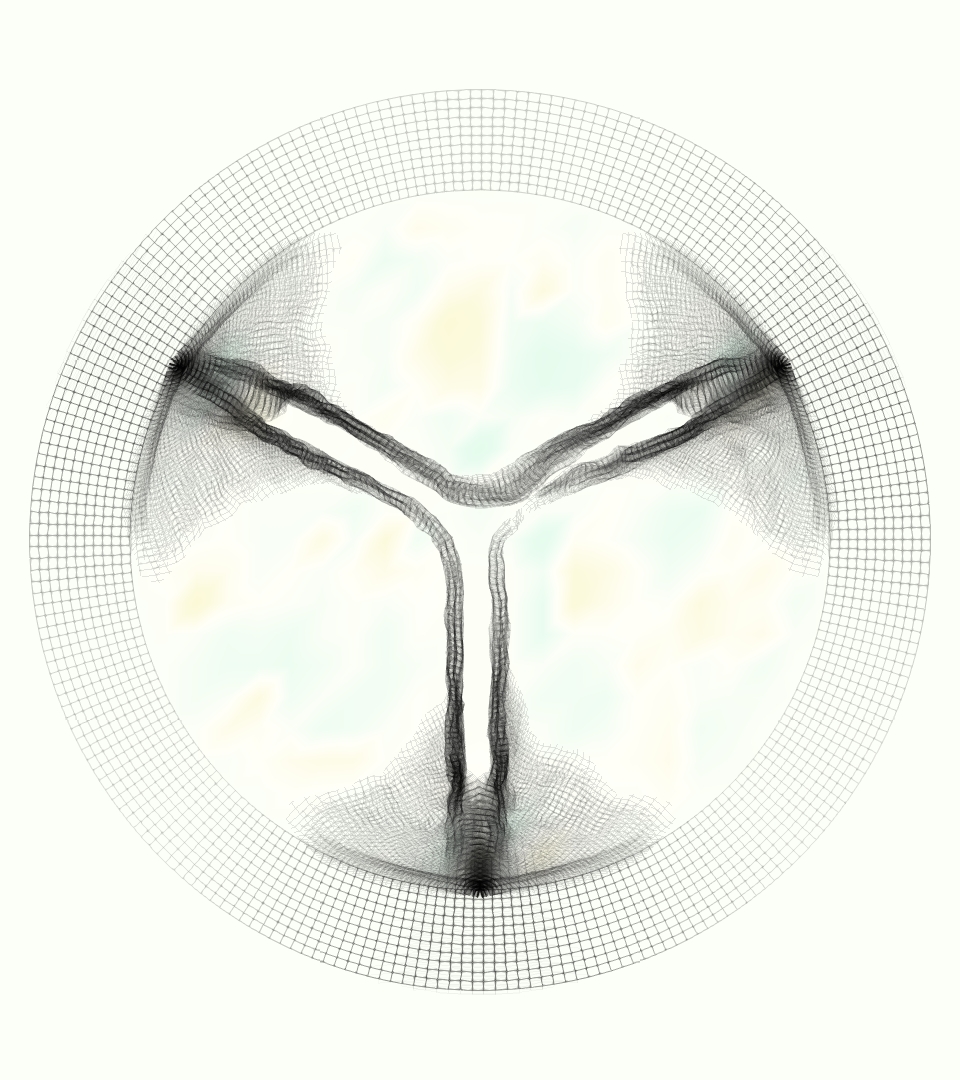} & 
\includegraphics[width=.14\textwidth]{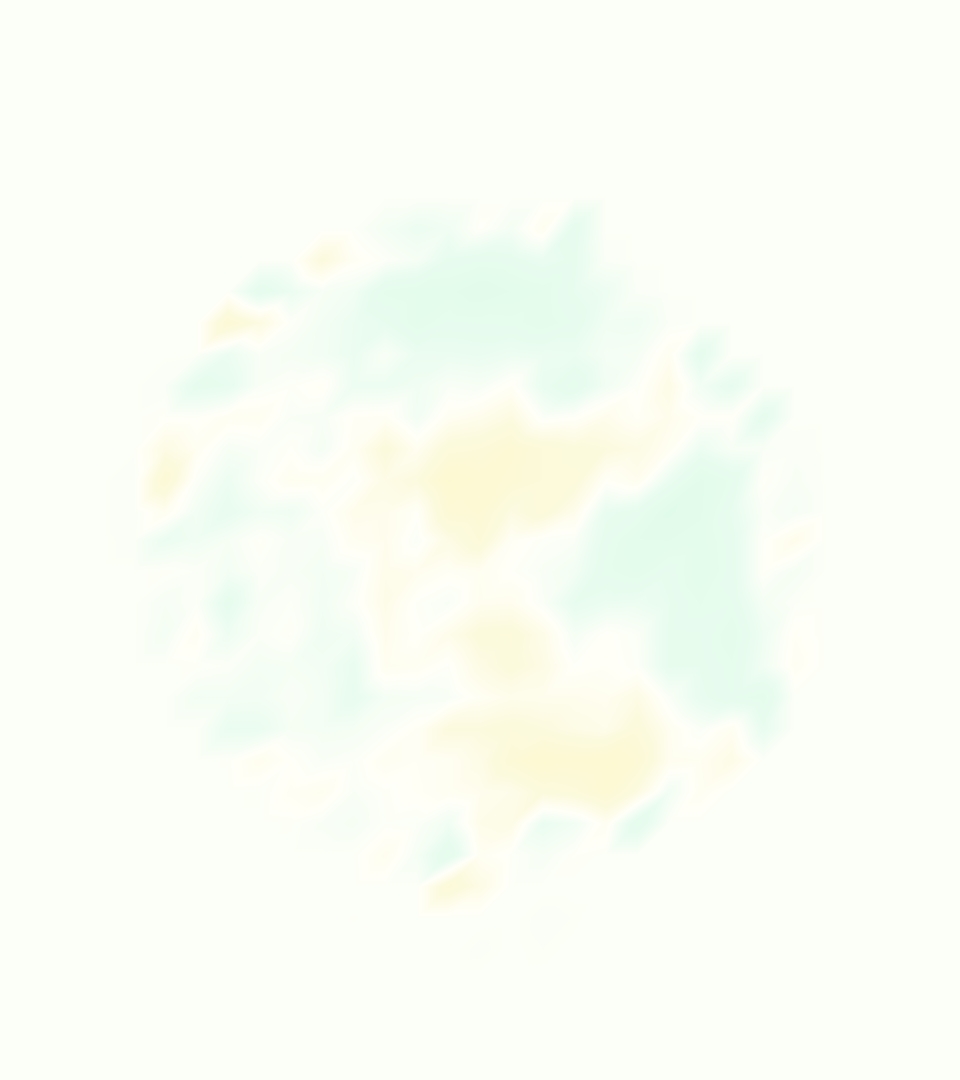} & 
\includegraphics[width=.14\textwidth]{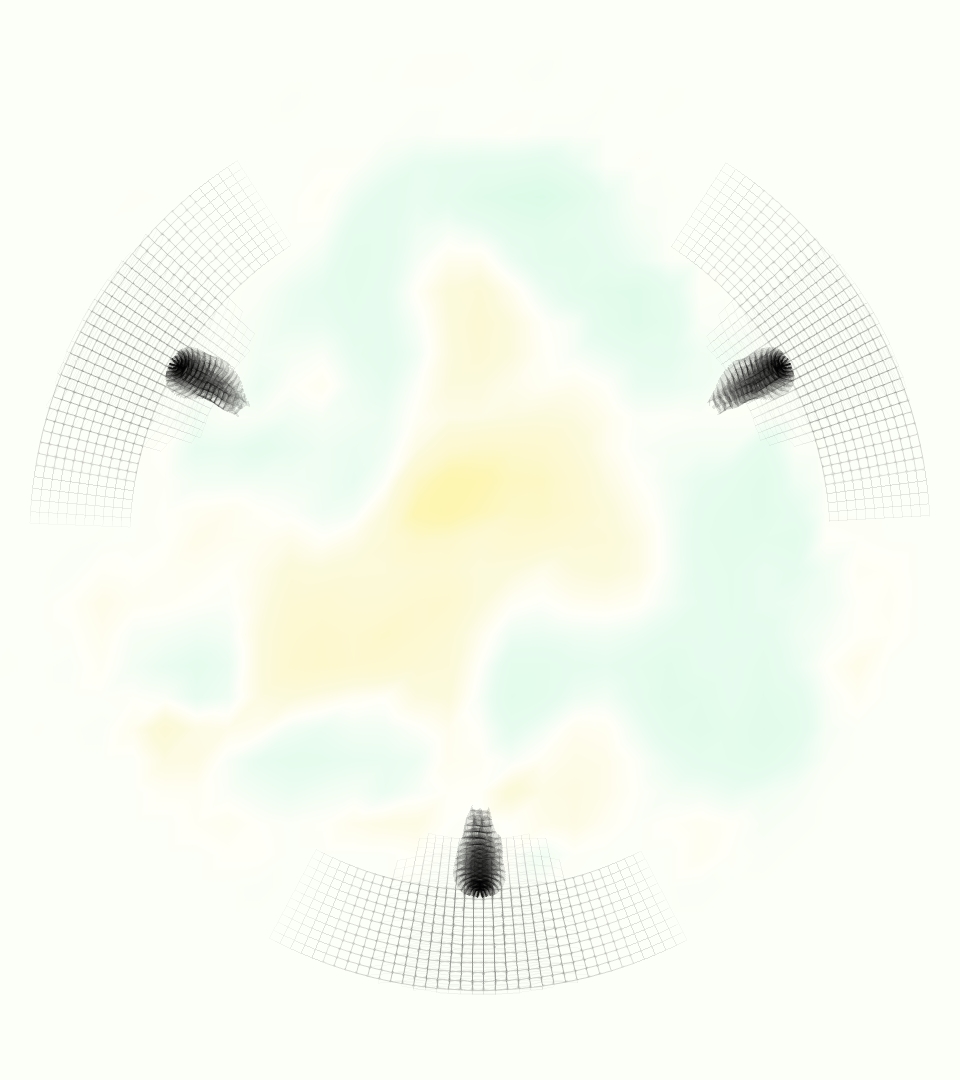} & 
\includegraphics[width=.14\textwidth]{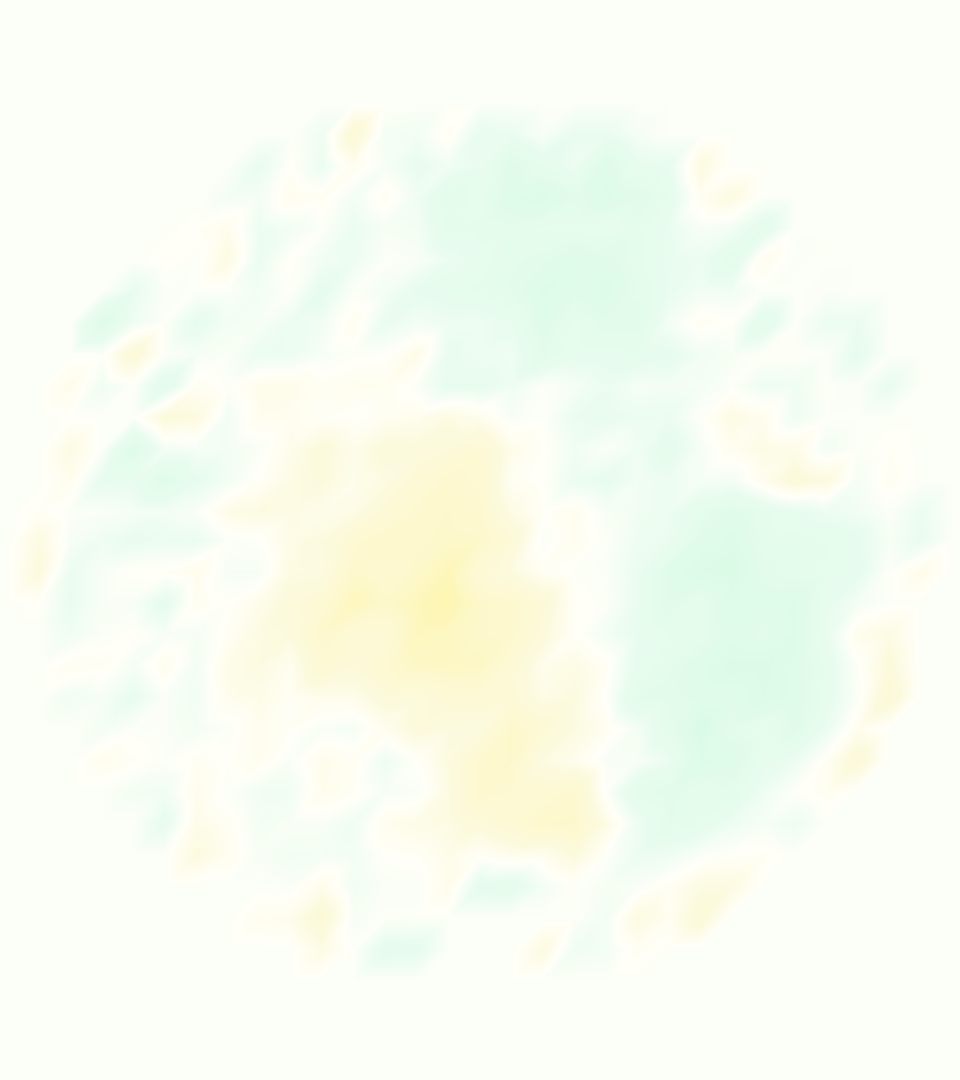} & 
\includegraphics[width=.14\textwidth]{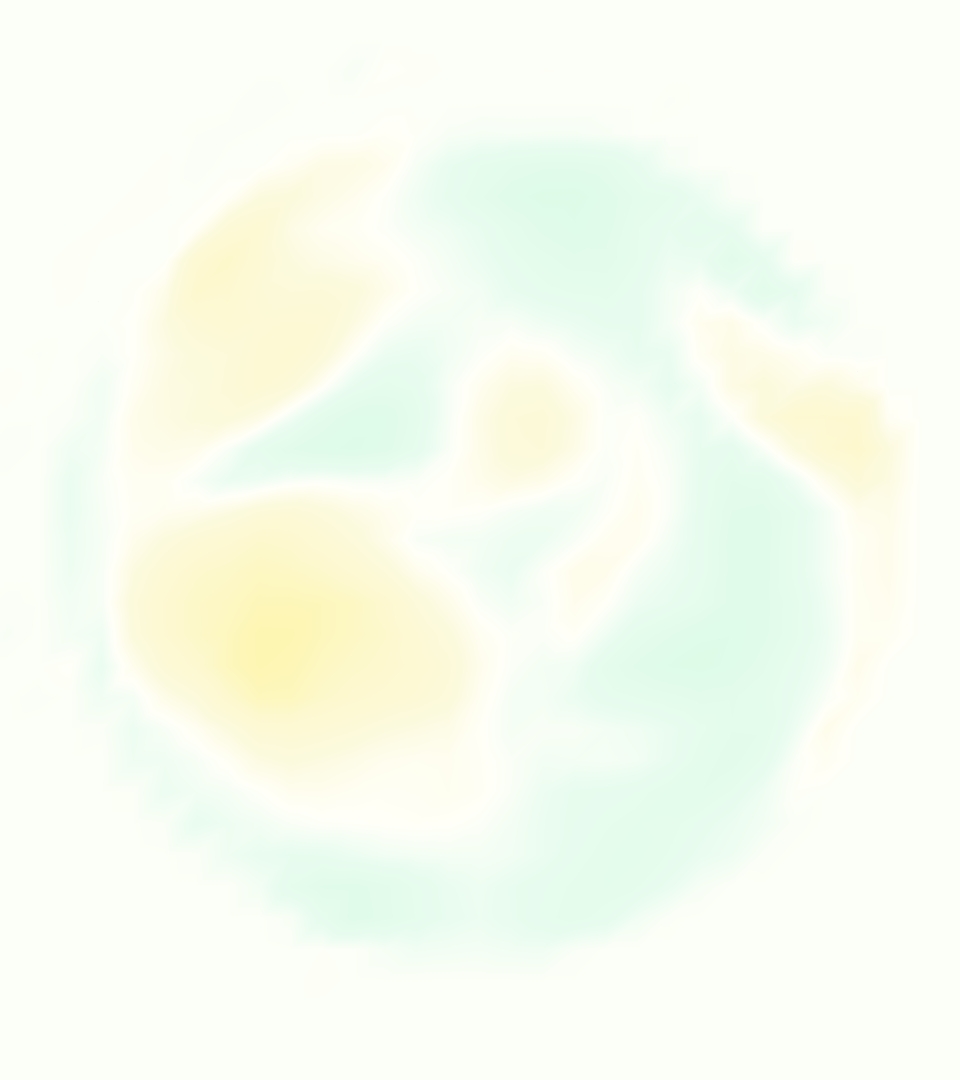} & 
\includegraphics[width=.14\textwidth]{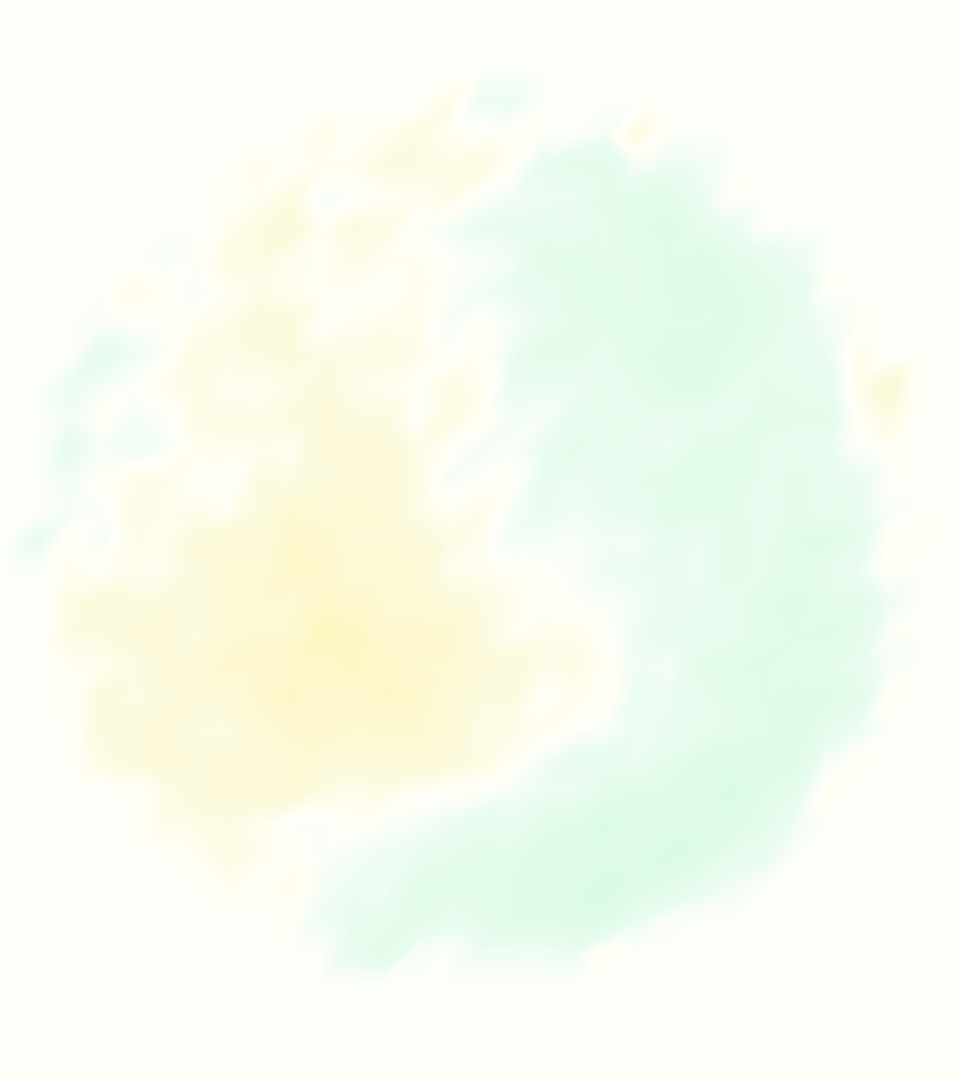}  \\ 
\multicolumn{1}{c}{ } & \multicolumn{2}{c}{ $x$ = 0 } &   \multicolumn{2}{c}{ $x$ = 0.625 cm} &   \multicolumn{2}{c}{ $x$ = 1.25 cm }  \\ 
\end{tabular}
\caption{Axial view of the axial component of velocity in simulation and experiment.}
\label{axial}
}
\end{figure*}

Slice views of the $x$ (axial) component of fluid velocity in the axial plane, orthogonal to the flow direction, show that the simulation replicated the development of the jet over the cardiac cycle. 
At each phase, the simulation matched the forward speed of the core of the jet and the locations of reverse flow back towards the valve annulus. 
While there are some differences in the shape of the jet at certain phases and locations, the simulation clearly produced the general dynamics of  the cardiac cycle that are seen in the experiment.

At $x$ = 0, the axial slice directly at the valve annulus shown in columns 1 and 2 in Figure \ref{axial}, there was excellent agreement between the simulation and experiment over the cardiac cycle in the speed and shape of the jet through the valve. 
In both cases, the axial velocity increased as the flow accelerated during systole and the valve leaflets opened, then decreased during diastole with slight negative velocity before the valve leaflets were fully closed. 
The forward flow through the valve annulus did not form a full circle, but rather developed a triangular shape with a point of the triangle forming along the interior curve of the vessel, at the bottom of the axial slices. 
At $x$ = 0, the points of this triangular jet shape aligned with the commissures of the valve. 
This shape persisted during peak systole and was well-matched by the simulation.

The axial slice $x$ = 0.625 cm, shown in columns 3 and 4 in Figure \ref{axial}, cut through the support scaffolding of the valve and the leaflets when they are open. 
In the experimental data, the shape of the jet changed as it moved downstream. 
A triangular shape occurred, but the points were then aligned with the middle of each open leaflet as opposed to the commissures. 
Those points were also more rounded than they were at $x$ = 0. 
The peak velocity of the jet was faster at $x$ = 0.625 cm than at $x$ = 0, as the flow accelerated through the open valve leaflets. 
The simulation produced these features at $x$ = 0.625 cm. 
The triangular shape of the jet shifted similarly, and its speed increased compared to the upstream slice. 
As the flow decelerated into diastole, the jet shape remained roughly triangular but diminished in intensity before disappearing after valve closure. 

The jet continued to develop at $x$ = 1.25 cm, an axial slice immediately downstream of the valve scaffolding and open leaflets, shown in columns 5 and 6 in Figure \ref{axial}. 
In the experimental data, the points of the triangular jet shape extended further towards the vessel wall. 
In addition, regions of reversed flow developed in the locations downstream of the commissures, resulting in curved sides to the shape of the jet. 
Each tip of the jet was unique, due to variations in the individual leaflets in the physical bioprosthetic valve. 
These variations are apparent in the velocity fields, possibly because the jet edges are similar enough cycle to cycle that irregularities are still being captured even with phase averaging. 
Further discussion of these features can be found in Schiavone et al. \cite{schiavone2021vitro}, which showed that the jet tip shapes occurred in different pulmonary anatomies, demonstrating that they were likely due to inherent properties of each leaflet. 
The leaflets in the mathematical model of the valve are identical, so these nuances in leaflet variation could not be replicated. 
The simulation did capture some of the extension of the tips of the jet, as they were closer to vessel wall at slice $x$ = 1.25 cm than $x$ = 0.625 cm. 
The curves in the triangular sides of the jet were also present in the simulation, though they were less pronounced than the experimental data. 
At both $x$ = 0.625 cm and $x$ = 1.25 cm, the jet shape in the simulation was smoother than the jet in experiment. 
It is possible that the free edges of the leaflets in the mathematical model are not fully replicating the behavior of the physical leaflets of the bioprosthetic valve, in particular the amplitude or frequency of leaflet flutter, leading to the variations seen in the jet shape at $x$ = 1.25 cm downstream of the leaflet edges. 
The simulation, however, does capture the key features of the triangular shape and speed of the jet. 
Overall, qualitative comparisons demonstrated that the simulation reproduced key features of the flow during systole and diastole.

\begin{figure*}[t!]  
\centering
\begin{tabular}{cc}
\includegraphics[width=.48\textwidth]{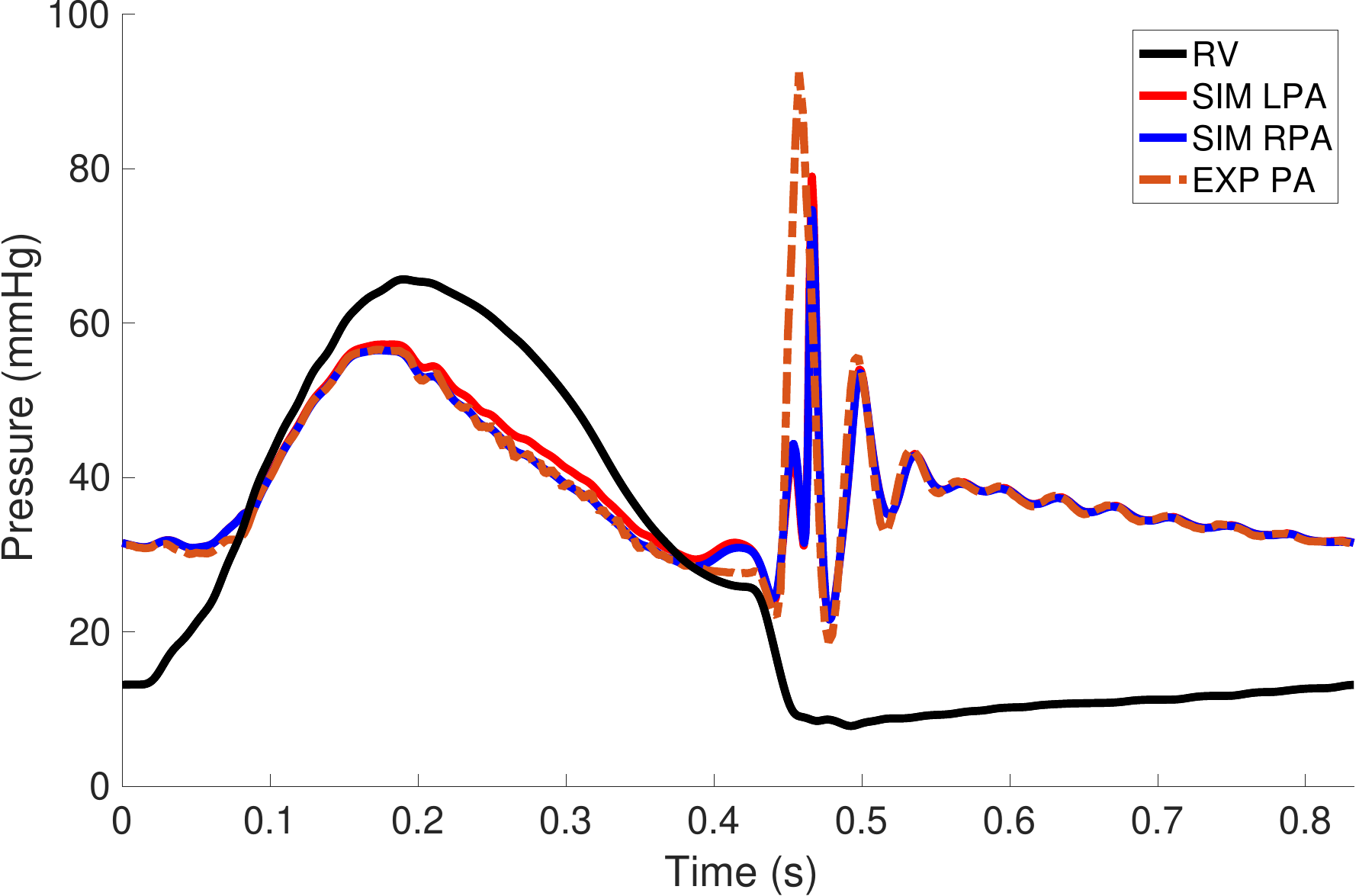} 
\includegraphics[width=.48\textwidth]{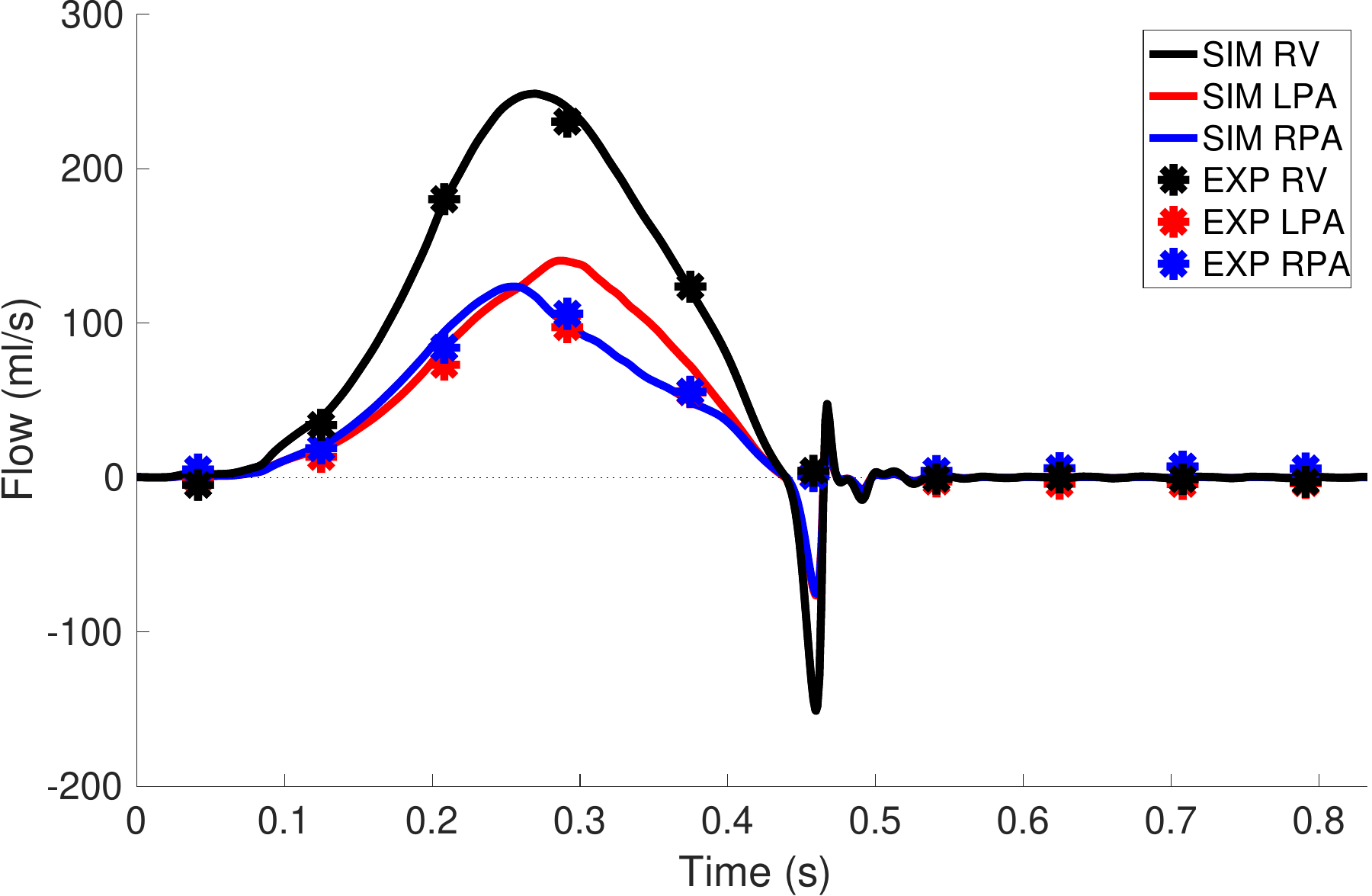} \\ 
\end{tabular}
\caption{Pressure and flow waveforms from the simulation and experiment.
The RPA and LPA pressures prescribed at the outlets in the simulation are determined via a variable resistance boundary condition connected to the experimental pressure (see eqn. \eqref{resistance_eqn}).
The RV pressure is prescribed directly from experiment. 
The RV flow waveform shows excellent agreement between simulation and experiment.  
}
\label{flow_pressure}
\end{figure*}

Pressure and flow waveforms from the second cycle in the simulation and the experiment are shown in Figure \ref{flow_pressure}.
During systole, the pressures were prescribed. 
During valve closure, the pressures showed an oscillation as determined by the interaction of the prescribed pressure and resistance (Equation \ref{resistance_eqn}). 
The experimental pulmonary artery pressure had a similar oscillation, but not with precisely the same magnitude or frequency. 
The RV flow rate closely matched that of the experimental flow. 
The RPA flow was somewhat greater than the experimental RPA flow, and the LPA flow was somewhat less than the experimental LPA flow. 
The total experimental flow was not precisely conserved, which is expected given pointwise uncertainty in the 4D flow MRI velocity measurements of $\pm$4.7 cm/s, uncertainty in gating of the cardiac cycles and variation from cycle to cycle.
The total experimental inflow minus total outflow was up to 27.0 ml/s, which is nearly all experimental uncertainty, as the vessel maintains approximately constant volume. 
Therefore, it was not possible to simultaneously match all three experimental flow rates. 
All three simulation flow rates showed a substantial oscillation at the beginning of diastole. 
Similar oscillations in the experimental flow rates almost surely occurred but could not be captured due to the time resolution of the 4D flow MRI data.
We estimated the stroke volume to be 46.11 ml and cardiac output to be 3.32 L/min. 
Using the maximum flow rate of $ Q_{max} = 248.59$ ml/s and diameter of the valve, 2 cm, as the length scale, we estimated the peak Reynolds number of the flow as 
\begin{align} 
\text{Re}_{max} = \frac{\rho (Q_{max}/A) 2r}{\mu} \approx 4400, 
\end{align}
where $A = \pi r^{2}$ is the valve orifice area. 
The Reynolds number is much greater than one and indicates that the flow was in a physically unstable regime.

\begin{figure*}[t!]  
\hfill \hfill \includegraphics[width=.25\textwidth]{colorbar.jpeg}
{
\centering
\setlength{\tabcolsep}{2.0pt}        
\begin{tabular}{ c | c}        
phase averaged, resampled & instantaneous \\ 
\includegraphics[width=.42\textwidth]{sagittal_fine_resampled_points0003_cropped.jpeg} & \includegraphics[width=.42\textwidth]{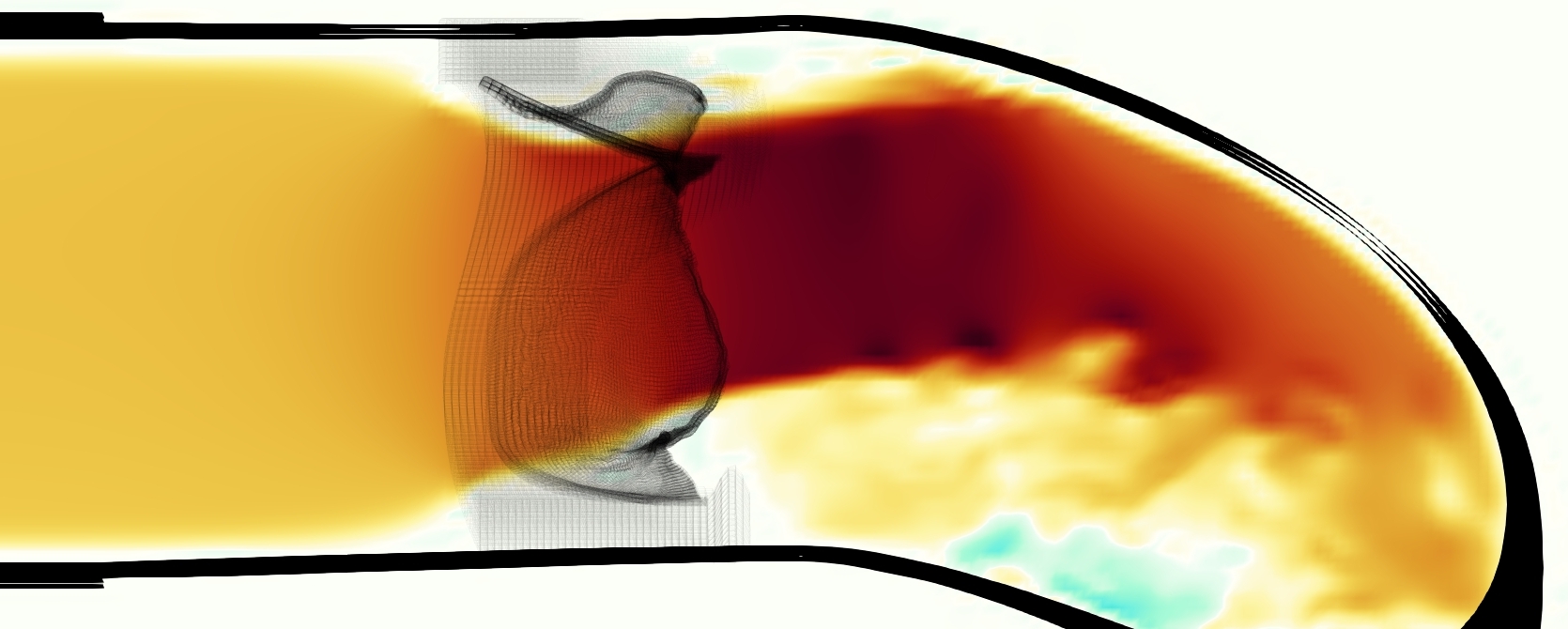} \\ 
 phase averaged & 4D MRI   \\  
\includegraphics[width=.42\textwidth]{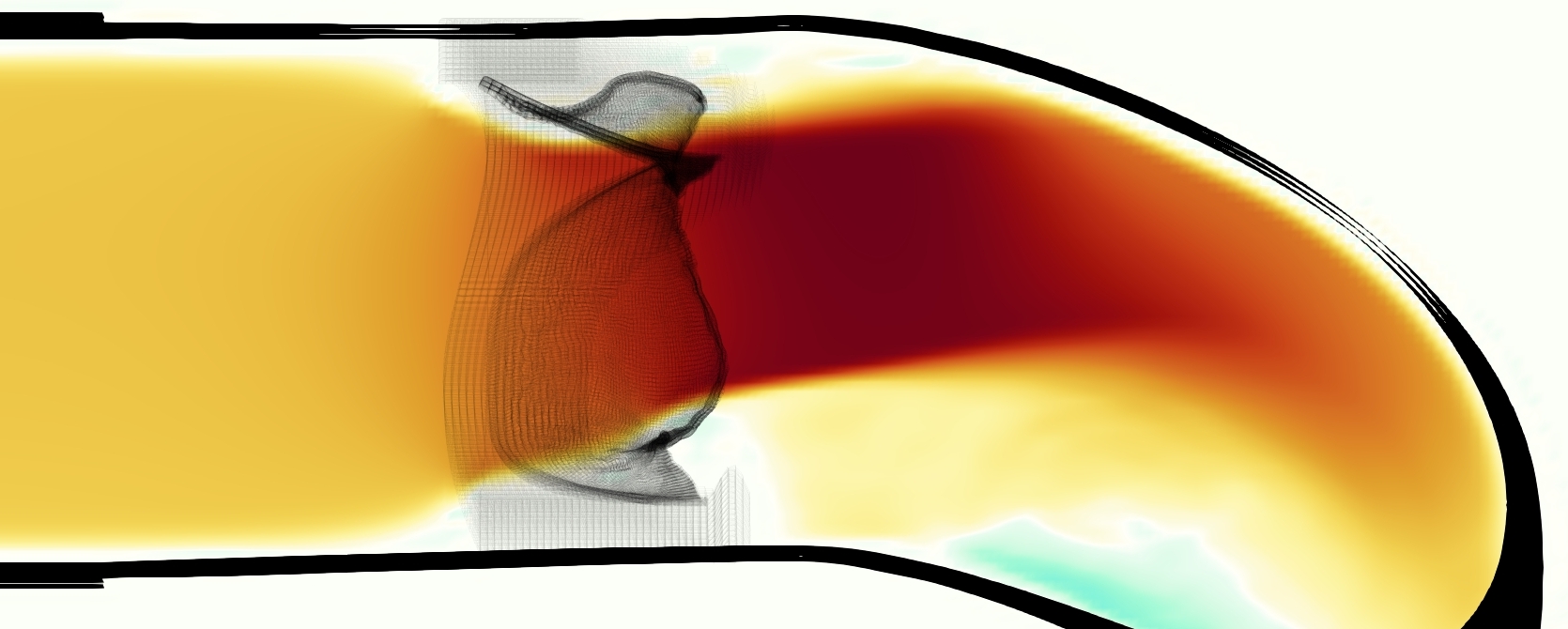} & 
\includegraphics[width=.42\textwidth]{sagittal_mri_fine_resampled_points0003_cropped.jpeg} \\ 
\end{tabular}

\vspace{10pt}

\begin{tabular}{ c c | c | c | c |}
& phase averaged, & & & \\ 
& resampled 
& phase averaged
& instantaneous 
& 4D MRI \\ 
\rotatebox[origin=l]{90}{$x$ = 0 cm} & 
\includegraphics[width=.14\textwidth]{axial_fine_resampled_points_x0_0003.jpeg} & 
\includegraphics[width=.14\textwidth]{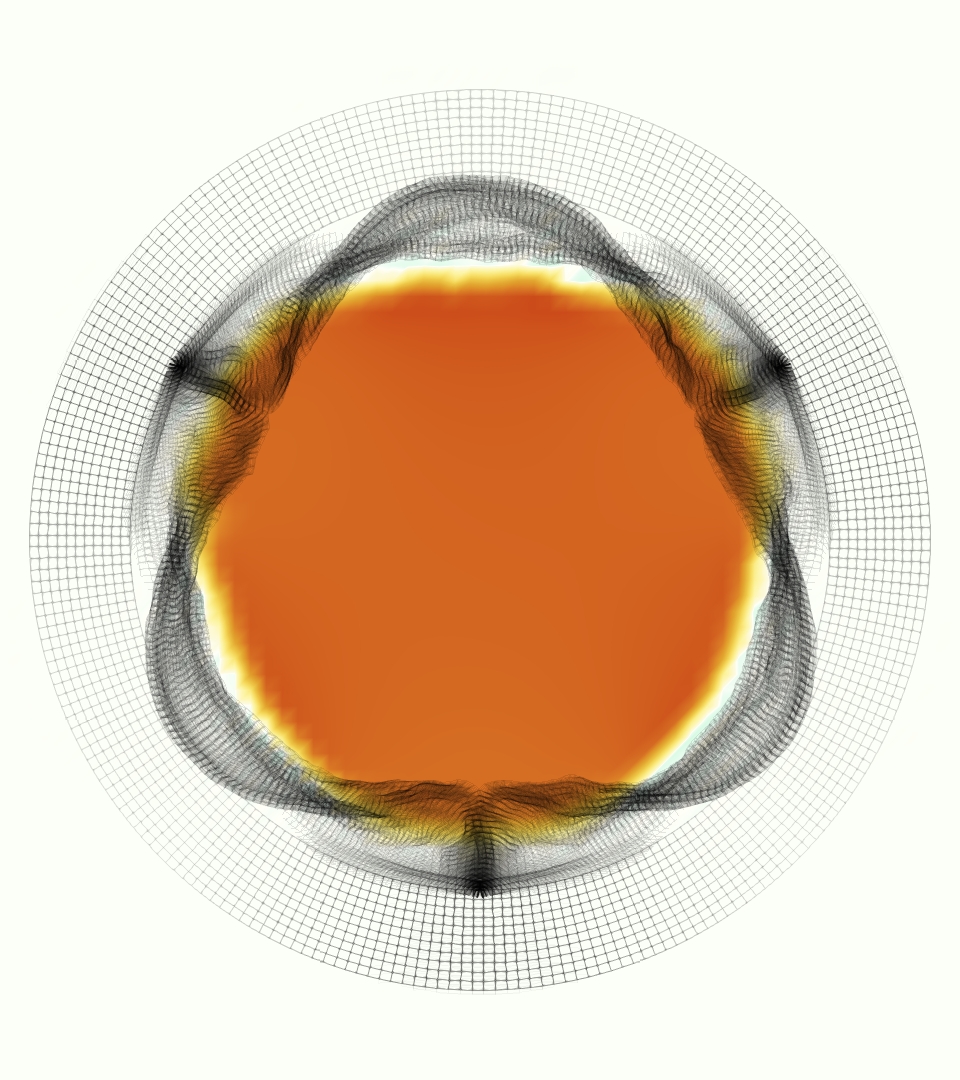} & 
\includegraphics[width=.14\textwidth]{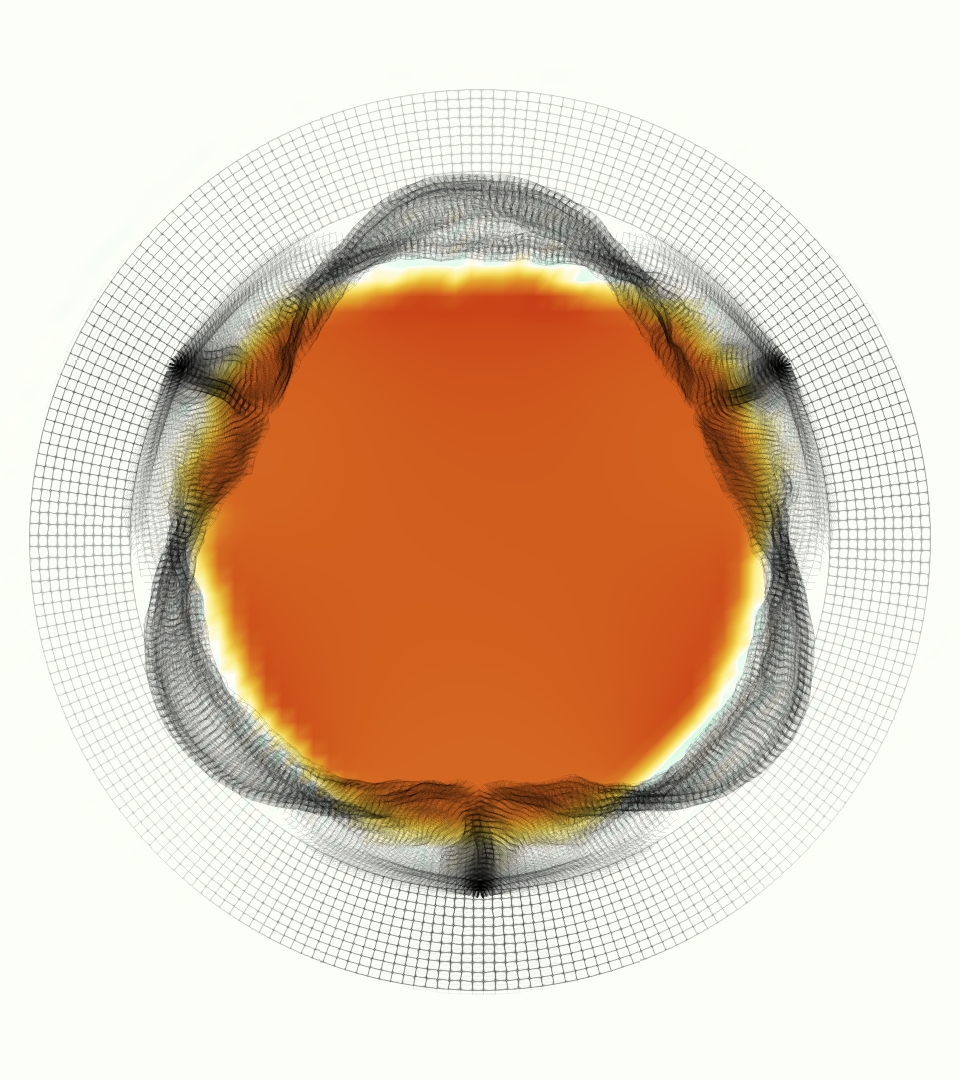} &  
\includegraphics[width=.14\textwidth]{axial_mri_resampled_points_x0_0003.jpeg}  \\ 
\rotatebox[origin=l]{90}{$x$ = 0.625 cm} & 
\includegraphics[width=.14\textwidth]{axial_fine_resampled_points_xpt625_0003.jpeg} & 
\includegraphics[width=.14\textwidth]{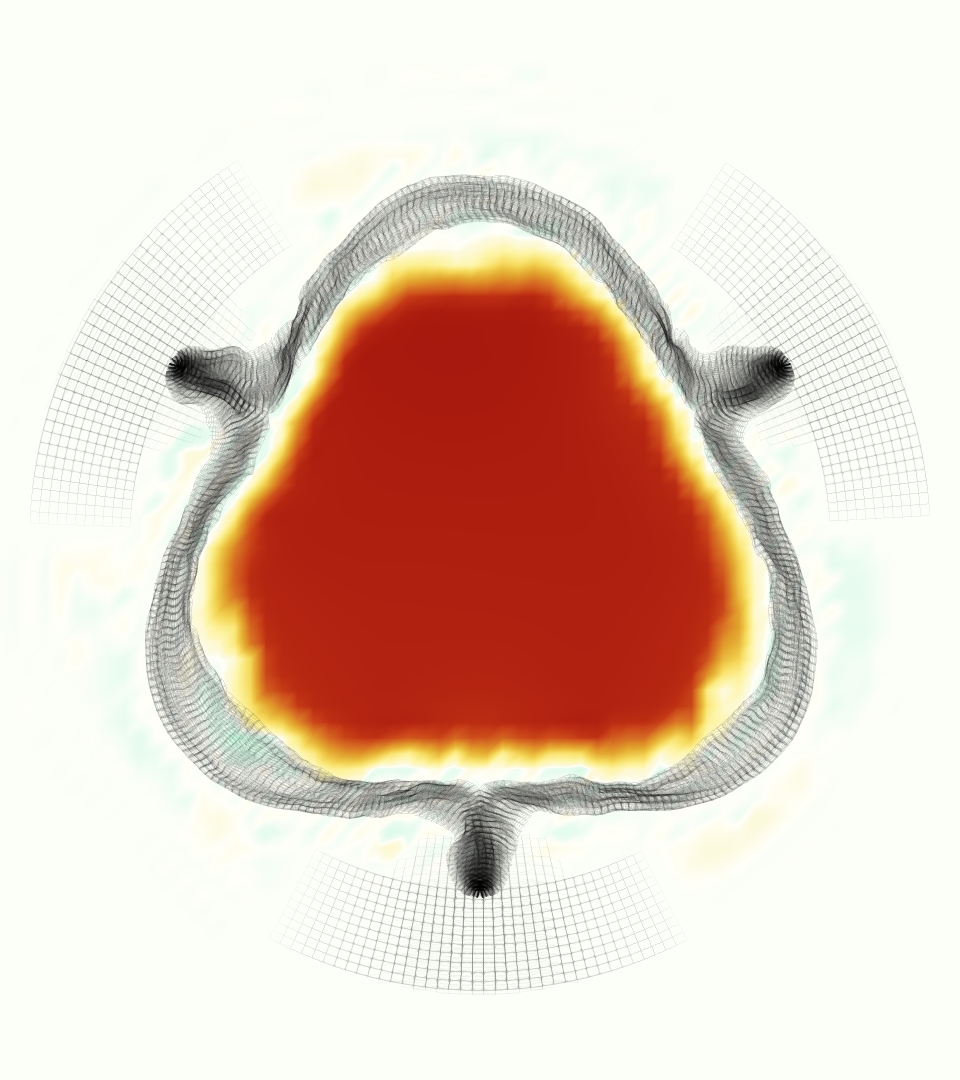} & 
\includegraphics[width=.14\textwidth]{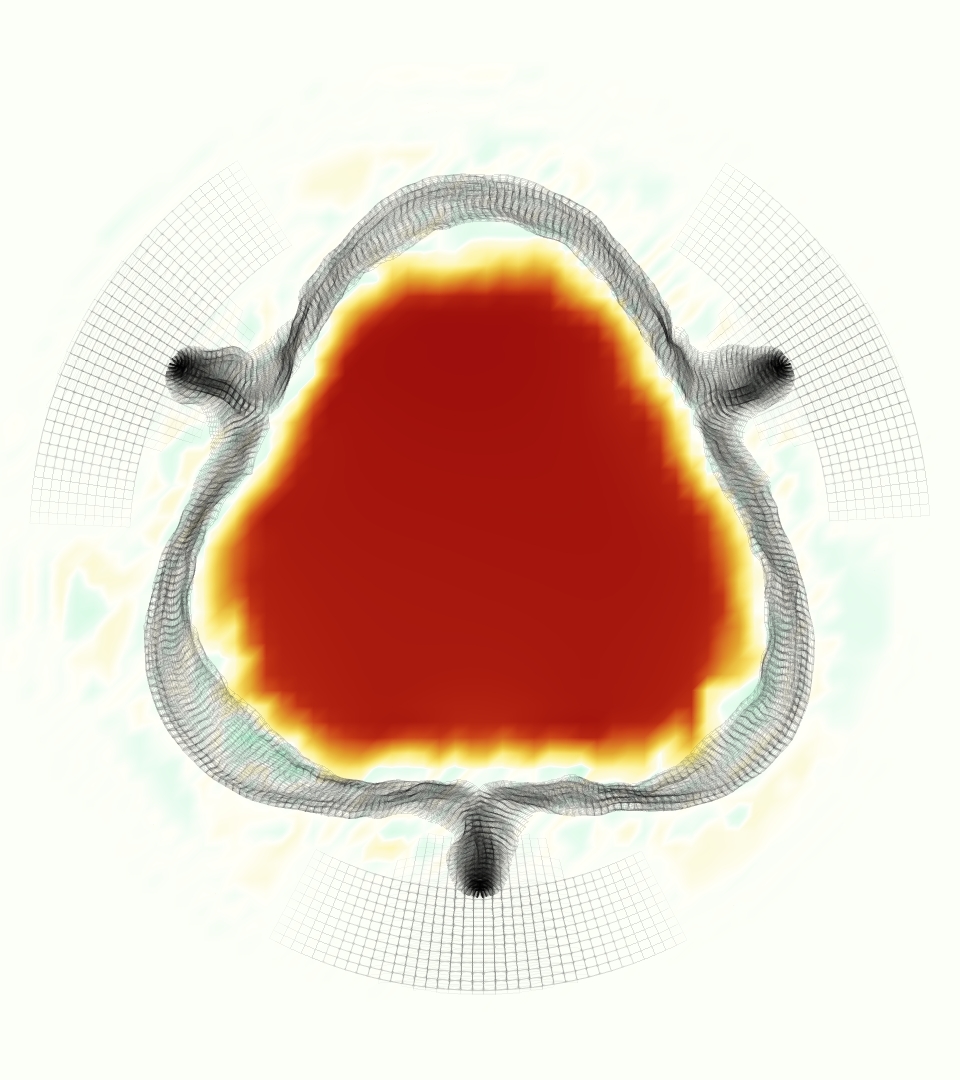} & 
\includegraphics[width=.14\textwidth]{axial_mri_resampled_points_xpt625_0003.jpeg} \\
\rotatebox[origin=l]{90}{ $x$ = 1.25 cm} & 
\includegraphics[width=.14\textwidth]{axial_fine_resampled_points_x1pt25_0003.jpeg} &
\includegraphics[width=.14\textwidth]{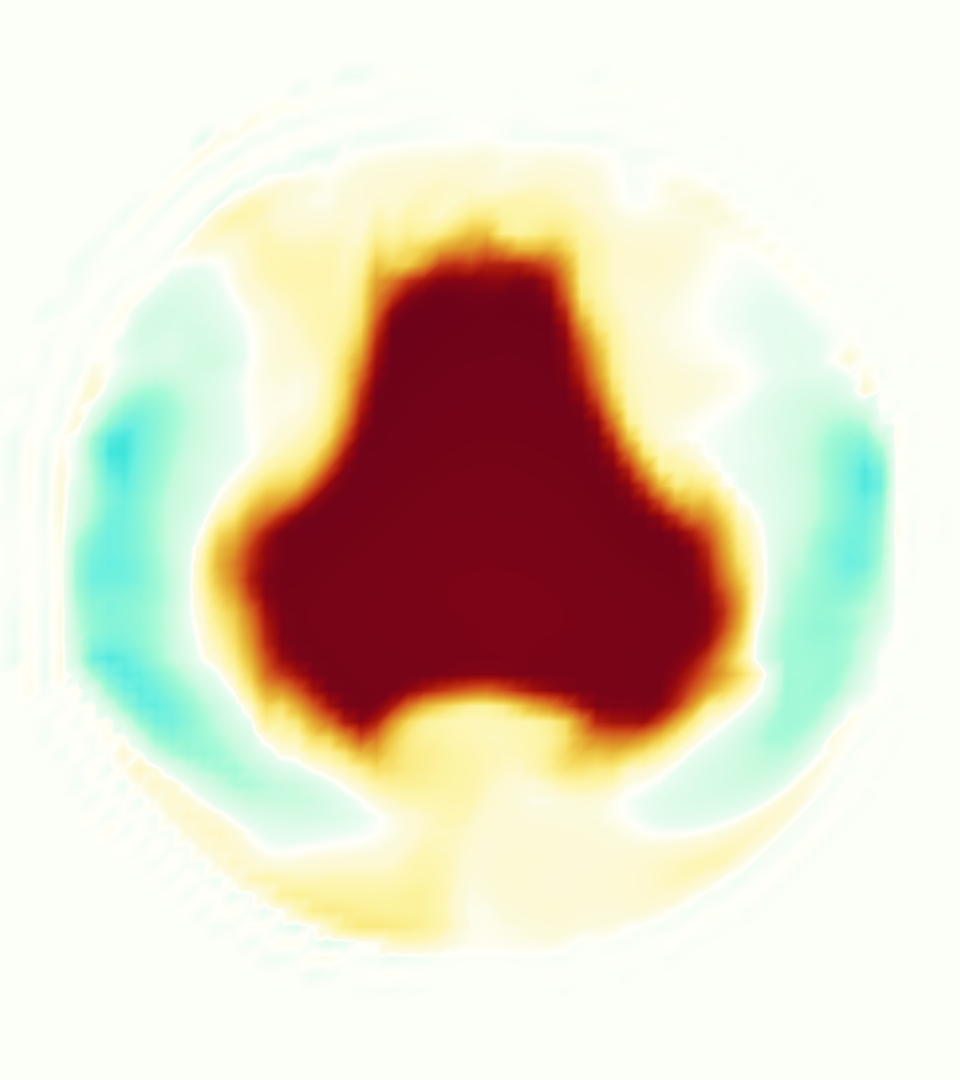} & 
\includegraphics[width=.14\textwidth]{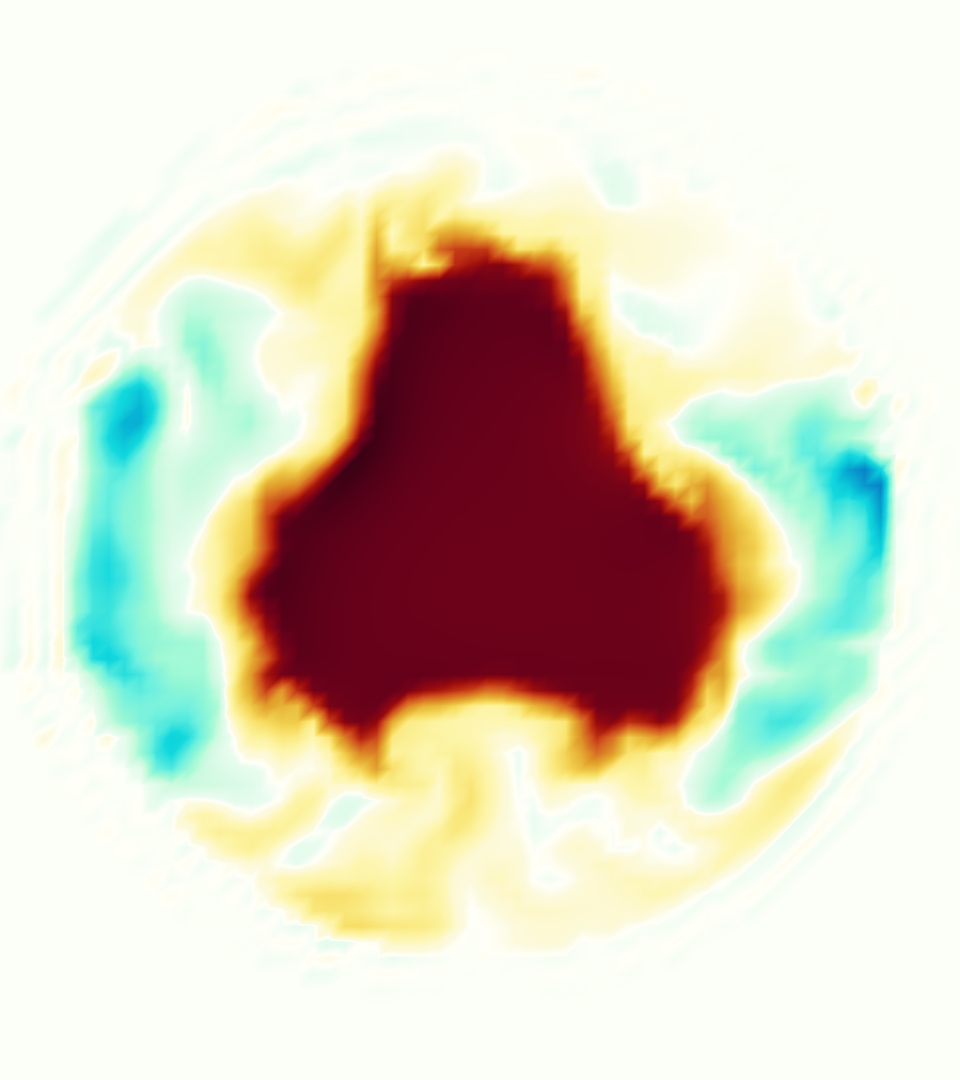}  & 
\includegraphics[width=.14\textwidth]{axial_mri_resampled_points_x1pt25_0003.jpeg}  \\ 
\end{tabular}

\caption{Slice views of the axial component of phase-averaged, resampled velocity, phase-averaged velocity with no resampling, instantaneous velocity from the second cardiac cycle at the simulation resolution, and the 4D MRI velocity field.}
\label{fine_res_comparison}
}
\end{figure*}

A comparison between the phase-averaged velocity resampled onto the MRI grid, the phase-averaged velocity without resampling, the instantaneous flow field without resampling and the MRI velocity field is shown in Figure \ref{fine_res_comparison}. 
The coarser resolution is visible as a ``stair-stepping'' or ``pixelated'' effect on the phase-averaged resampled velocity, whereas the edges of the jet are more smooth in the phase-averaged velocity with no resampling. 
Otherwise, the resampled velocity appears qualitatively similar to the phase-averaged velocity with no resampling. 
Fine scale features were present in the instantaneous field that were lost in the phase-averaging and resampling process. 
There was more variation in the local velocity of the center of the jet and local areas of high velocity at the edges of the jet that suggested vortices.
Vortical structures adjacent to the jet and in the areas of reversed flow were apparent. 
These comparisons suggest that 4D flow MRI data may lose substantial fine-scale flow features in the flow field via phase-averaging.

The $L^1$ and $L^{2}$ relative errors on the velocity magnitude and the $x$ component of velocity indicated quantitative agreement in the velocity field between simulation and experiment during systole (Table \ref{integrals}). 
We computed the relative error on the full domain and on the three slices $x = $ 0, 0.625 and 1.25 cm. 
The $L^1$ relative error of velocity magnitude on the full domain ranged from 0.20-0.34. 
At all three slices, the relative error was below 0.37 through systole, and at peak flow was 0.25, 0.25 and 0.32 at $x = $ 0, 0.625 and 1.25 cm, respectively. 
The $L^{1}$ relative error of the $x$ component on the entire domain was 0.36 at peak flow. 
At $x = 0$ and $x$ = 0.625 cm, the relative error was below 0.29 through systole with values of 0.25 and 0.21 at peak flow, respectively. 
At $x = 1.25$ cm, the relative error was 0.44 at peak flow. 
This error was larger than the error of the other slices, likely because there was more reverse flow that was not aligned between simulation and experiment at that slice. 
The $L^{2}$ relative error of velocity magnitude ranged from 0.27-0.41 over the full domain, and from 0.25-0.40 on the slices. 
At peak flow, errors were 0.38 on the full domain, and 0.29, 0.25 and 0.38 at slices at $x = $ 0, 0.625 and 1.25 cm, respectively. 
On the $x$ component, errors ranged from 0.31-0.50 on the full domain, and 0.23-0.48 on the slices.

\begin{figure*}[h]  
\centering
\begin{tabular}{c}
\includegraphics[width=.48\textwidth]{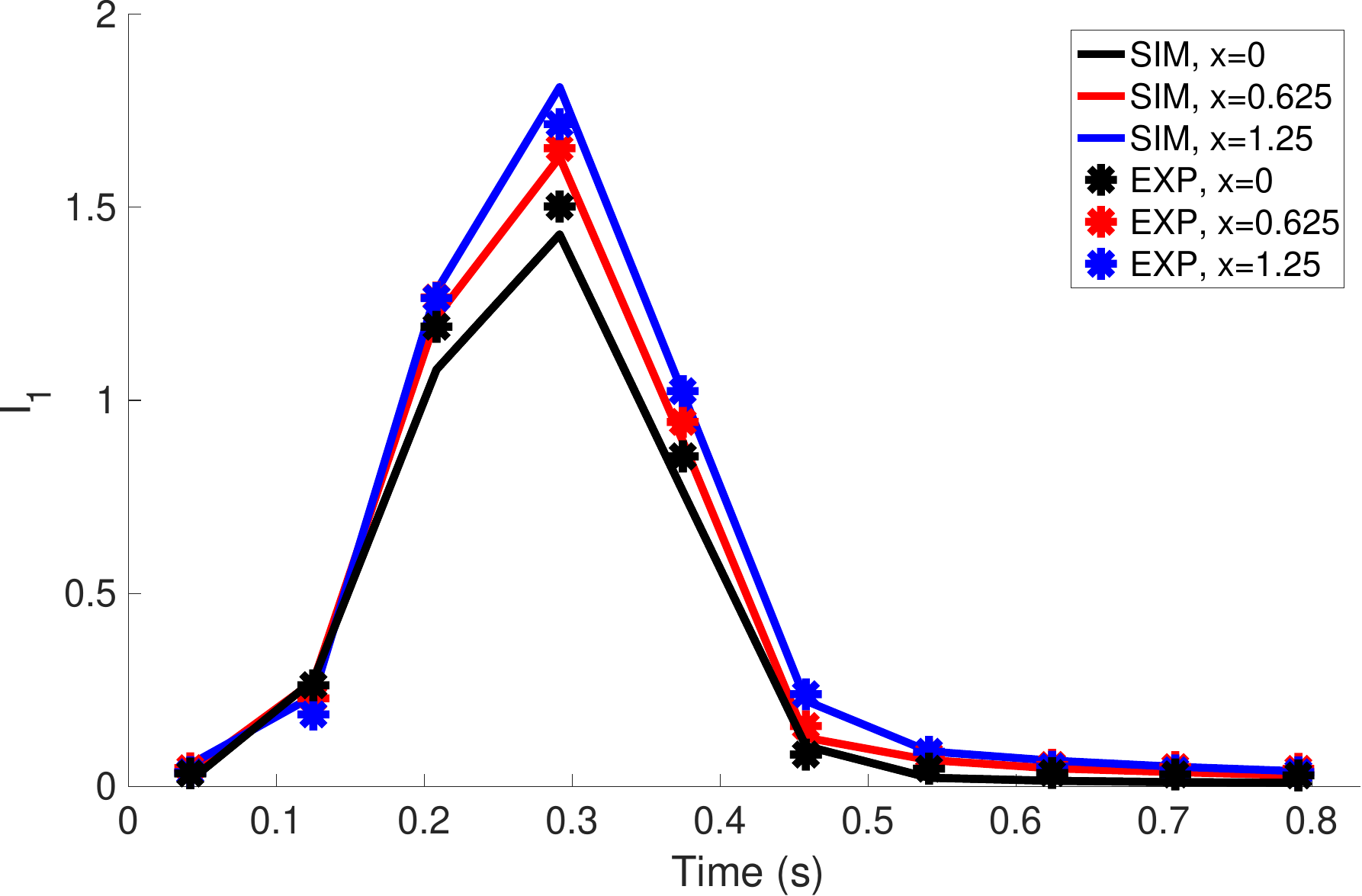}
\end{tabular}
\caption{Values of the integral metric $I_1$, which represents the non-dimensional streamwise momentum on slices at $x = $ 0, 0.625 and 1.25 cm; $I_{1}$ was evaluated on the same three slices. 
}
\label{I1_figure}
\end{figure*}

The integral metric $I_1$ indicated that the streamwise momentum of the jets in both simulation and experiment are very similar. 
The value of $I_1$ over the cardiac cycle is also shown in Figure \ref{I1_figure} on the three slices  $x = $ 0, 0.625 and 1.25 cm. 
Values remained close between simulation and experiment throughout systole and diastole. 
In both simulation and experiment, the value of $I_1$  increases monotonically from the annulus slice at $x = 0$ cm to the most distal slice at $x = 1.25$ cm.

\begin{table*}[t!]  
\centering
\begin{tabular}{| c | c|c|c|c |c|c|c|c|}
\multicolumn{9}{c}{ $L^{1}$  }   \\ 
\cline{2-9}
 \multicolumn{1}{c|}{ } & \multicolumn{4}{c|}{ magnitude }  &  \multicolumn{4}{c|}{ $x$ component }   \\ 
\hline 
time (s) & full & $x$ = 0 & $x$ = 0.65 & $x$ = 1.25  & full & $x$ = 0  & $x$ = 0.65  & $x$ = 1.25  \\ 
\hline 
0.21 & 0.20 & 0.29 & 0.27 & 0.30 & 0.24 & 0.29 & 0.24 & 0.34 \\ 
\hline 
0.29 & 0.30 & 0.25 & 0.25 & 0.32 & 0.36 & 0.25 & 0.21 & 0.44 \\ 
\hline 
0.37 & 0.34 & 0.27 & 0.27 & 0.37 & 0.44 & 0.27 & 0.25 & 0.49 \\ 
\hline
\end{tabular}

\begin{tabular}{| c | c|c|c|c |c|c|c|c|}
\multicolumn{9}{c}{  }   \\ 
\multicolumn{9}{c}{ $L^{2}$  }   \\ 
\cline{2-9}
 \multicolumn{1}{c|}{ } & \multicolumn{4}{c|}{ magnitude }  &  \multicolumn{4}{c|}{ $x$ component }   \\ 
\hline 
time (s) & full & $x$ = 0 & $x$ = 0.65 & $x$ = 1.25  & full & $x$ = 0  & $x$ = 0.65  & $x$ = 1.25  \\ 
\hline 
0.21 & 0.27 & 0.31 & 0.25 & 0.34 & 0.31 & 0.32 & 0.23 & 0.36 \\ 
\hline 
0.29 & 0.38 & 0.29 & 0.25 & 0.38 & 0.43 & 0.31 & 0.23 & 0.46 \\ 
\hline 
0.37 & 0.41 & 0.28 & 0.26 & 0.40 & 0.50 & 0.30 & 0.25 & 0.48 \\
\hline 
\end{tabular}

\caption{Relative error during systole in the $L^1$ and $L^{2}$ norms of the $x$ component of velocity and velocity magnitude on the full domain interior to the vessel and slices at $x = $ 0, 0.625 and 1.25 cm.
}
\label{integrals}
\end{table*}

\section{Discussion}

We developed a model of a physical experiment in an in vitro pulse duplicator to validate an immersed boundary method simulation. 
We simulated flows through a model pulmonary valve in a test chamber designed to model that of the 3D printed physical pulmonary anatomy. 
This model valve was constructed with a design-based approach to elasticity, in which the valve geometry and material properties were designed and derived from the requirements that the valve supports a pressure. 
The simulations themselves were robust and reliable. 
The simulated valve opened freely and closed repeatedly over multiple cardiac cycles. 
Using our pulse duplicator, we directly measured the local fluid velocity with 4D flow MRI in the physical experiment. 
We then performed direct comparisons between simulated and experimental flow data.

The simulated velocity fields showed excellent qualitative agreement with the experimental velocity. 
Many features of the flow that were present in the experimental measurements also appeared in the simulations, including a large jet which angled up downstream of the valve and the appearance of a triangle-like shape in the jet. 
A recirculation zone appeared downstream, though the precise region of recirculation showed only some agreement at some points in time. 
We prescribed the experimentally measured pressures (subject to minor processing as described in Section \ref{fsi}), and the emergent flow rates matched those of the experiments. 

The simulated velocity fields also showed quantitative agreement with the experimental velocity. 
The entire flow domain interior to the vessel showed $L^{1}$ relative error of 30\% on velocity magnitude and 36\% on the $x$ component of velocity during mid-systole.
While this certainly leaves room for improvement, we are not aware of any other studies comparing simulated flow around heart valves to experiments that compute relative error on the entire flow field. 
Similar results were found at $x = 0$ and 0.625 cm, and slightly higher relative error downstream at $x = $ 1.25 cm. 
The nondimensional streamwise momentum, represented by $I_1$, was extremely similar in simulation and experiment.

Of potential future interest, the phase-averaged, resampled velocity showed substantial differences from the instantaneous velocity fields, with more variation in local fluid velocity and more fine scale vortical structures visible in the simulation (Figure \ref{fine_res_comparison}).  
This points to an advantage for simulations: since 4D flow MRI data is fundamentally phase-averaged, once validated, simulations provide a richer dataset when these instantaneous fluid features are of interest.

Our methods can be used to test scenarios of pulmonary valve placement, and further developed to study other congenital or acquired heart diseases such as tetralogy of Fallot, hypertrophic cardiomyopathy and aortic regurgitation or stenosis. 
Of potential utility to the modeling community, we provided a high-quality experimental dataset with pointwise velocity data on a Cartesian grid that may be used as a benchmark in future FSI studies for validation including direct velocity comparisons. 

To achieve this level of agreement required careful modeling of the leaflet and valve scaffold morphology.
We also extensively post processed the data by taking the phase average analogously to the phase averaging performed by the MRI scanner, then resampled the flow fields to align exactly with MRI data points. 
Achieving this level of agreement, however, did not require any new modeling methods beyond those we developed in previous work \cite{kaiser2019modeling,kaiser2020designbased,kaiser2022controlled}.

We made efforts to model many aspects of the flow loop, but nevertheless limitations and room for future work remain. 
Precise, global pointwise agreement of the velocity field was not achievable, as there were differences in the jet and recirculation zones by more than the MRI resolution of 0.9 mm in some locations. 
Physical instabilities in this inertial flow are present, which make pointwise comparisons challenging to impossible, though their effects are reduced by the phase averaging. 
We did not model the precise material properties of the prosthetic leaflets, as we could find very little literature addressing its material properties, which likely influences leaflet kinematics and may in turn influence jet shape. 
A model of the scaffold could include more precise geometry and local material texture, for example obtained via micro-CT scan or other imaging method. 
Despite these differences, the simulations realistically represented the physical flows, including some fairly subtle details such as the triangle-like jet shape downstream of the valve.

\section{Conclusions}

In immersed boundary simulations of a physical experiment of flow through a heart valve, we achieved excellent qualitative agreement and acceptable quantitive agreement between the simulations and experimental data. 
This work illustrates the construction of a comparator between simulations and experiment, and provides experimental data for use by other researchers. 
Despite the presence of limitations and potential for improvement in future work, we considered this strong confirmation that our methods are realistic and reliable.

\section{Acknowledgements}

ADK was supported in part by a grant from the National Heart, Lung and Blood Institute (1T32HL098049), Training Program in Mechanisms and Innovation in Vascular Disease. 
ADK and ALM were supported in part by the National Science Foundation SSI (Grant \#1663671). 
ADK and NS were supported in part by American Heart Association Transformational Project Award (Grant \# 19TPA34910000).
NS was supported in part by the Stanford Bio-X Bowes Fellowship.
Computing for this project was performed on the Stanford University's Sherlock cluster with assistance from the Stanford Research Computing Center. 
Simulations were performed using the open-source solver package IBAMR, \url{https://ibamr.github.io}.

\section*{Appendix} 
\label{appendix}

We performed a convergence study to evaluate the sensitivity of the phase-averaged velocity fields and flow rates to changes in simulation resolution. 
The fluid resolutions were set to $\Delta x = $ 0.9, 0.68, 0.45, 0.34 and 0.28 mm.
We ran each simulation ran for two cardiac cycles, then performed phase averaging on the second cycle and sampled onto the MRI mesh for comparisons, as described in Section \ref{post_processing}. 
For $\Delta x = $ 0.9 and 0.68 mm, we used a coarsened structure mesh with a target edge length of $\Delta s = 0.45 $ mm, for  $\Delta x = $ 0.45 and 0.34 mm we used the target edge length of $\Delta s = 0.225$ mm, as used in the remainder of this work,  and for $\Delta x = $ 0.28 mm we use the leaflet mesh targeted to $\Delta s = 0.225$ mm, and a slightly finer mesh for the vessel targeted to $\Delta s = 0.2$ mm.

Exact convergence in IB method simulations is challenging to achieve due to diffuse coupling of the fluid-structure interface and the physical instability of the flow. 
The radius of the support of the discrete $\delta$-function is $2.5\Delta x$, which depends on the fluid mesh width, meaning that the effective thickness of all objects in the structure depends on the fluid resolution.
Thus, IB method coupling reduces the effective radius of the valve and vasculature proportional to $\Delta x$, meaning that the effective radius converges to the true radius in a first order manner as $\Delta x \to 0$.  
Resistance to forward flow depends more than linearly on radius (to the fourth power in the case of Poiseuille flow), and thus slow convergence of the flow rate under a specified pressure difference is expected. 
Alternative schemes could include compensatory measures to adjust the boundary conditions to achieve a more constant flow rate, but this would not reduce numerical error at the fluid-structure interfaces. 
Additionally, the Reynolds number of the flow is much greater than one, the flow is inertially dominated and flow structures are unstable in time. 
This effect is partially mitigated by phase-averaging the flow.

\begin{figure*}[t!]  

\hfill \hfill \includegraphics[width=.25\textwidth]{colorbar.jpeg}

\begin{center}
{
\centering
\setlength{\tabcolsep}{2.0pt}     
\begin{tabular}{  c | c | c | c | c | }
& axial &  axial &   axial & sagittal  \\ 
& $x$ = 0 cm &  $x$ = 0.625 cm &   $x$ = 1.25 cm &   \\ 
\hline 
\rotatebox[origin=l]{90}{0.9  mm}&
\includegraphics[width=.14\textwidth]{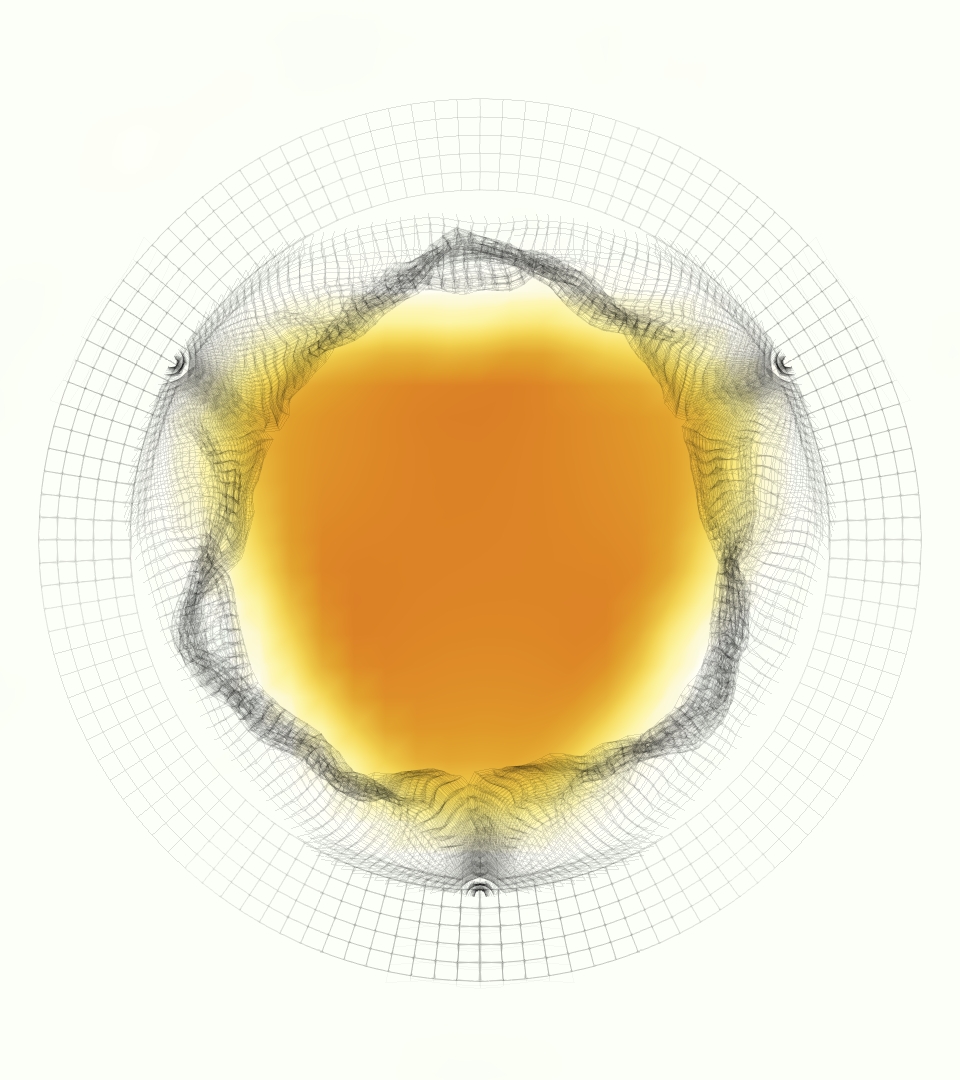} & 
\includegraphics[width=.14\textwidth]{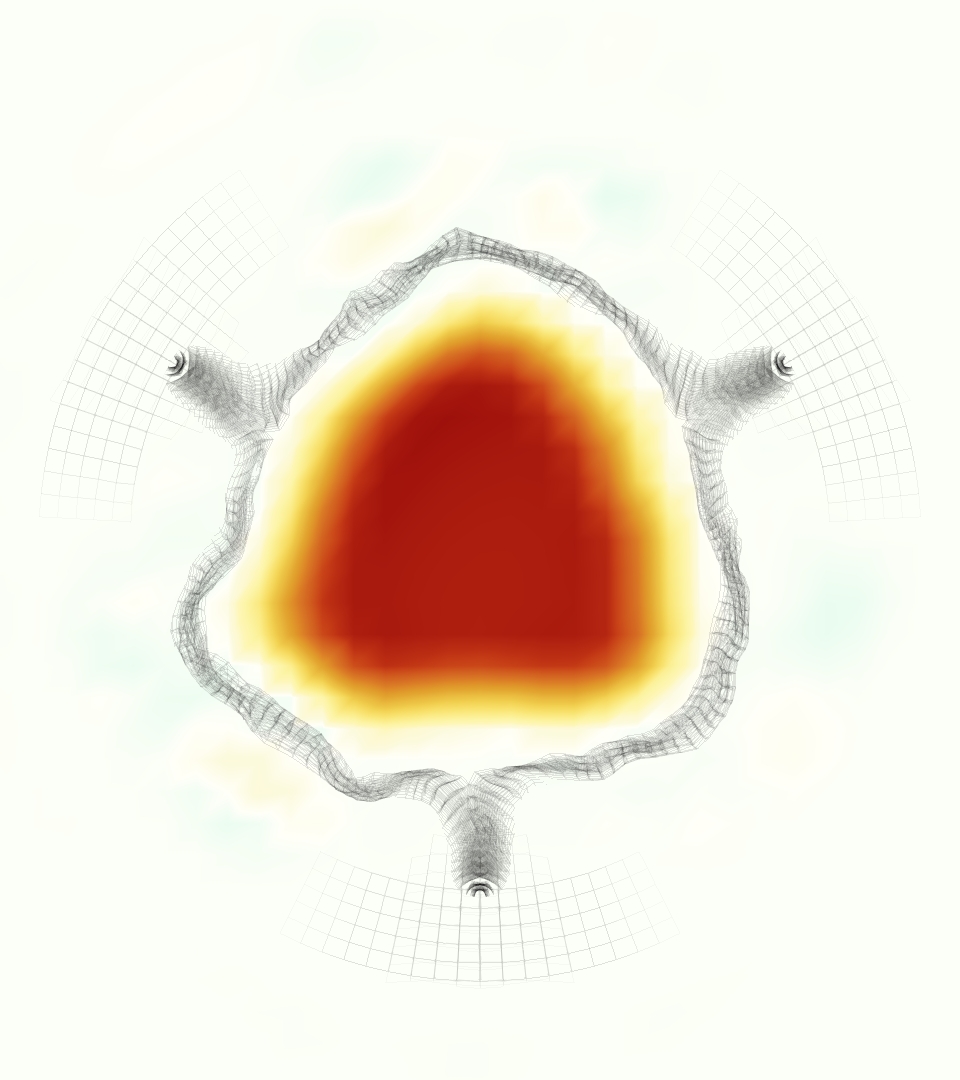} &
\includegraphics[width=.14\textwidth]{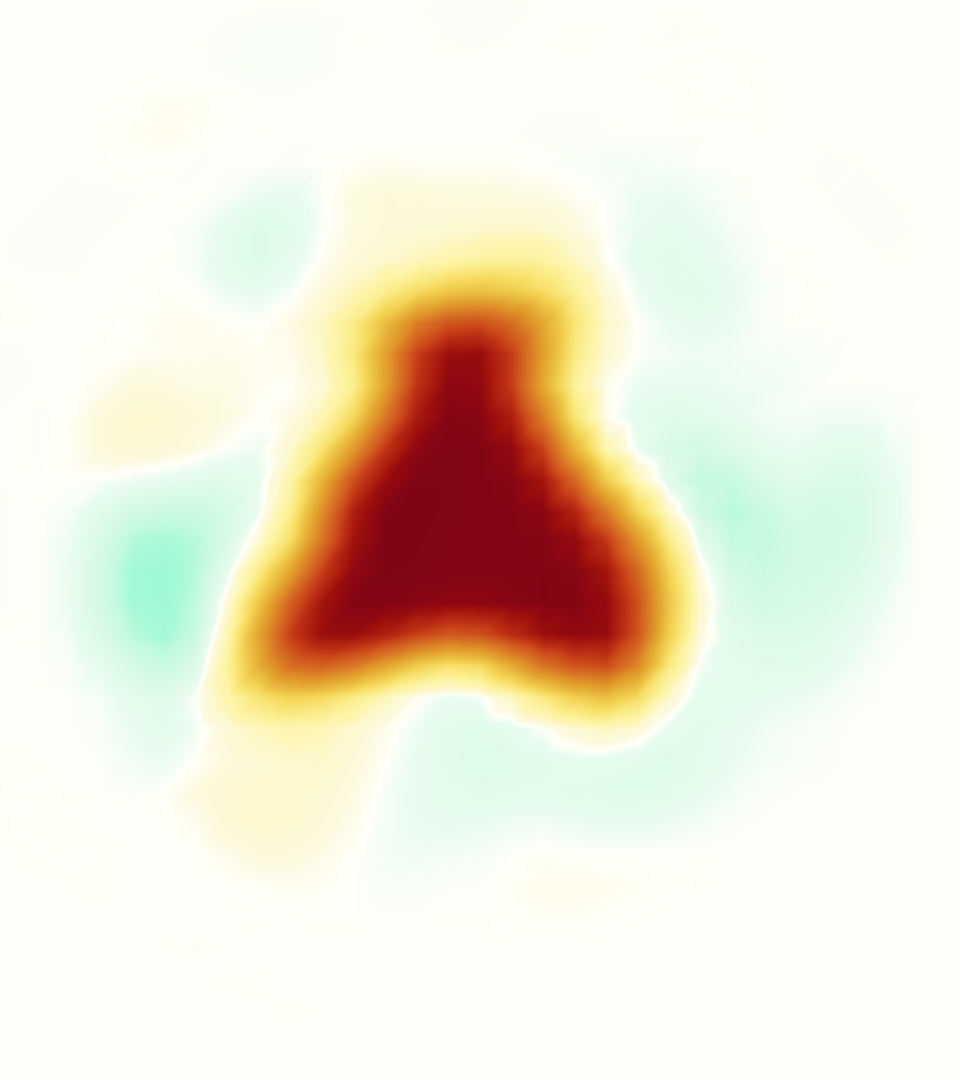} &
\includegraphics[width=.42\textwidth]{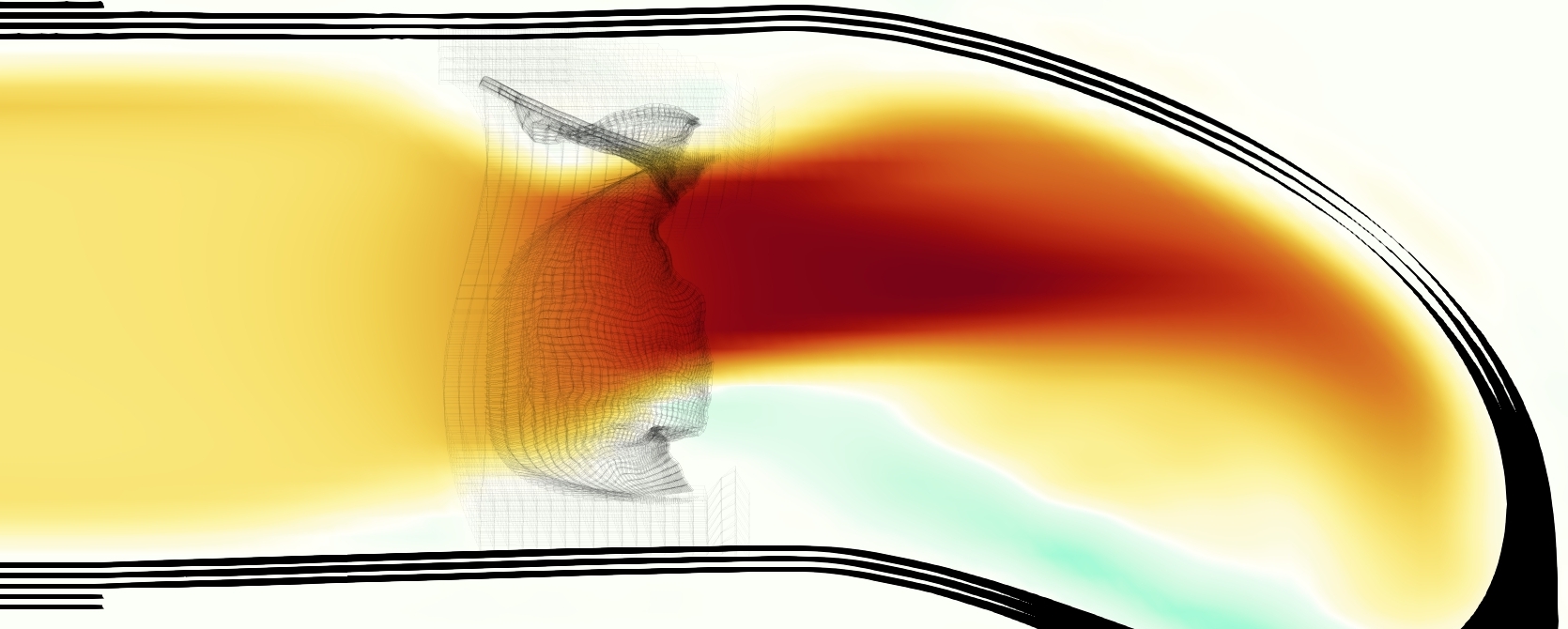} \\ 
\hline 
\rotatebox[origin=l]{90}{ 0.68  mm}&
\includegraphics[width=.14\textwidth]{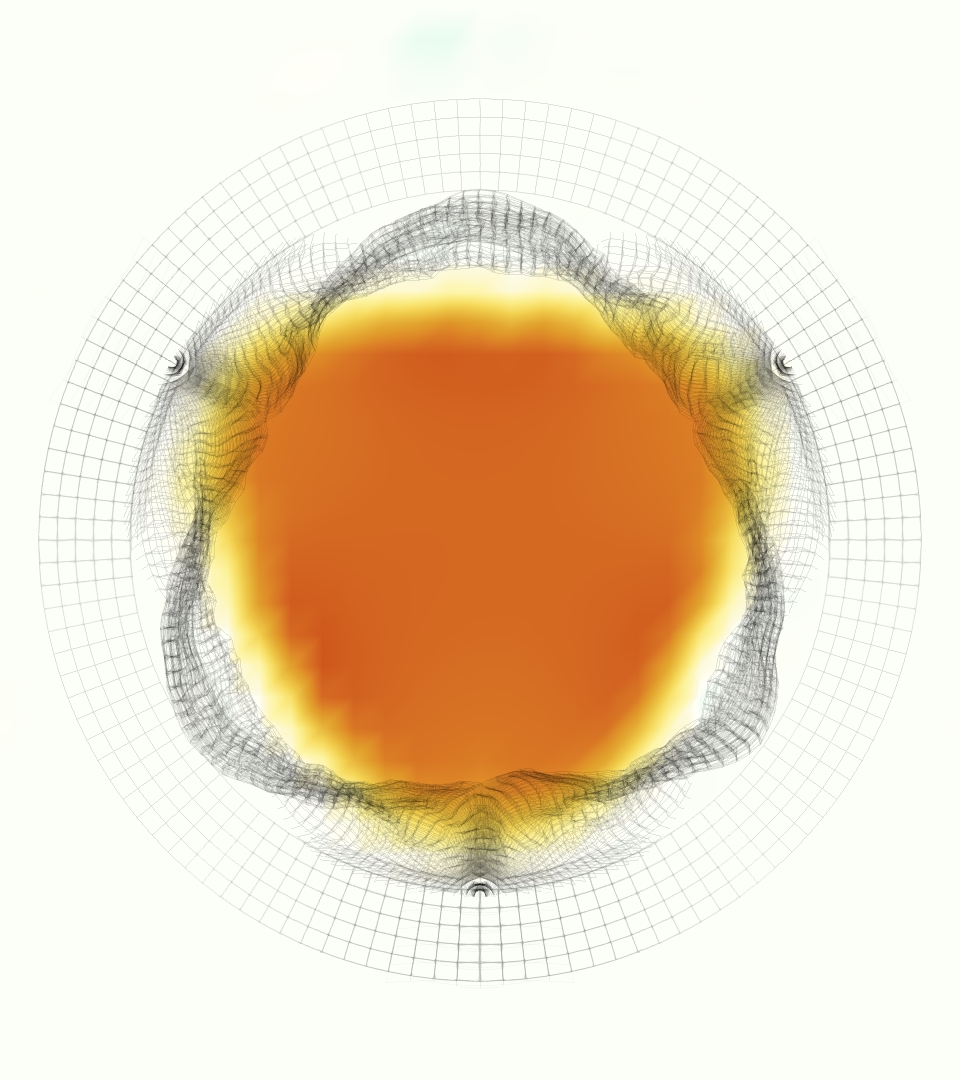} & 
\includegraphics[width=.14\textwidth]{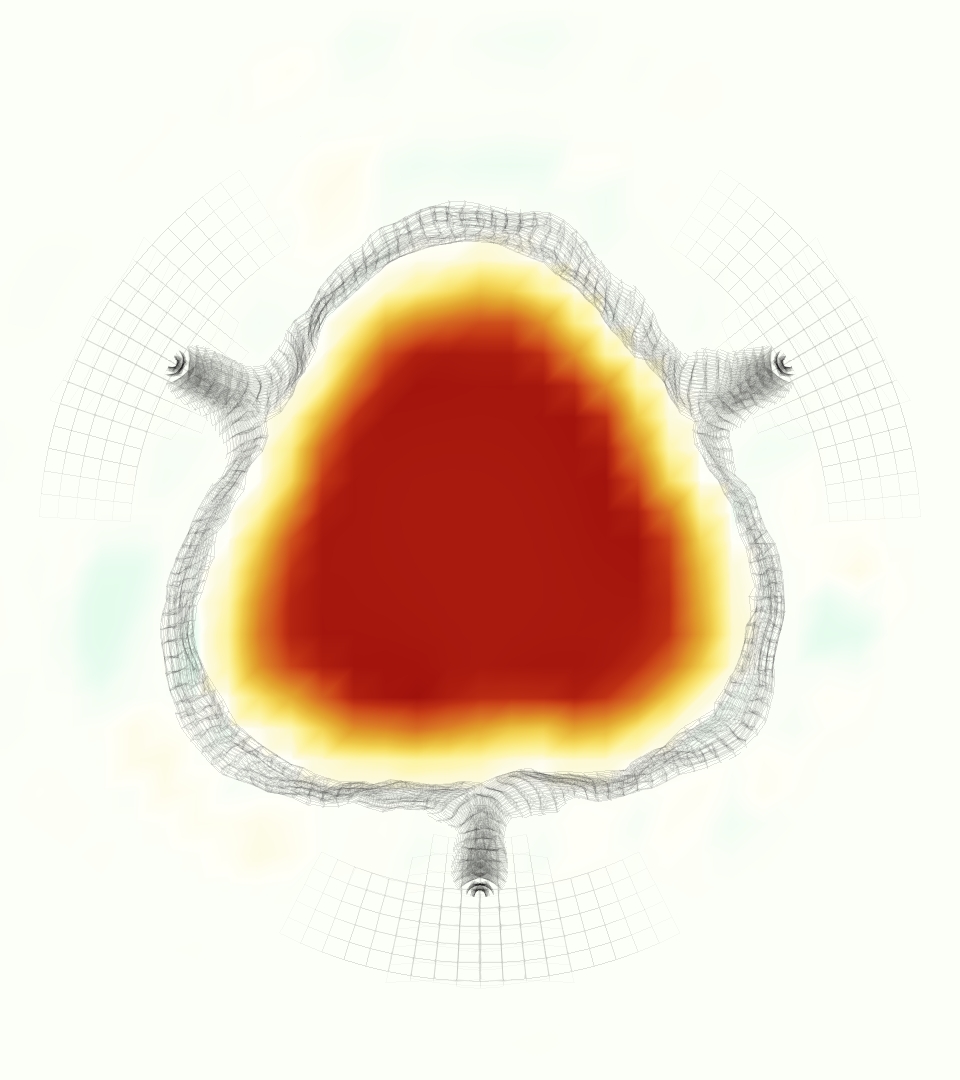} &
\includegraphics[width=.14\textwidth]{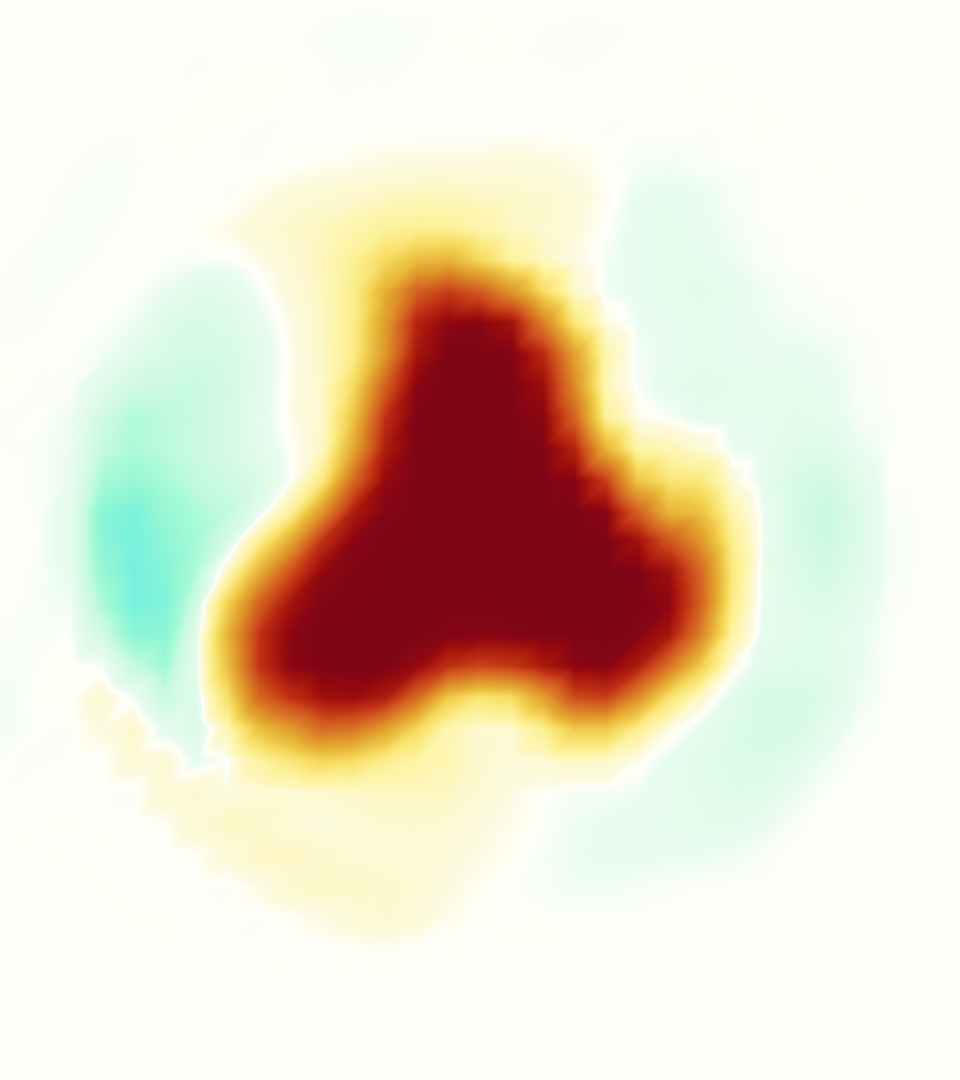} &
\includegraphics[width=.42\textwidth]{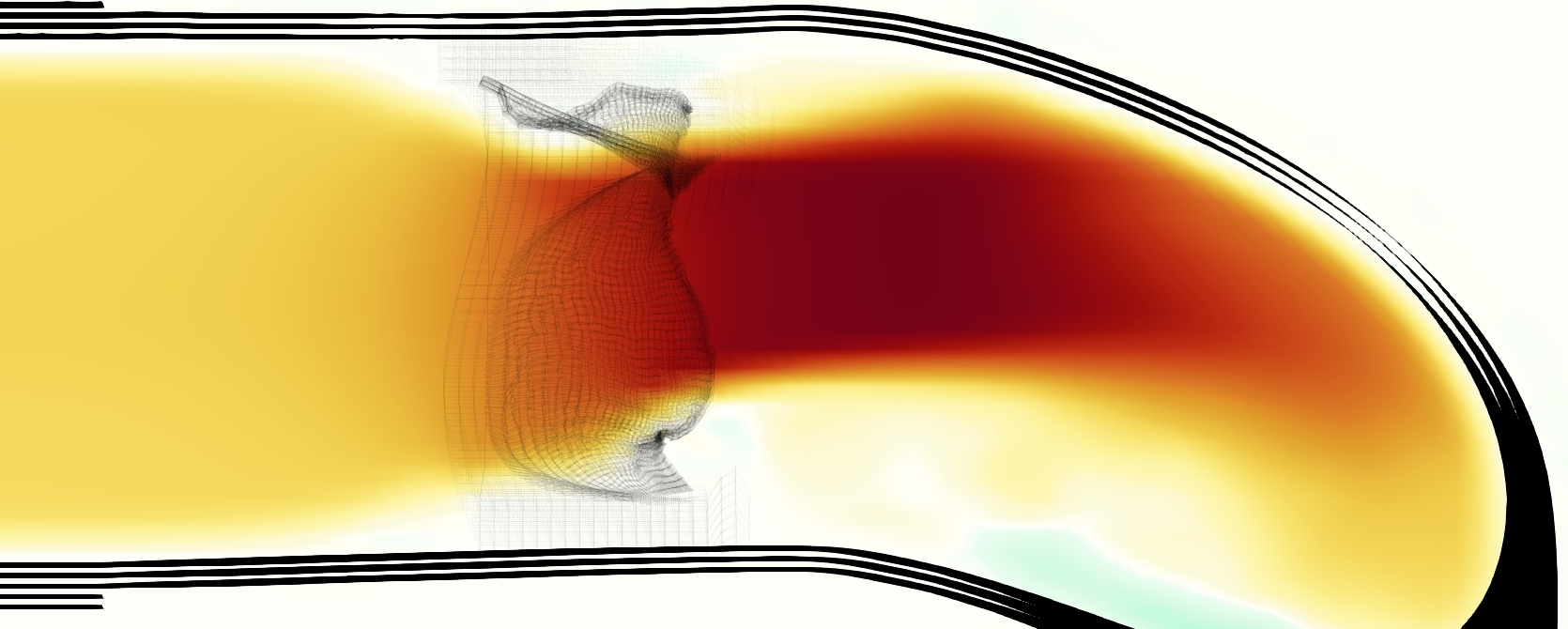} \\ 
\hline 
\rotatebox[origin=l]{90}{ 0.45 mm}&
\includegraphics[width=.14\textwidth]{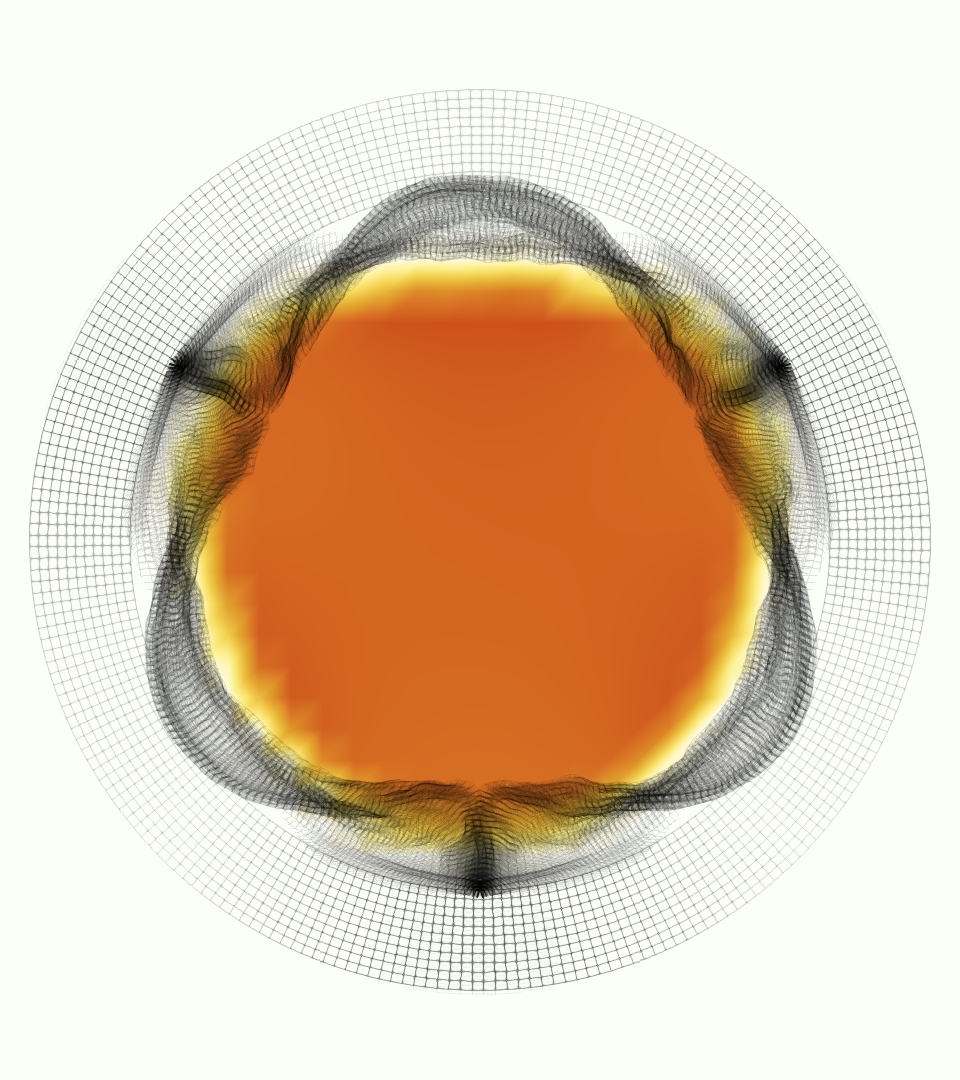} & 
\includegraphics[width=.14\textwidth]{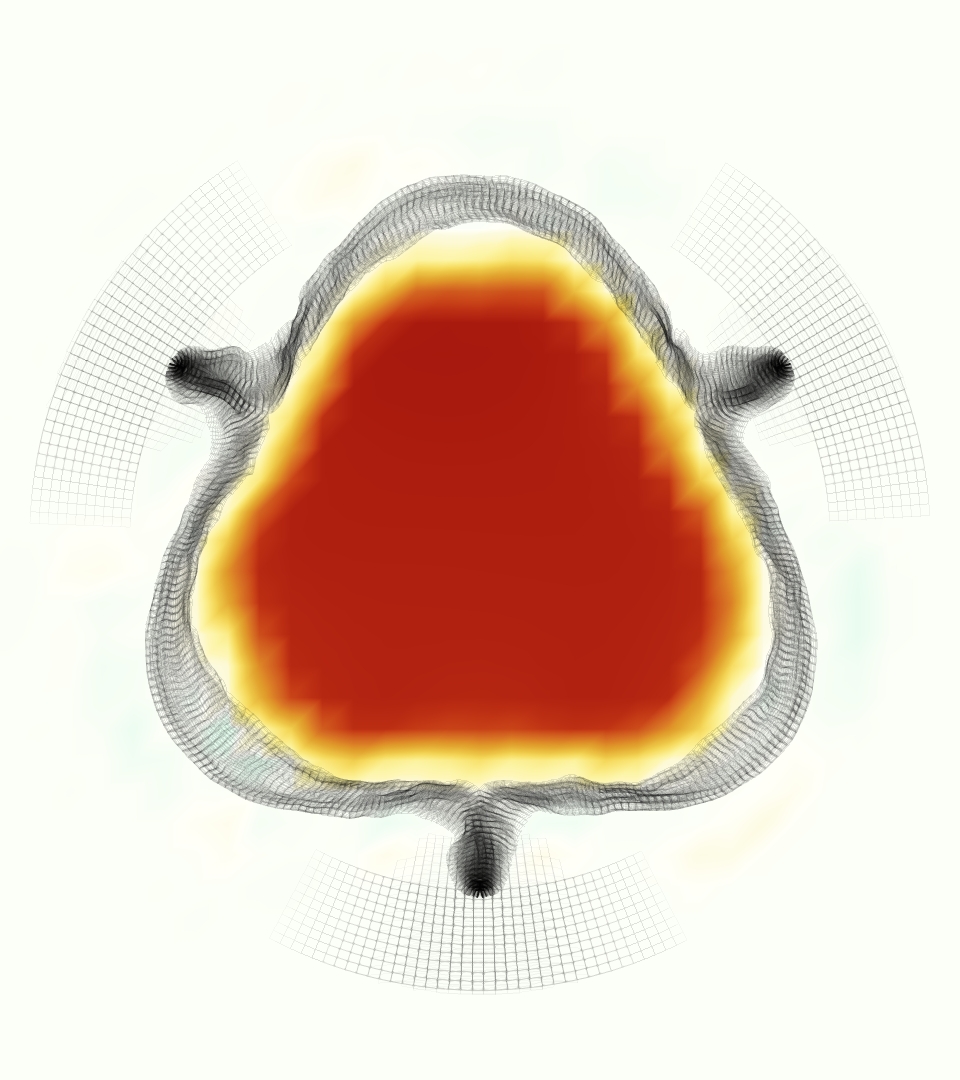} &
\includegraphics[width=.14\textwidth]{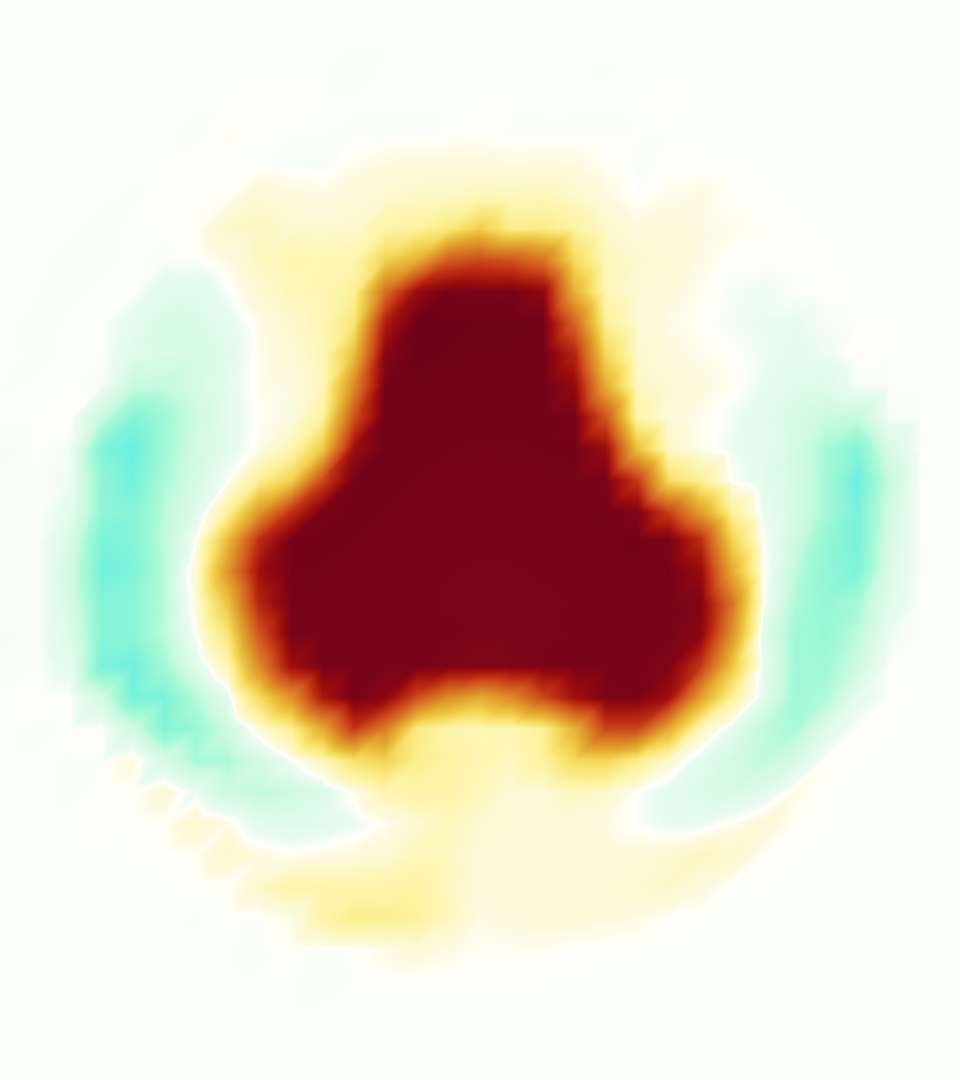} &
\includegraphics[width=.42\textwidth]{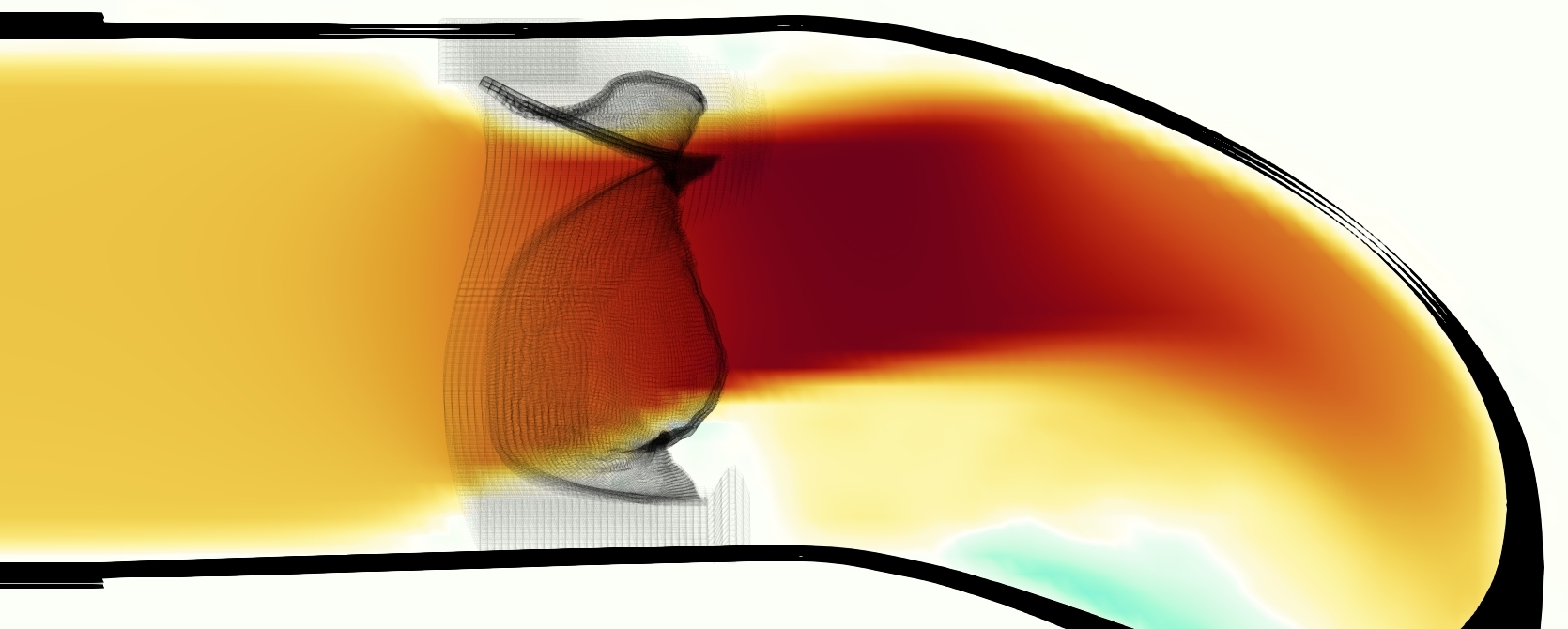} \\ 
\hline 
\rotatebox[origin=l]{90}{ 0.33  mm}&
\includegraphics[width=.14\textwidth]{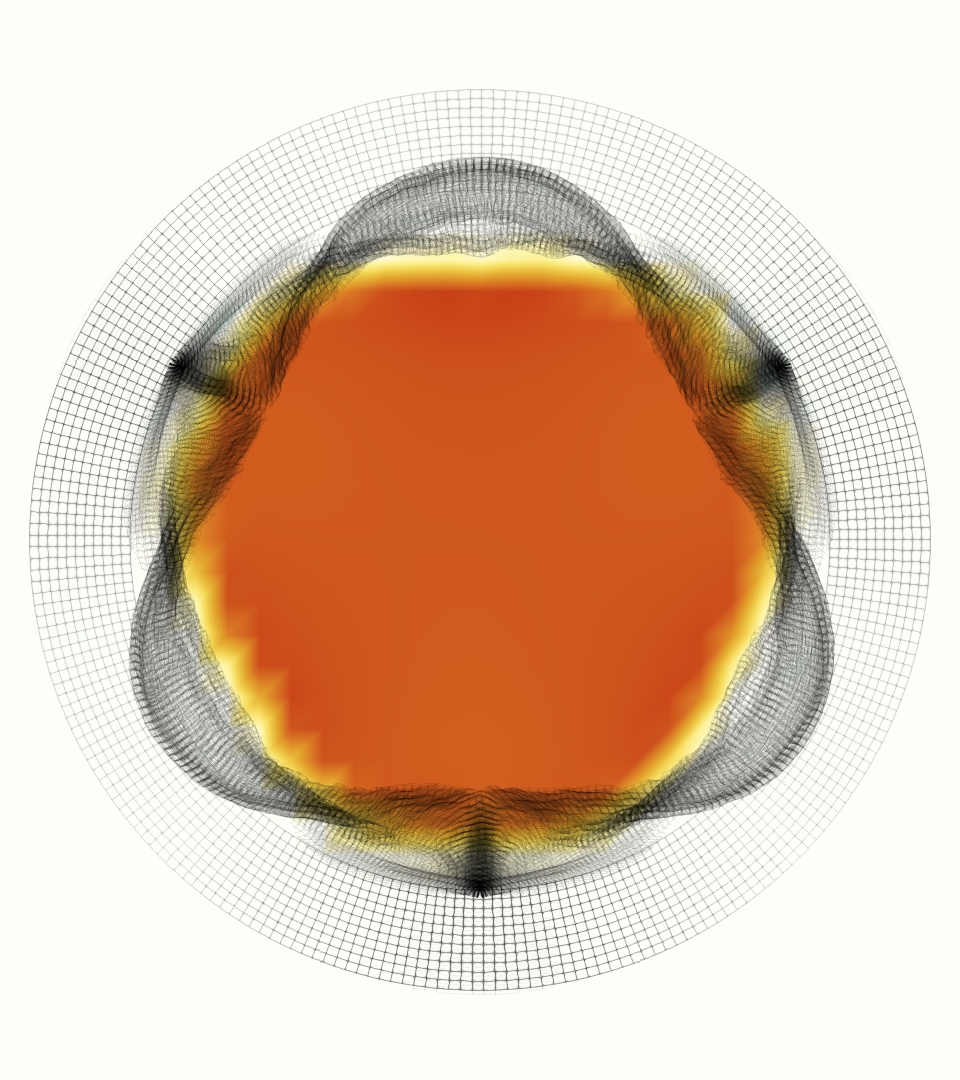} & 
\includegraphics[width=.14\textwidth]{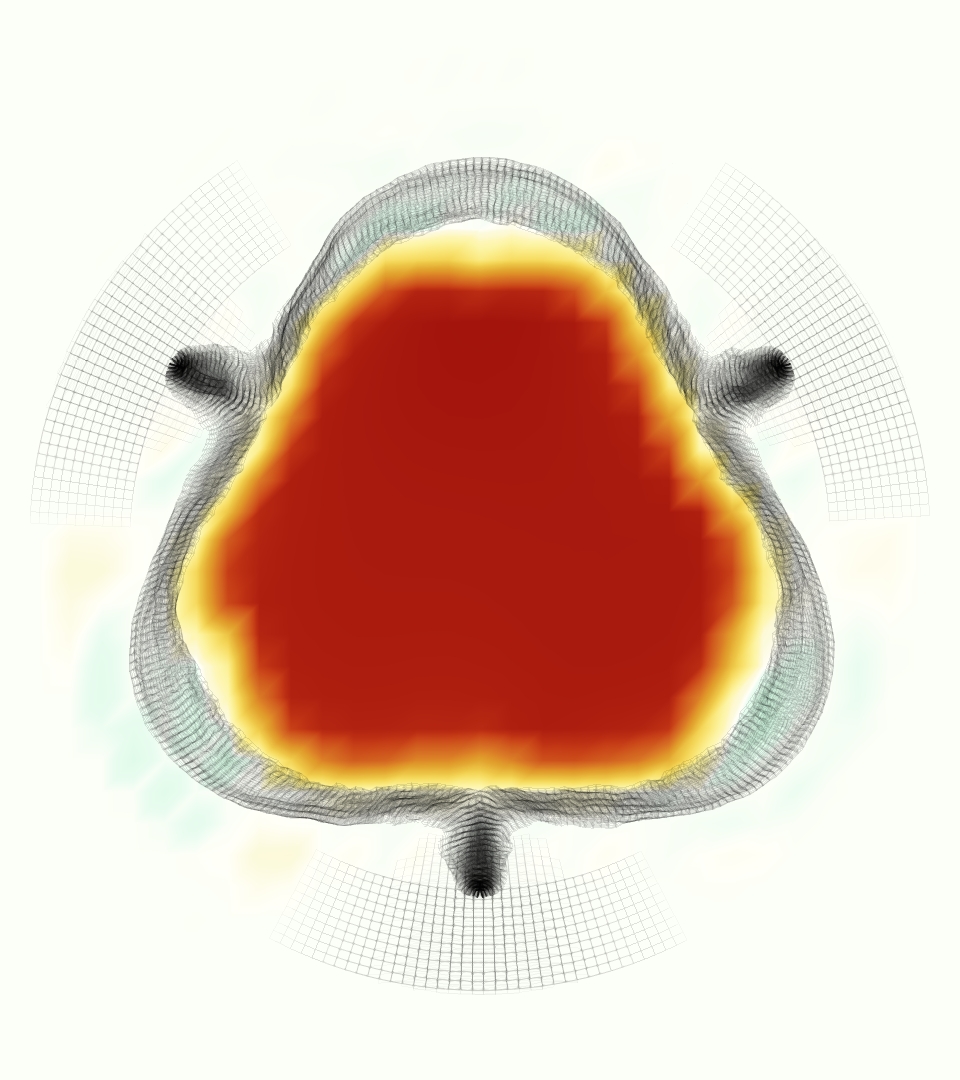} &
\includegraphics[width=.14\textwidth]{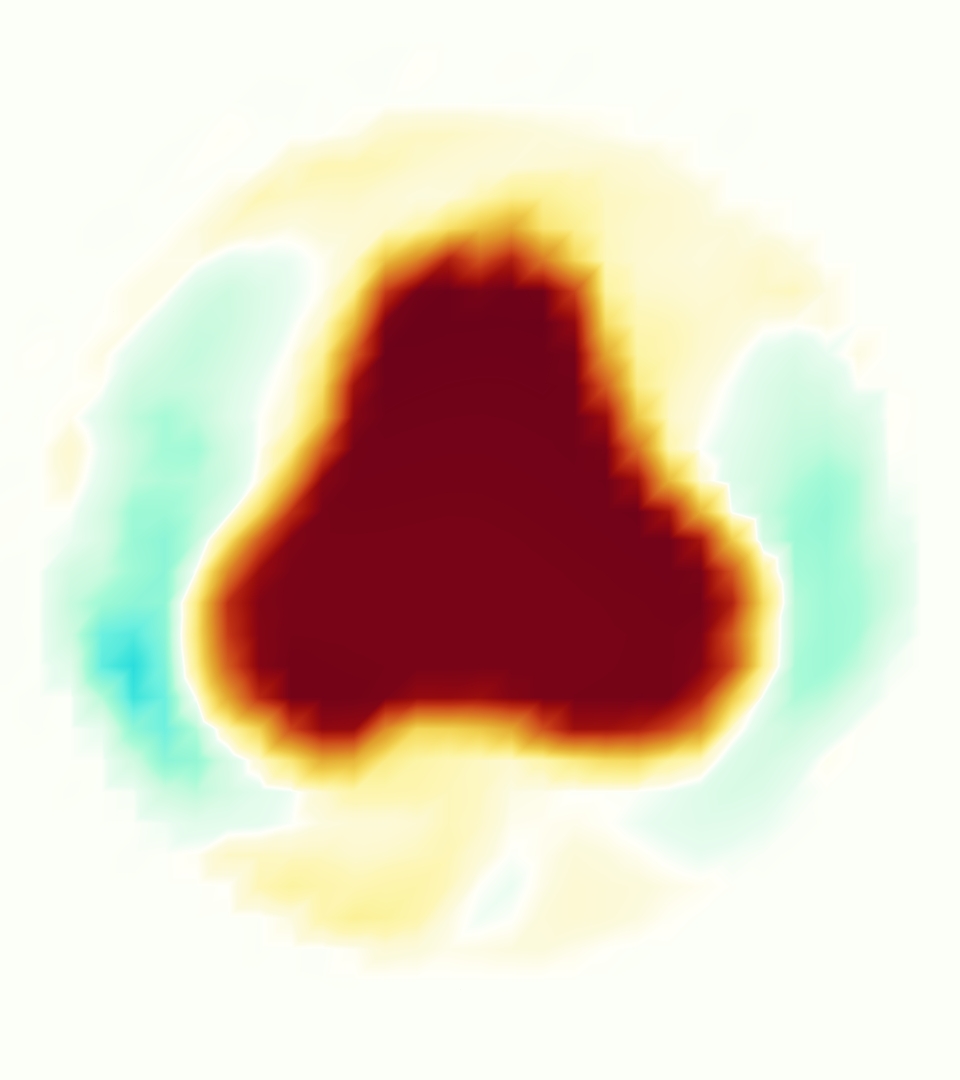} &
\includegraphics[width=.42\textwidth]{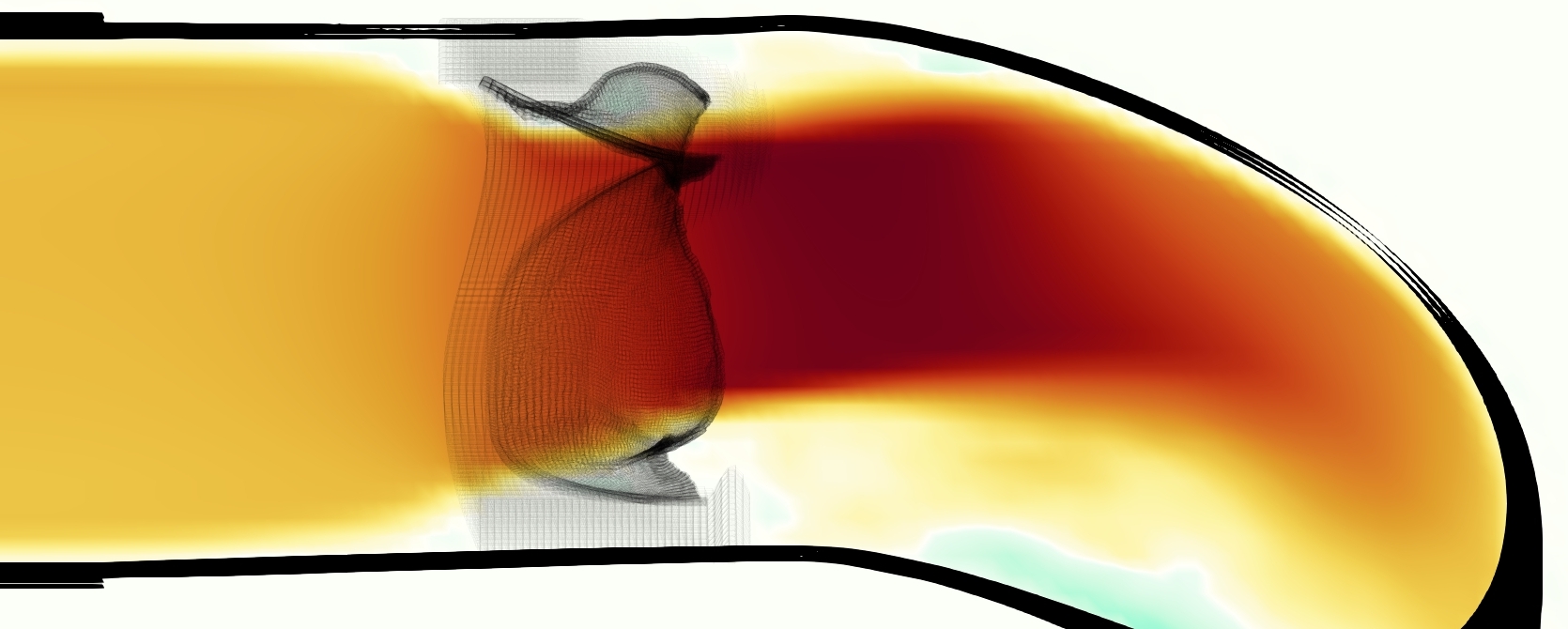} \\ 
\hline 
\rotatebox[origin=l]{90}{ 0.28  mm}&
\includegraphics[width=.14\textwidth]{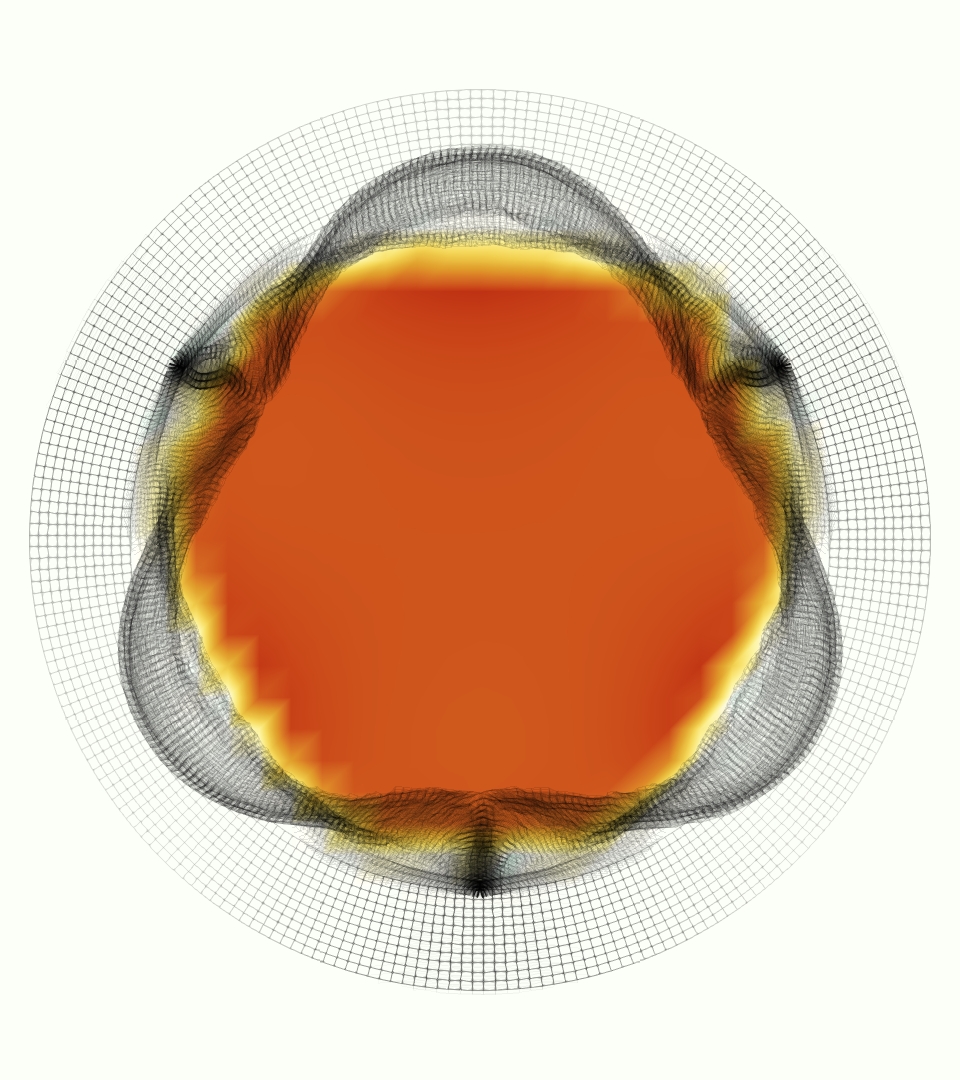} & 
\includegraphics[width=.14\textwidth]{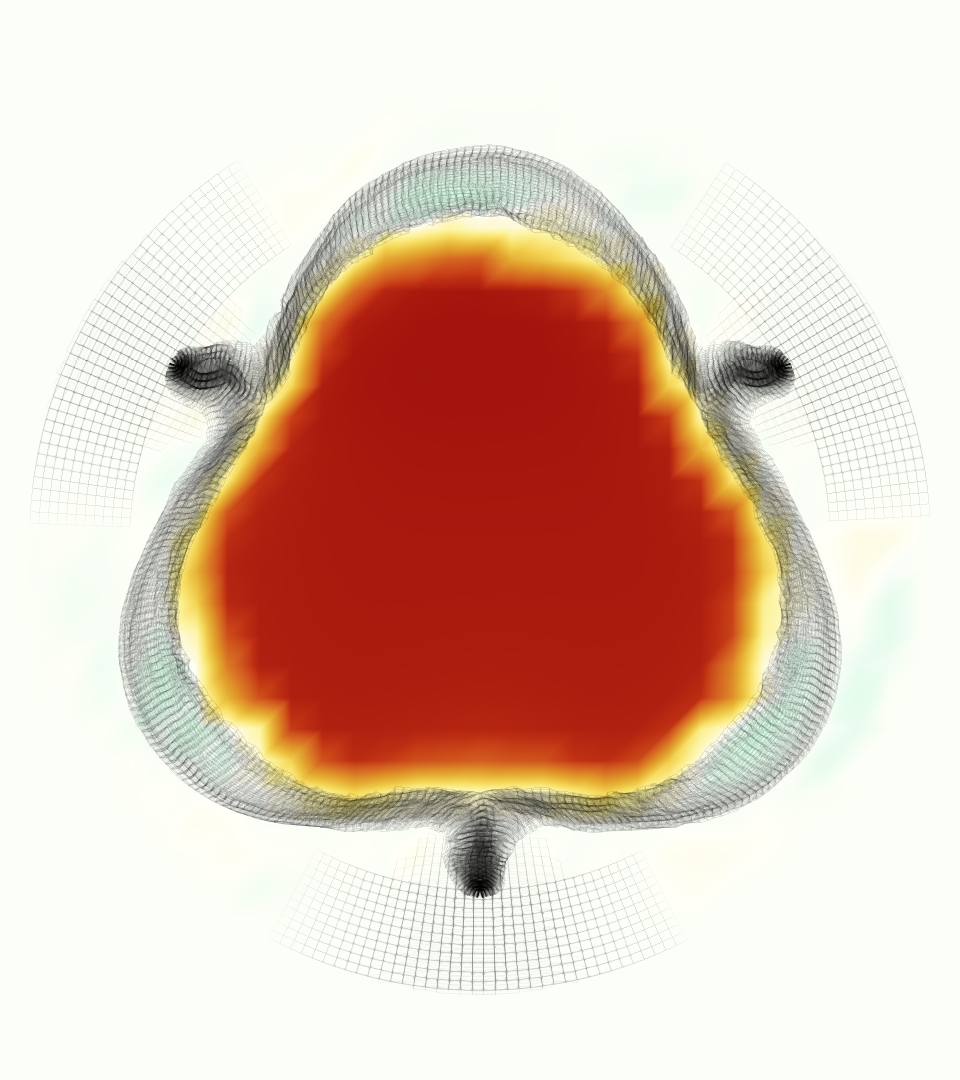} &
\includegraphics[width=.14\textwidth]{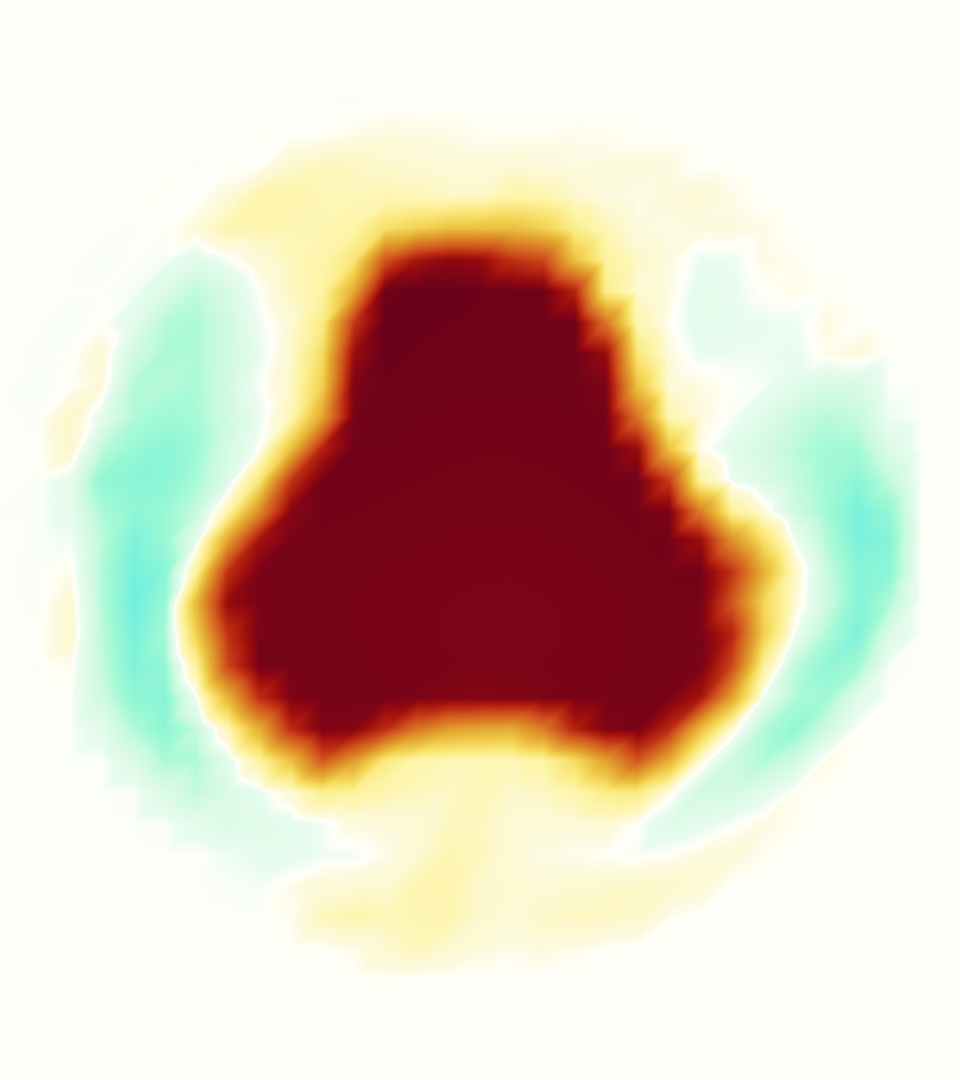} &
\includegraphics[width=.42\textwidth]{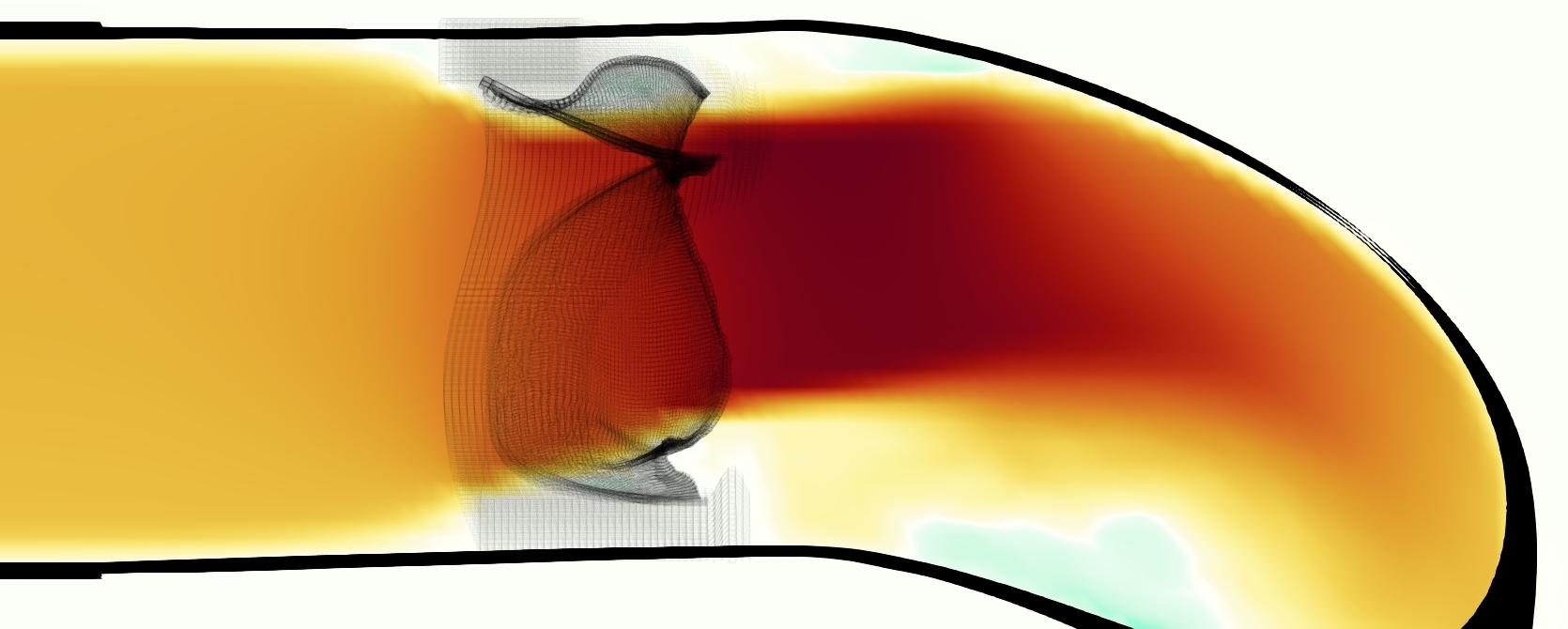} \\ 
\hline 
\end{tabular}
}
\end{center}

\caption{
Comparison of three axial view slices and one sagittal slice of phase-averaged velocity across multiple resolutions. 
}
\label{conv_figure}
\end{figure*}

\begin{figure*}[t!]  

\hfill \hfill \includegraphics[width=.25\textwidth]{colorbar.jpeg}

\begin{center}
{
\centering
\setlength{\tabcolsep}{2.0pt}     
\begin{tabular}{  c | c | c | c | c | }
& axial &  axial &   axial & sagittal  \\ 
& $x$ = 0 cm &  $x$ = 0.625 cm &   $x$ = 1.25 cm &   \\ 
\hline 
\rotatebox[origin=l]{90}{0.9  mm}&
\includegraphics[width=.14\textwidth]{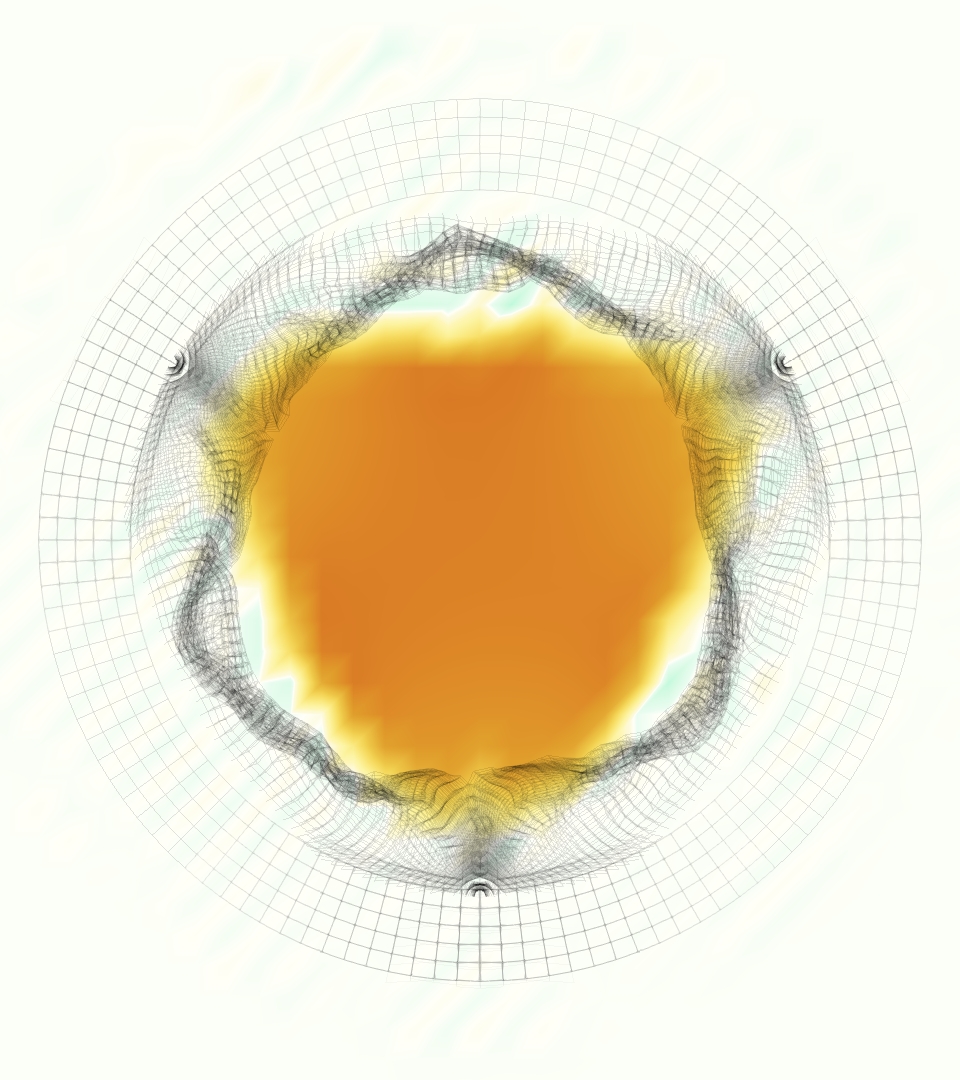} & 
\includegraphics[width=.14\textwidth]{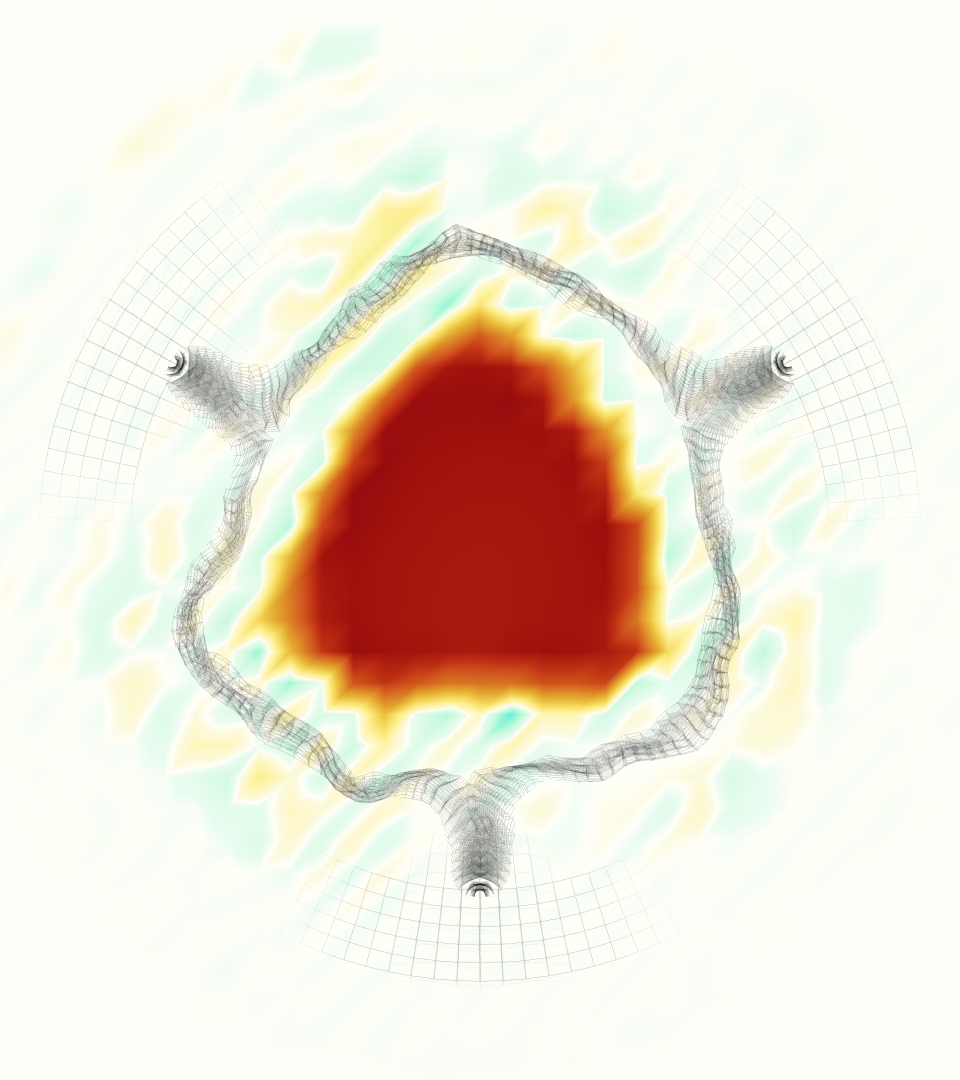} &
\includegraphics[width=.14\textwidth]{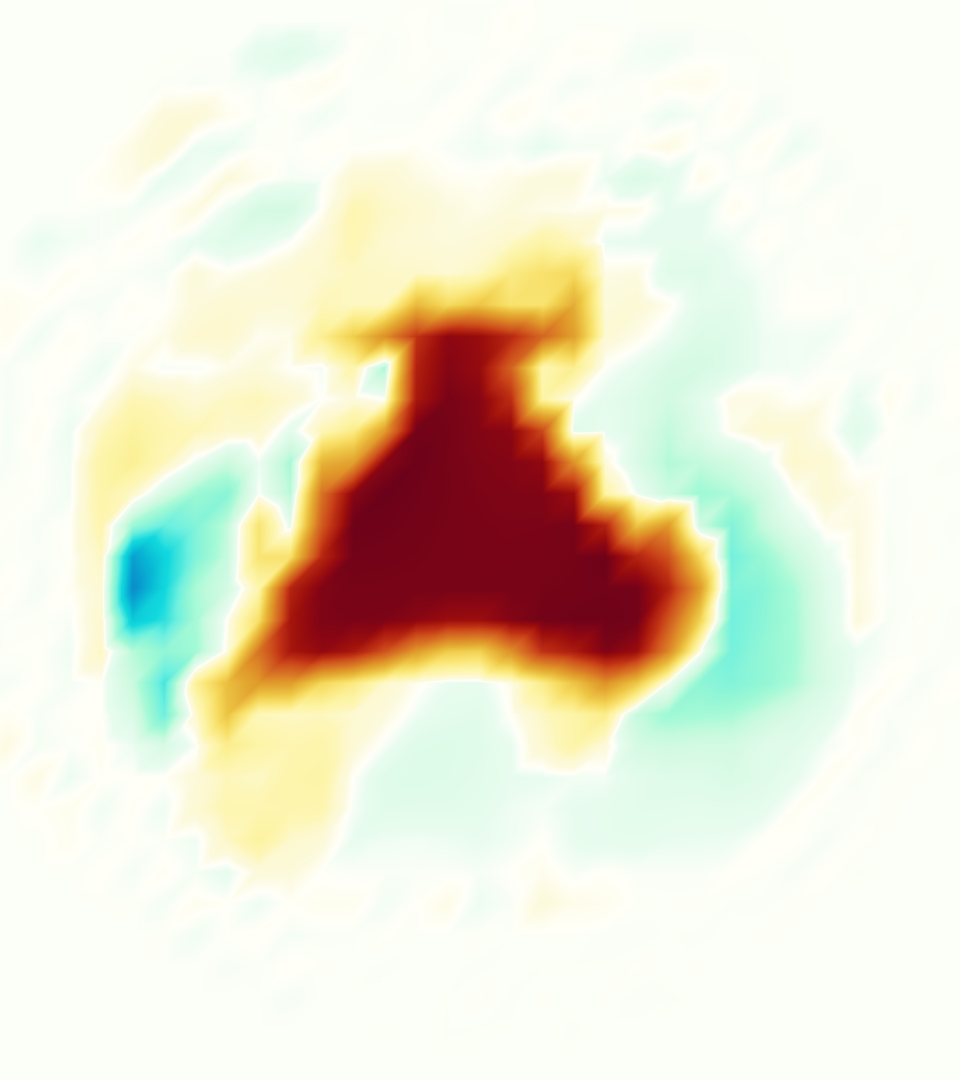} &
\includegraphics[width=.42\textwidth]{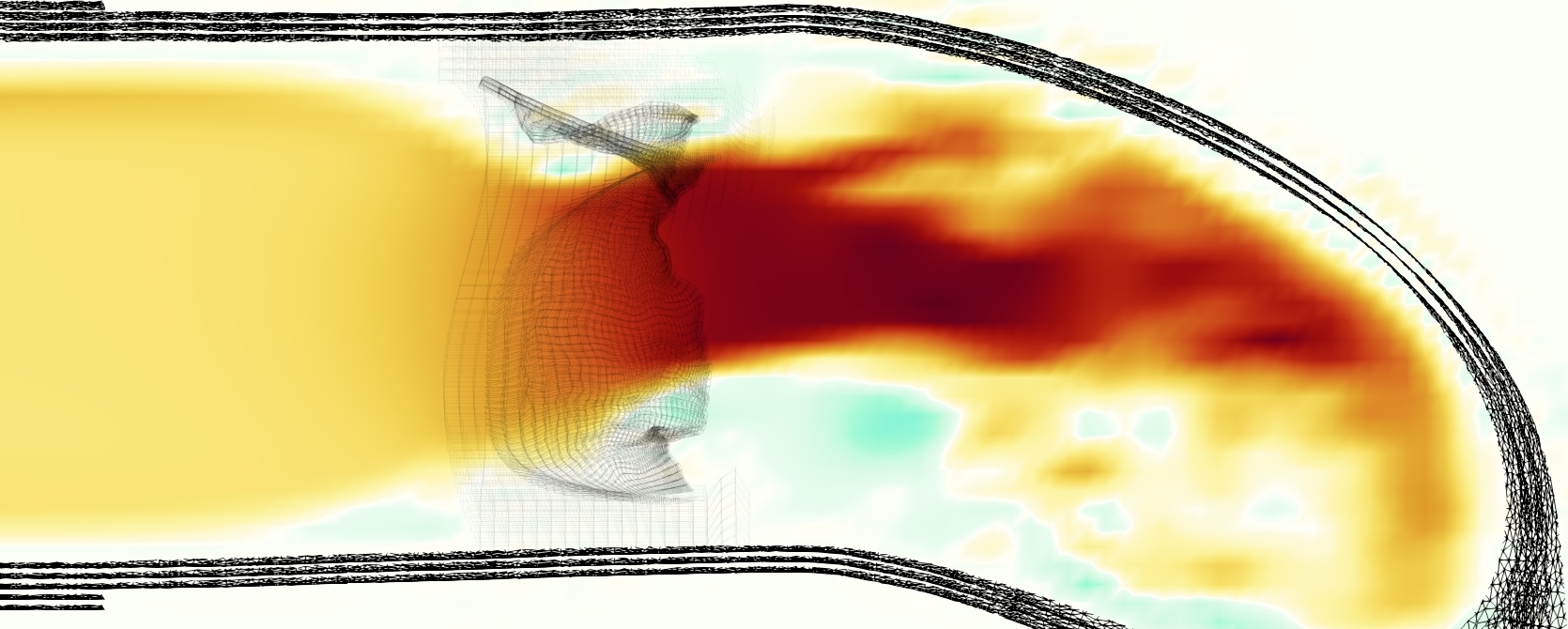} \\ 
\hline 
\rotatebox[origin=l]{90}{ 0.68  mm}& 
\includegraphics[width=.14\textwidth]{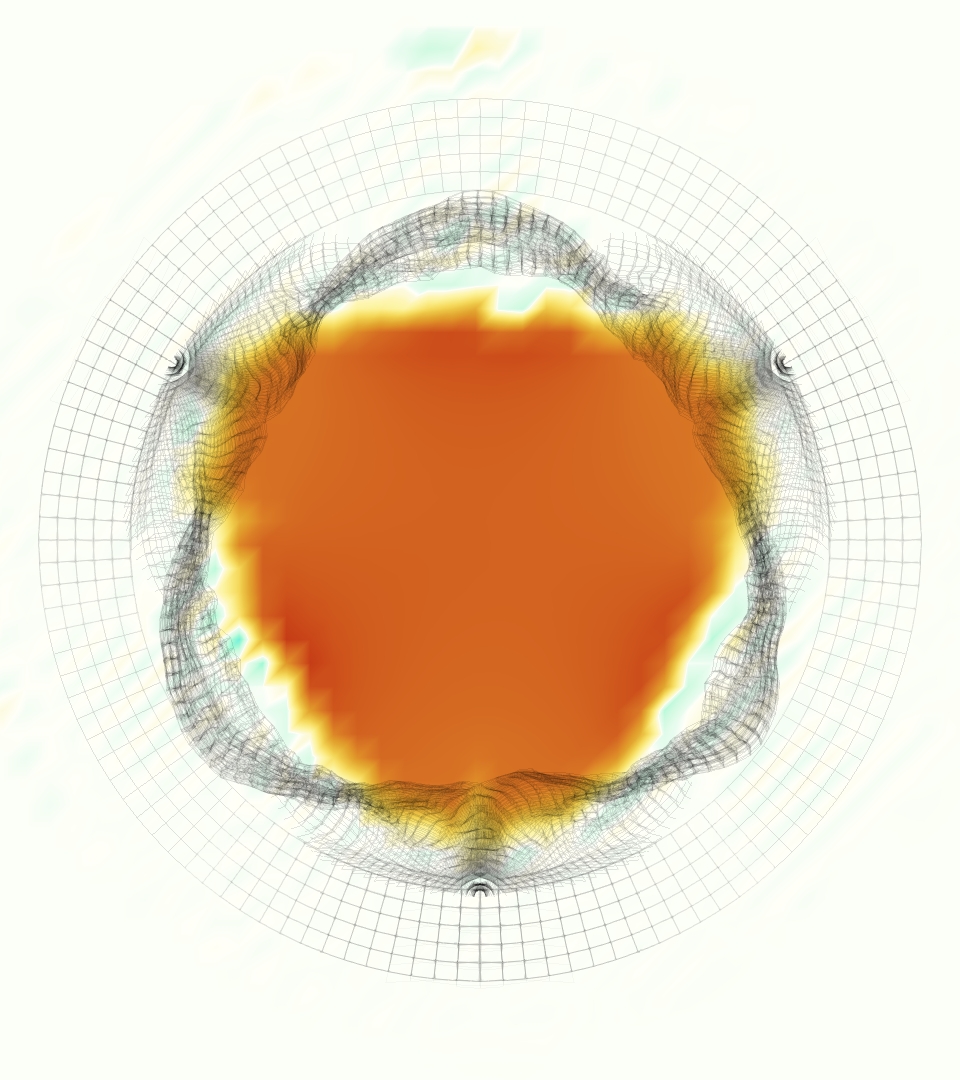} & 
\includegraphics[width=.14\textwidth]{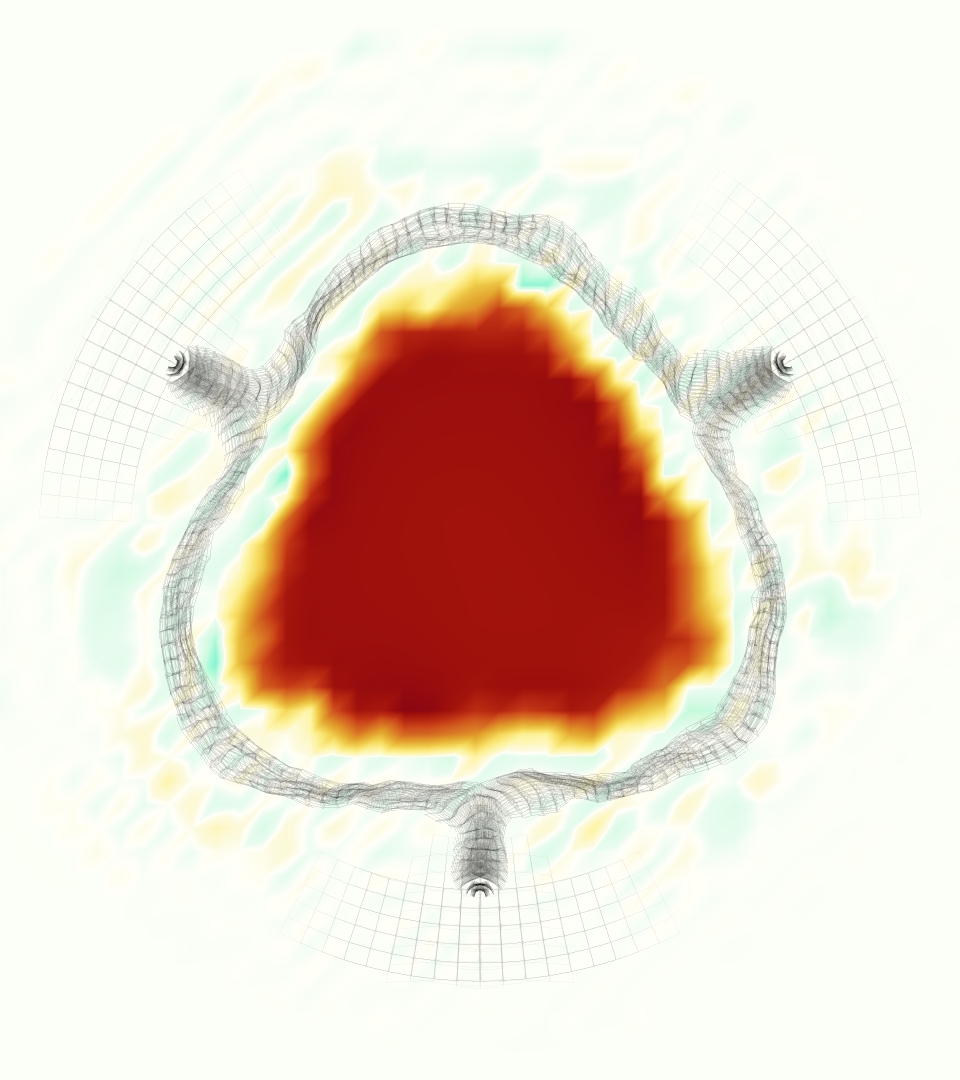} &
\includegraphics[width=.14\textwidth]{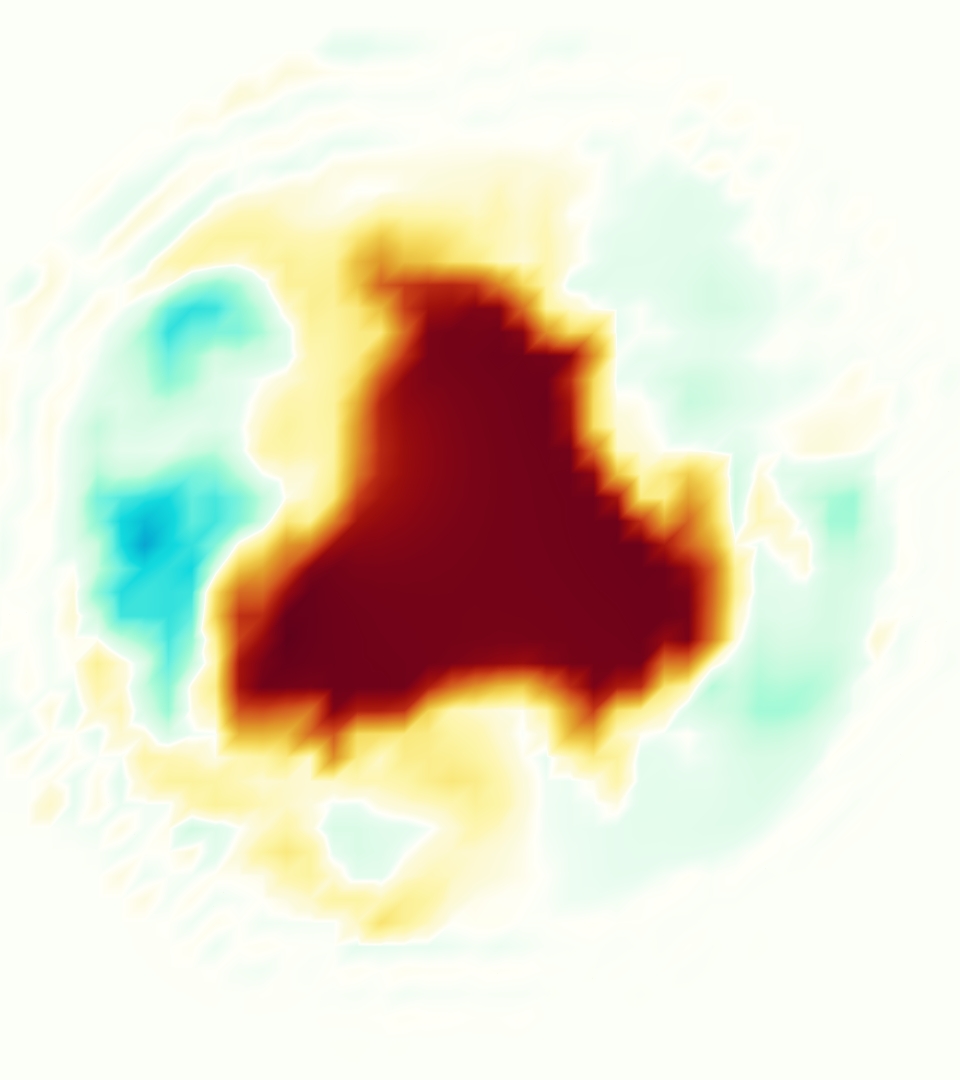} &
\includegraphics[width=.42\textwidth]{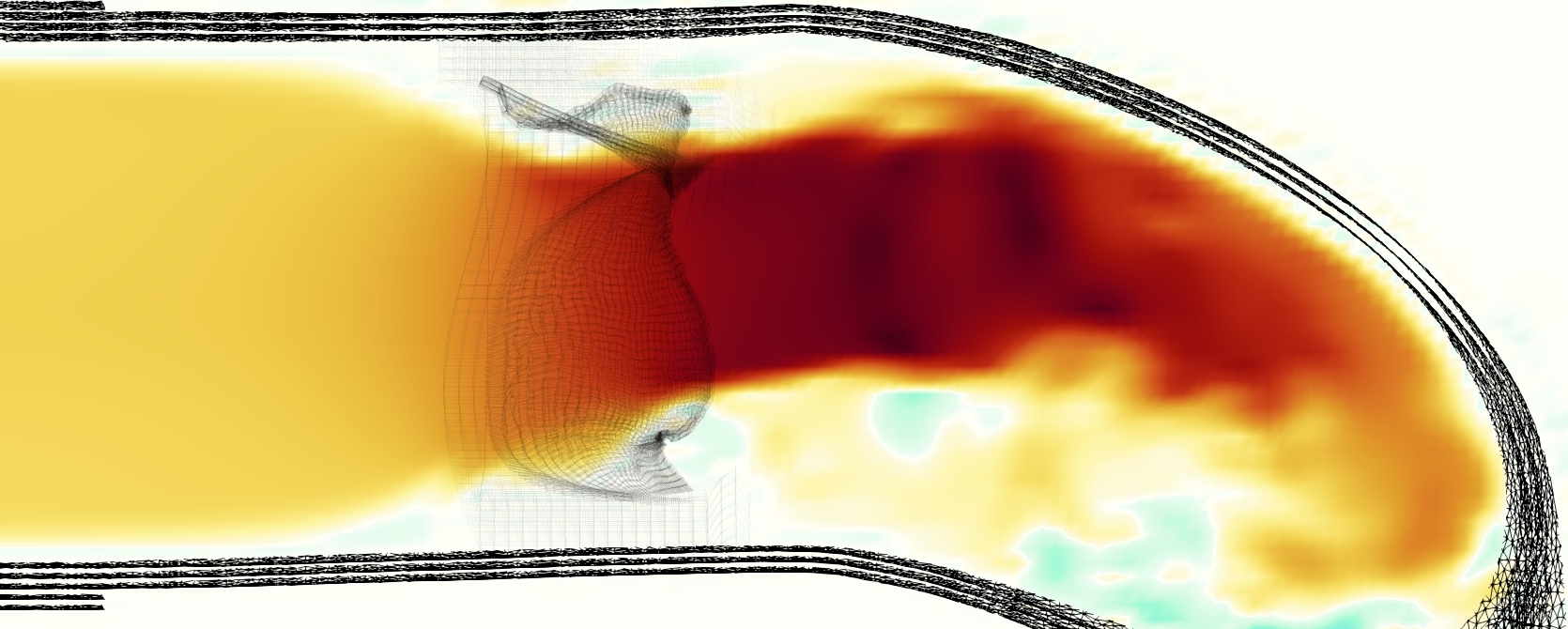} \\ 
\hline 
\rotatebox[origin=l]{90}{ 0.45 mm}&
\includegraphics[width=.14\textwidth]{axial_fine_x0_1350.jpeg} & 
\includegraphics[width=.14\textwidth]{axial_fine_xpt625_1350.jpeg} &
\includegraphics[width=.14\textwidth]{axial_fine_x1pt25_1350.jpeg} &
\includegraphics[width=.42\textwidth]{sagittal_fine1350_cropped.jpeg} \\ 
\hline 
\rotatebox[origin=l]{90}{ 0.33  mm}&
\includegraphics[width=.14\textwidth]{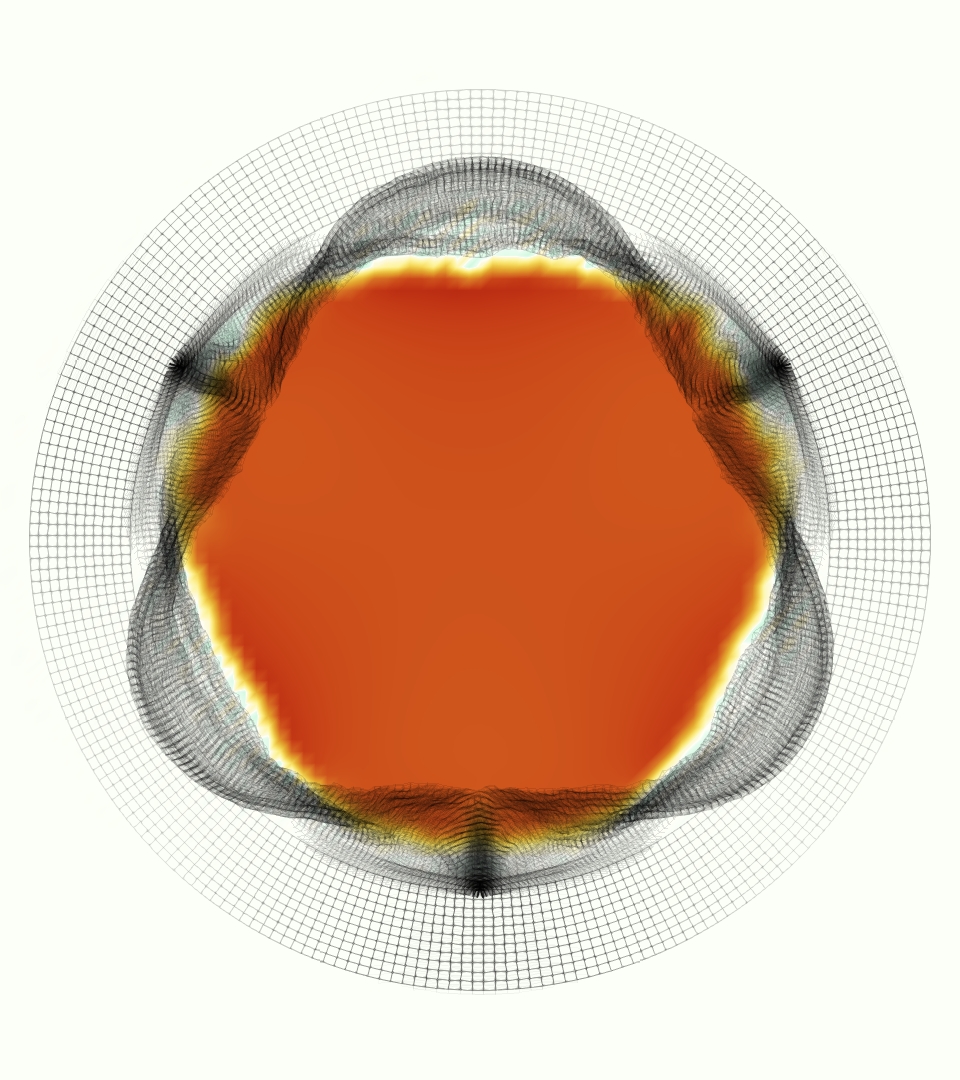} & 
\includegraphics[width=.14\textwidth]{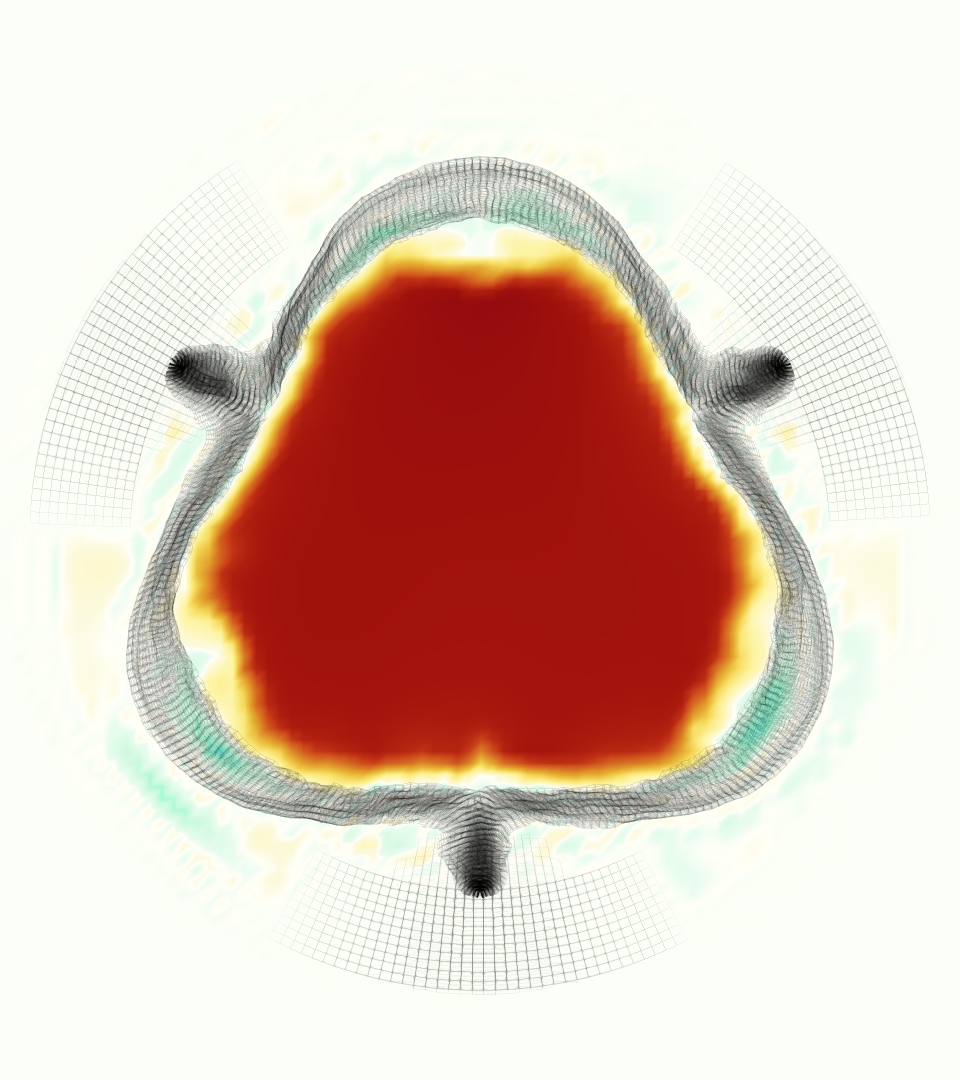} &
\includegraphics[width=.14\textwidth]{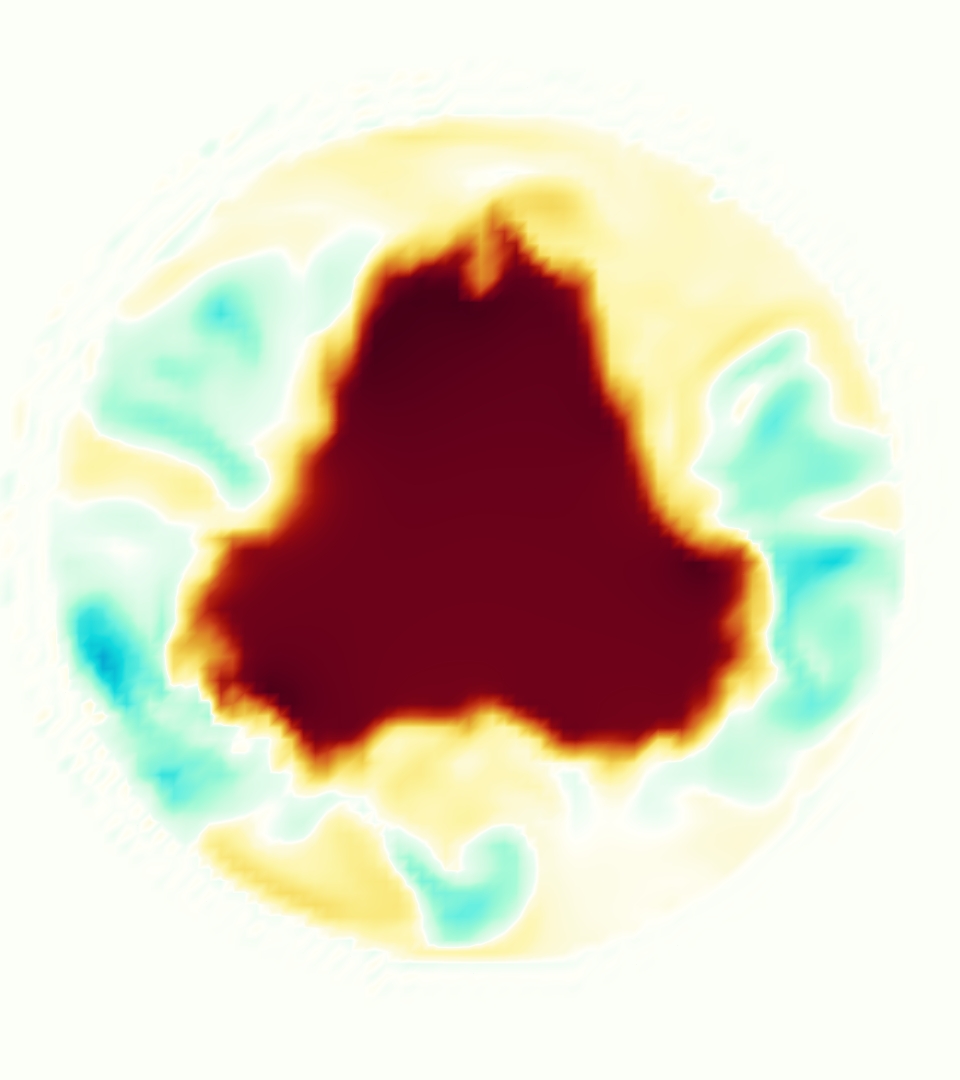} &
\includegraphics[width=.42\textwidth]{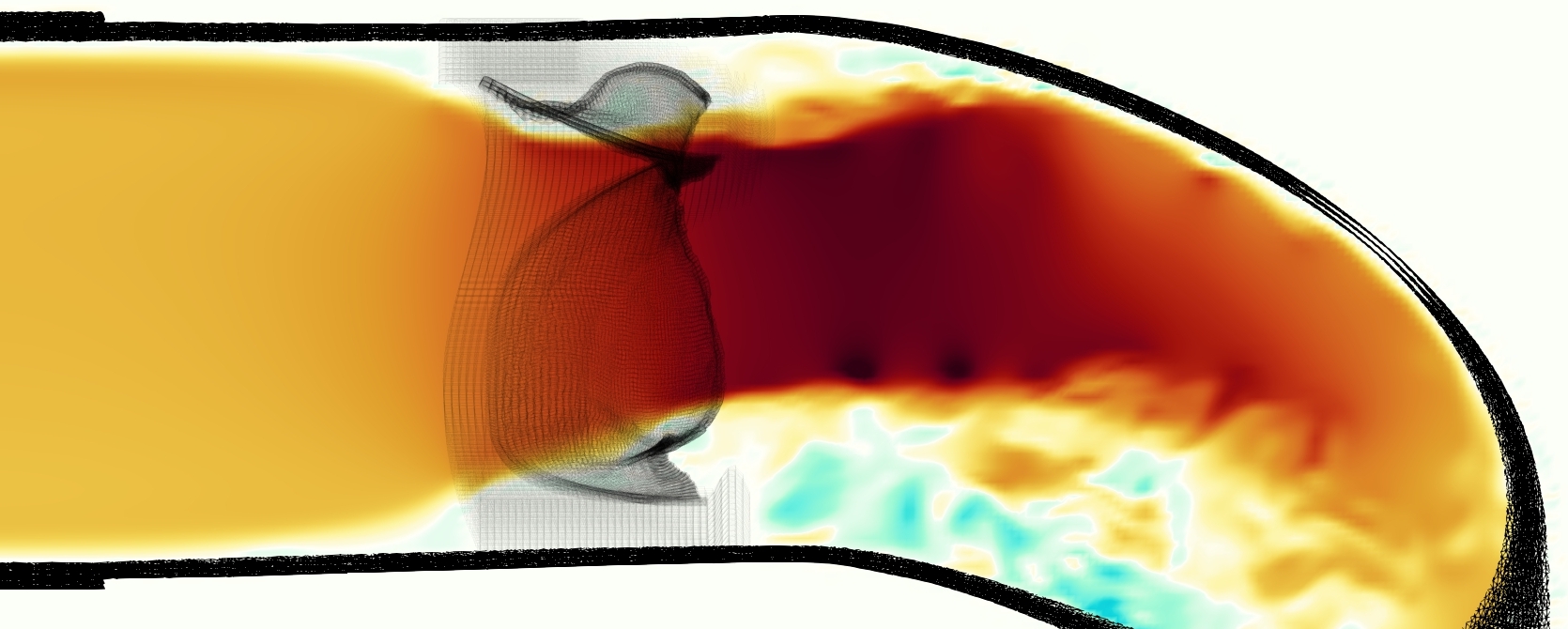} \\ 
\hline 
\rotatebox[origin=l]{90}{ 0.28  mm}&
\includegraphics[width=.14\textwidth]{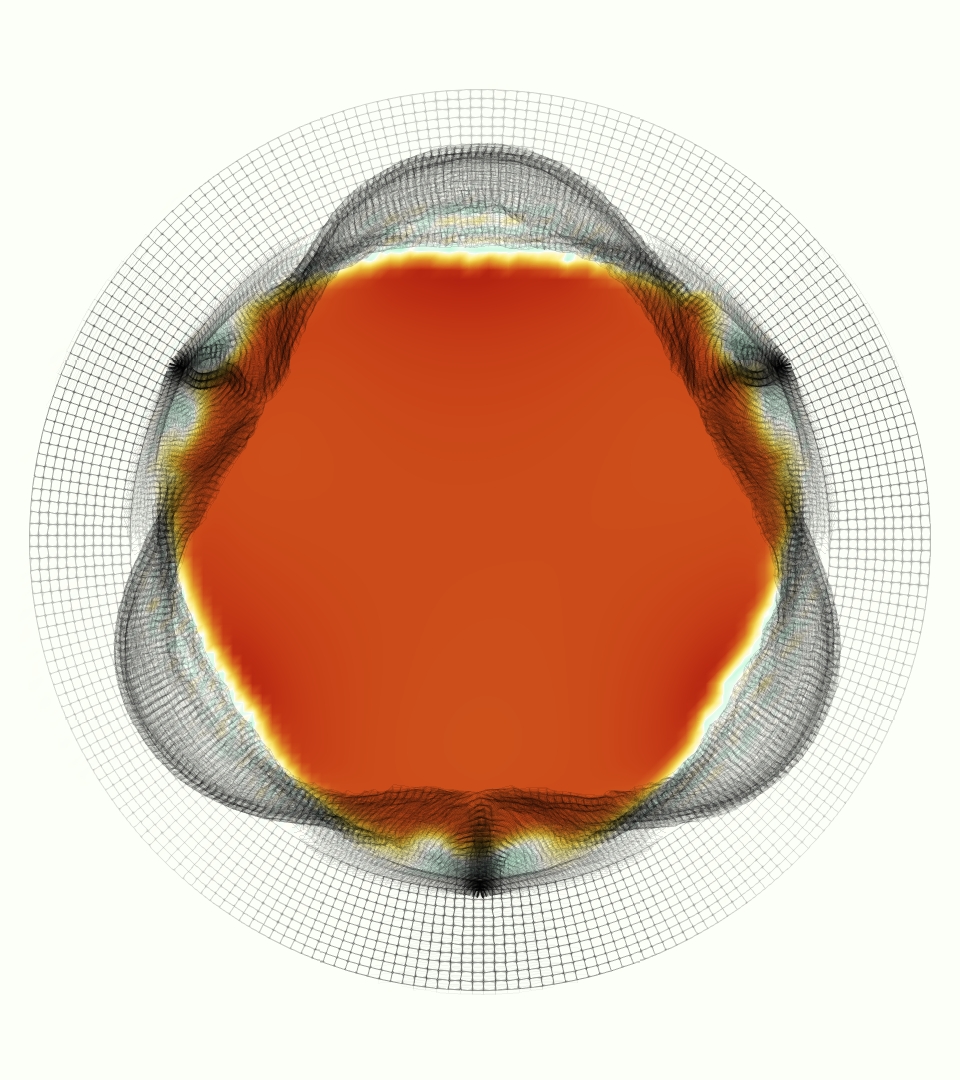} & 
\includegraphics[width=.14\textwidth]{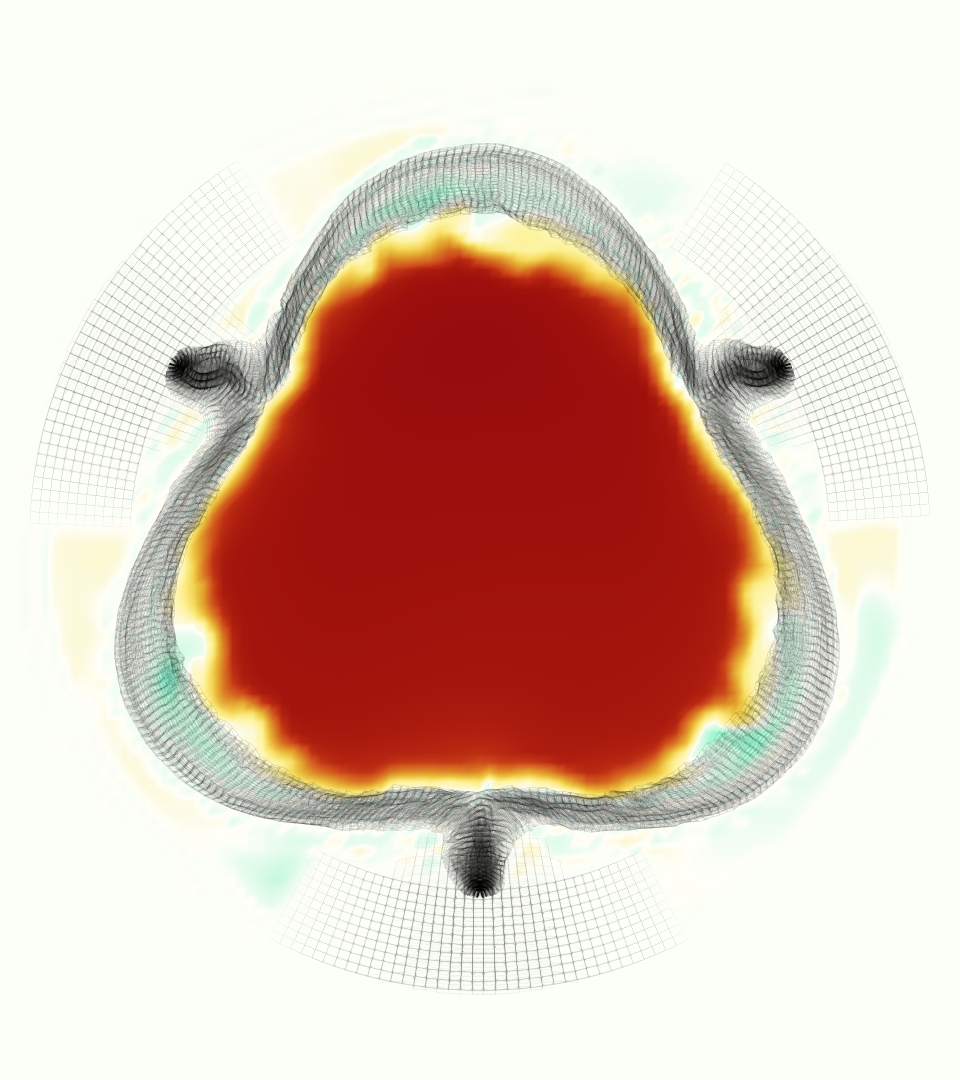} &
\includegraphics[width=.14\textwidth]{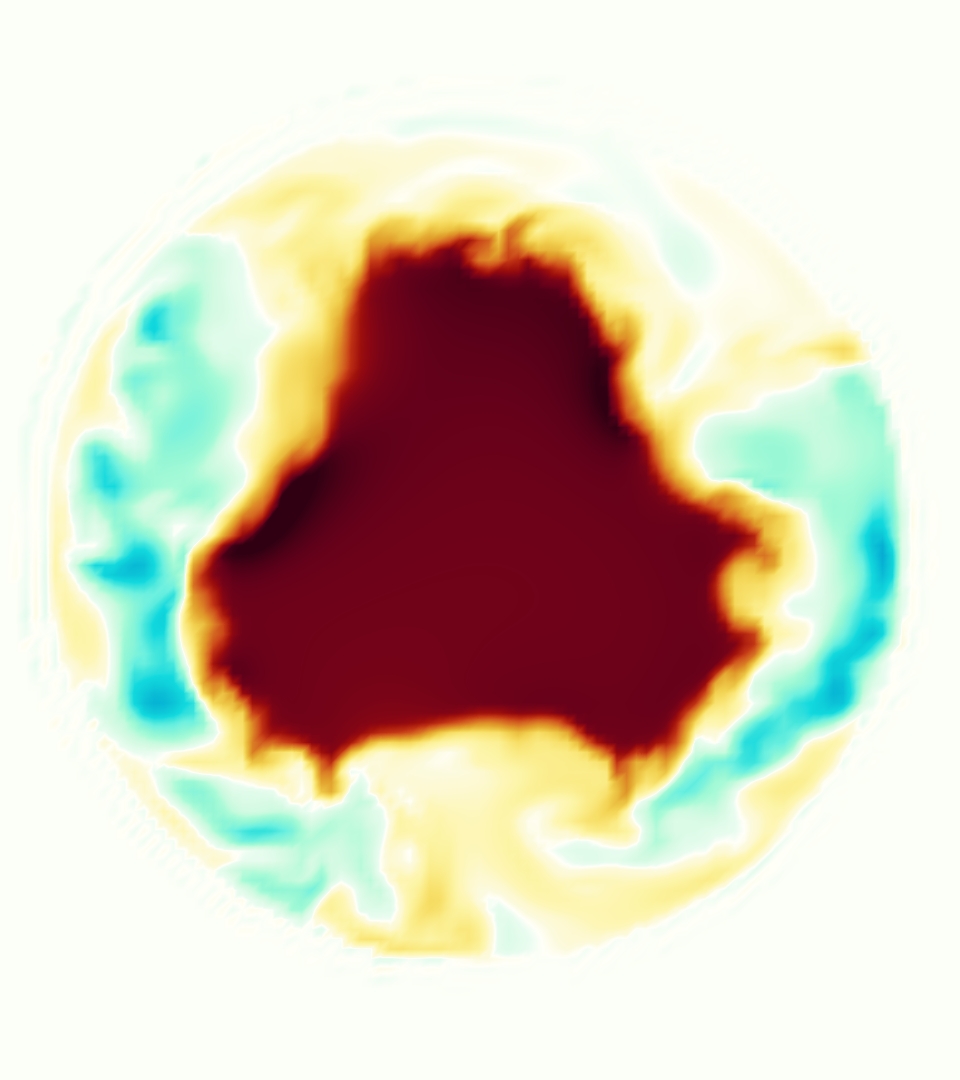} &
\includegraphics[width=.42\textwidth]{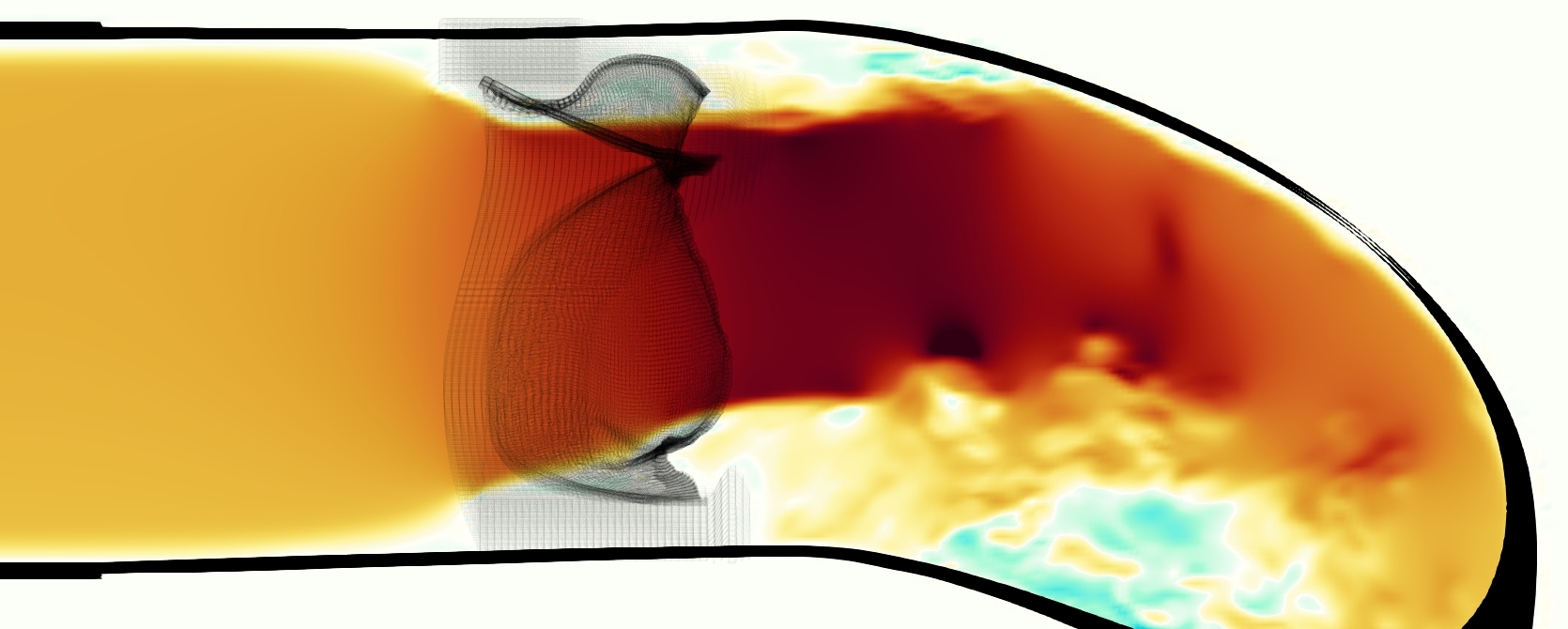} \\ 
\hline 
\end{tabular}
}
\end{center}

\caption{
Comparison of three axial view slices and one sagittal slice of instantaneous velocity across multiple resolutions. }
\label{conv_figure_instantaneous}
\end{figure*}

The phase-averaged, resampled velocity fields during peak systole and flow rates at each resolution are shown in Figure \ref{conv_figure}. 
Despite the limitations discussed above, we observe similar qualitative trends in the flow field at all resolutions. 
At all resolutions, a jet formed and angled up downstream of the valve orifice, as shown in the sagittal view. 
The jets showed a triangle-like cross section at $x = 0$ with points aligned with the commissures. 
At $x = 0.625$ cm, the jet appears like a rounded triangle in the opposing orientation, with its points aligned with the center of the leaflets. 
At $x = 1.25$ cm, the jet is narrower downstream of the commissures, and wider downstream of the leaflets, again with a triangle-like cross section. 
The area of the jet increased with resolution, as expected given the IB method thickening of the valve structure. 
The narrowed jets at the two more coarse resolutions show locally elevated velocities relative to the two more fine resolutions. 
Figure \ref{conv_figure_instantaneous} shows the instantaneous velocity fields at each resolution in the same axial and sagittal views. 
At $\Delta x = $ 0.9 mm, the sagittal view shows a qualitatively different jet than at finer resolutions, with regions of lower velocity farther from the vessel wall, indicating insufficient resolution. 
At $\Delta x = $ 0.9 and 0.68 mm, the jet is visibly narrowed compared to higher resolutions. 
While some features are similar at these two coarse resolutions, we conclude that the narrower jets indicate these simulations are under-resolved. 
Flows in the three finest resolutions, $\Delta x = $ 0.45, 0.34 and 0.28 mm. appeared qualitatively similar, with slightly more fine structure detail in both the axial and sagittal views present at the edges of the jet. 
The jets in the axial views all showed a similar triangle-like cross section, slightly narrower downstream of the commissures, as in the phase-averaged fields. 
In both the phase-averaged and instantaneous fields, the three finest resolutions appear sufficiently similar that the conclusions of this study would be identical with any of these resolutions.

\begin{figure}[H]
\begin{center}
\includegraphics[width=.48\textwidth]{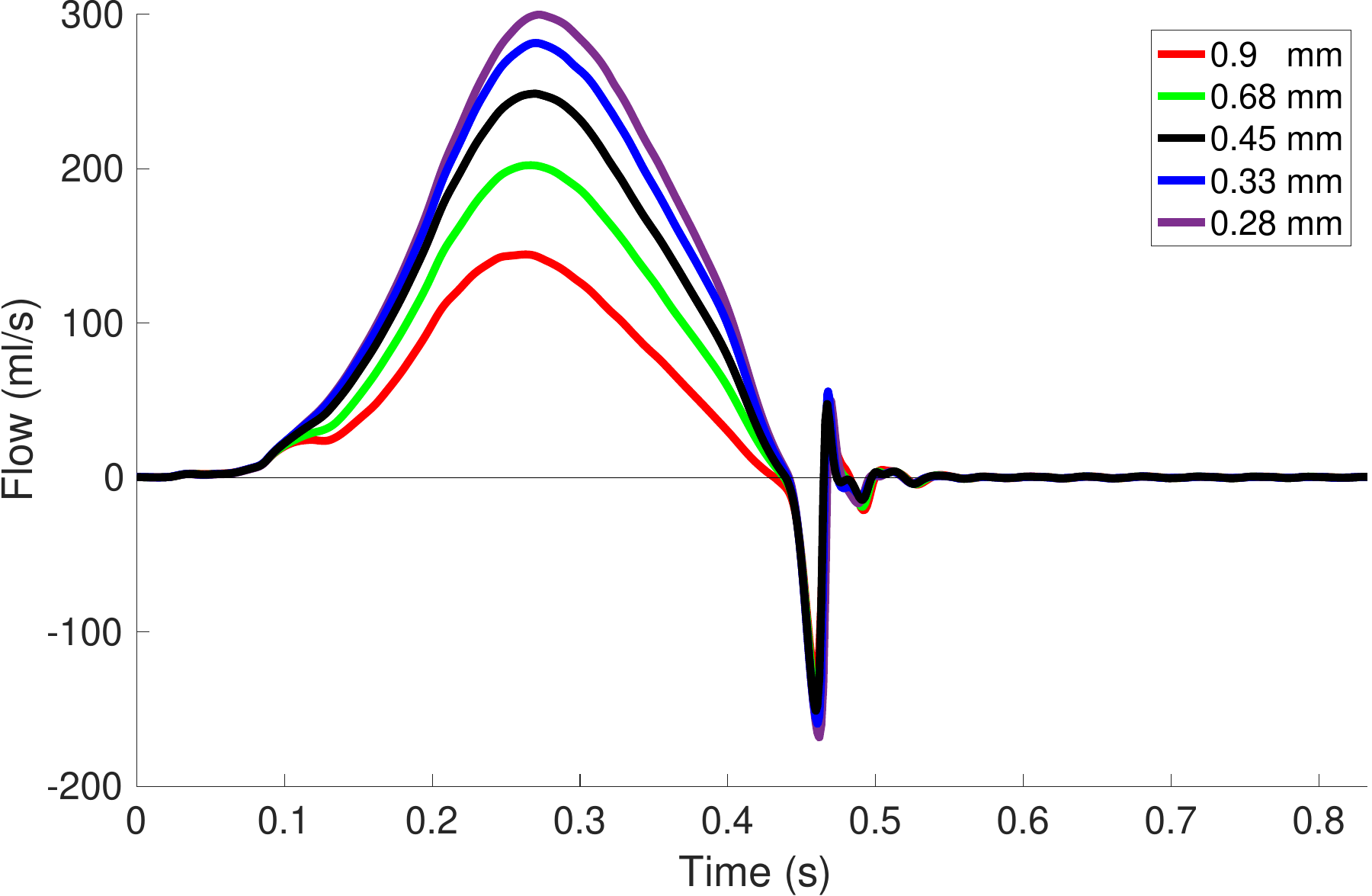}
\end{center}
\caption{Flow rates through the right ventricular inlet with various resolutions. }
\label{conv_figure_flows}
\end{figure}

The flow rates, depicted in Figure \ref{conv_figure_flows}, increase with increasing resolution, also as expected given the increase in effective valve orifice area and radius. 
The maximum flow rates at $\Delta x = $ 0.45 and 0.34 mm were 248.6 and 281.4 ml/s, respectively, a relative difference of 11.7\%. 
The mean flow rates at $\Delta x = $ 0.45 and 0.34 were 55.5 and 63.2 ml/s, respectively, a relative difference of 12.4\%. 
The maximum flow rates at $\Delta x = $ 0.34 and 0.28 mm were 281.3 and 299.8 ml/s, respectively, a relative difference of 6.1\%. 
The mean flow rates at $\Delta x = $ 0.34 and 0.28 mm were  63.2 and 67.4 ml/s, respectively, a relative difference of 6.2\%. 
Thus the flow rates show signs of converging with increasing resolution, though some minor differences remain.

\begin{figure}[H]
\begin{center}
\includegraphics[width=.48\textwidth]{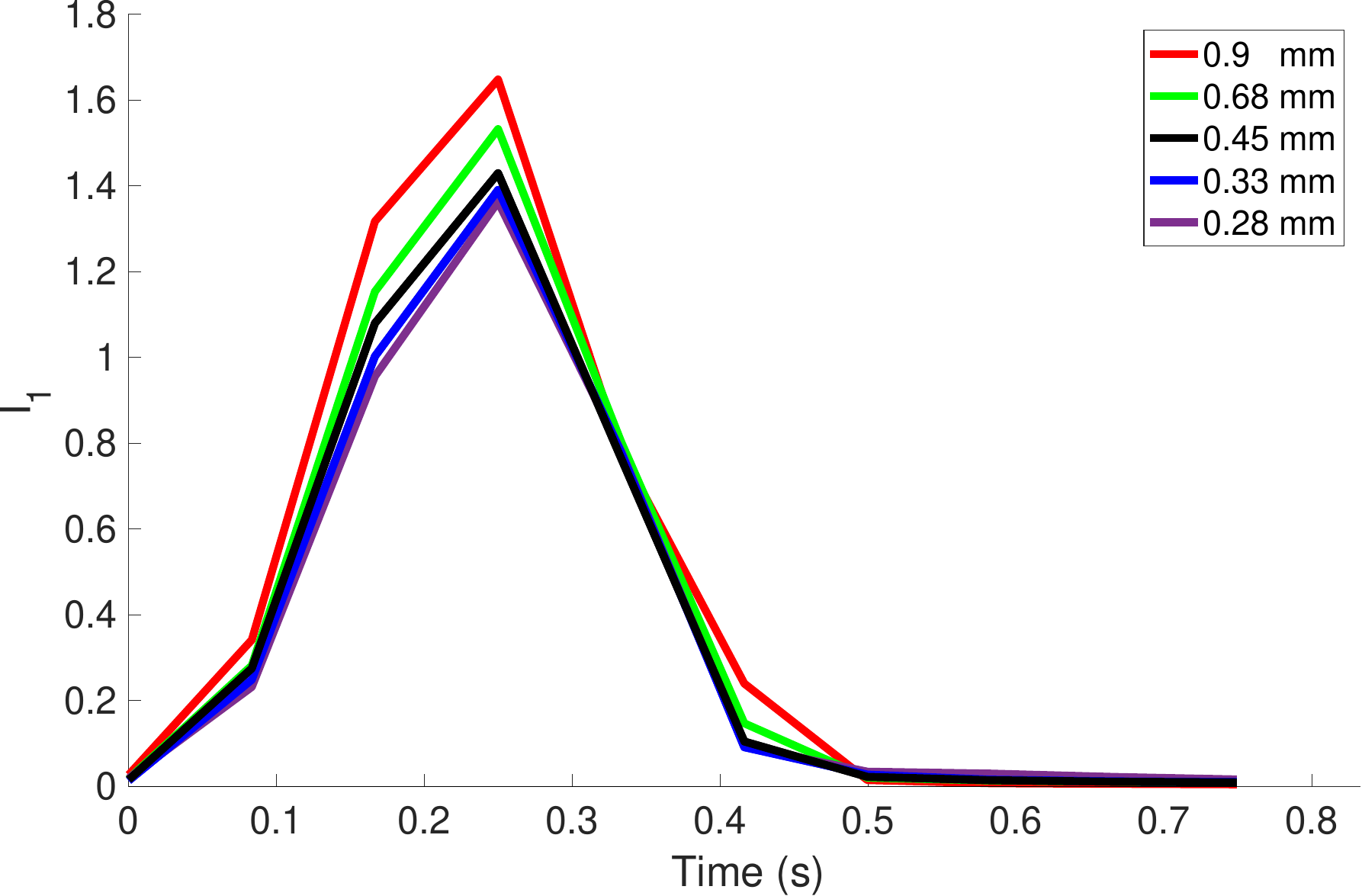}
\end{center}
\caption{ The integral metric $I_{1}$ evaluated at the origin with various resolutions. }
\label{conv_figure_I1}
\end{figure}

Figure \ref{conv_figure_I1} shows the integral metric $I_{1}$ evaluated at $x = 0$ cm with various resolutions, where each resolution uses its own velocity scale (see eqn. \eqref{velocity_scale}).
Values at $\Delta x = $ 0.9, 0.68 were elevated relative to finer resolutions, indicating under resolution. 
Values of $I_{1}$ at $\Delta x = $ 0.45, 0.34 and 0.28 mm are extremely similar, with slight decrease as resolution increases. 
Values of $I_{1}$ at $x = $ 0.625 and 1.25 cm showed similar trends and are not shown.

Given the overall qualitative similarity and relative changes in flow rates, we believed all conclusions in this work would be similar with $\Delta x = $ 0.45, 0.34 or 0.28 mm. 
We therefore selected  $\Delta x = $ 0.45 mm, twice the MRI resolution, for the main portion of this study.

Additionally, we compared averaging the second cycle with phase averaging the second, third and fourth cycles with a resolution of $\Delta x = 0.45 $ mm. 
We found the fields appeared qualitatively similar in both cases.  
Averaging the second, third and fourth cycles resulted in slightly worse relative error compared to experiment than the second alone. Comparing the two averaging methods at peak systole, relative error on velocity magnitude, the $x$-component of velocity was 0.07, and 0.09,  
respectively in $L^{1}$, and 0.10 and 0.11 in $L^{2}$. 
As these differences were substantially lower than relative differences between the simulation results and experiments, we ended all further simulations after the second cycle and use only the second cycle for phase averaging.

\clearpage

{\small
\bibliographystyle{acm}
\bibliography{ib_comparison_refs.bib}
}

\end{document}